%% file: AA_main.tex
\PassOptionsToPackage{dvipsnames,svgnames,x11names}{xcolor}
\documentclass[12pt,letterpaper]{article}
\usepackage{natbib}
\usepackage[margin=0.9in]{geometry}
\usepackage{setspace}
\usepackage{titling}
\include{preamble}

\title{{\Large{SLOPE and Designing Robust Studies for Generalization}}}
\date{\vspace{-1.5cm}}

\author[1]{Xinran Miao\thanks{Email: \href{mailto:xinran.miao@pennmedicine.upenn.edu}{xinran.miao@pennmedicine.upenn.edu}}}
\author[2,3]{Jiwei Zhao}
\author[2]{Hyunseung Kang}

\affil[1]{Department of Biostatistics, Epidemiology and Informatics, University of Pennsylvania}
\affil[2]{Department of Statistics, University of Wisconsin-Madison}
\affil[3]{Department of Biostatistics and Medical Informatics, University of Wisconsin-Madison}

\begin{document}
\maketitle

\vspace{-1cm}

\abstract{
A common task in generalization is to learn about a new target population using data from another source population. This task relies on conditional exchangeability, which assumes that differences between the source and target populations are fully captured by observable variables. However, this assumption is often untenable in practice due to remaining, unobservable differences, and it cannot be verified with data. These limitations warrant the development of robust study designs that are inherently less sensitive to violations of the assumption. We propose SLOPE (Sensitivity to LOcal Perturbations from Exchangeability), a simple and novel measure that quantifies sensitivity to local violations of conditional exchangeability. SLOPE combines ideas from sensitivity analysis in causal inference and derivative-based robustness measure from Hampel's influence function. To the best of our knowledge, SLOPE is the first metric to quantify the robustness of study designs with respect to violations of conditional exchangeability. Specifically, SLOPE measures the sensitivity of two design-level characteristics: (i) the functional of interest (e.g., the mean or the median) and (ii) the study distributions. We demonstrate how SLOPE can guide robust study designs through a re-analysis of a multinational randomized experiment.
}

\noindent%
\vfill

\newpage

\section{Introduction}\label{sec:intro}

\subsection{Background and Overview}
There has been a growing interest in generalizing or transporting information from an existing source population to a new, target population under the assumption of \emph{conditional exchangeability} and the setup can be formalized as follows.
Suppose each datum is represented as random variables $(\O,X)$ %
and the goal is to learn the target distribution $\QOX$ given (i) the ``full'' data $(\O,X)$ from the source distribution $\POX$ and (ii) the ``partial'' data $X$ %
from the target distribution $\QX$.
Conditional exchangeability states that conditional on $X$, the distribution of $O$ between the target and the source population is identical:
\begin{equation} \label{eq:condExch}
 \QOmidX(\cdot\mid X=x) = \POmidX(\cdot\mid X=x) \text{ almost everywhere in $\QX$. }
\end{equation}
Equation \eqref{eq:condExch} enables learning about the target distribution $\QOX$ based only on (i) $\POX$ and  (ii) $\QX$ and this phenomenon can be illustrated with a heuristic, yet simple equality:
\[
\QOX = \QOmidX \times \QX = \POmidX \times \QX.
\]
The first equality is from the definition of conditional probability and the second equality is from \eqref{eq:condExch}. By the same heuristic argument, we can learn a low-dimensional feature of the target distribution $\QOX$, denoted as $\psi(\QOX)$ and referred to as the target estimand or target functional, by $\psi(\QOX) = \psi(\POmidX \times \QX)$. Some popular target estimands include the mean of $O$ or the average treatment effect in the target population;
see Section \ref{subsec:review} for details and more examples.

Unfortunately, recent works \citep{allcott2015site,jin2024beyond} argued that conditional exchangeability is likely violated in practice due to unobservable differences between the source and the target population.
Worse, conditional exchangeability is inherently untestable because the variable $O$ is unavailable from the target distribution; under the setup above, we only have access to samples from (i) the joint distribution in the source,  $P_{O,X}$, and (ii) the marginal distribution in the target, $Q_{X}$ \citep{dahabreh2023sensitivity,zeng2025efficient,huang2024sensitivity}.
Taken together, these challenges underscore the need to have data collection processes and more broadly, study designs that are inherently less sensitive to violation of conditional exchangeability before analyzing data for generalization.

\begin{figure}[!h]
    \centering
    \includegraphics[width=1\linewidth]{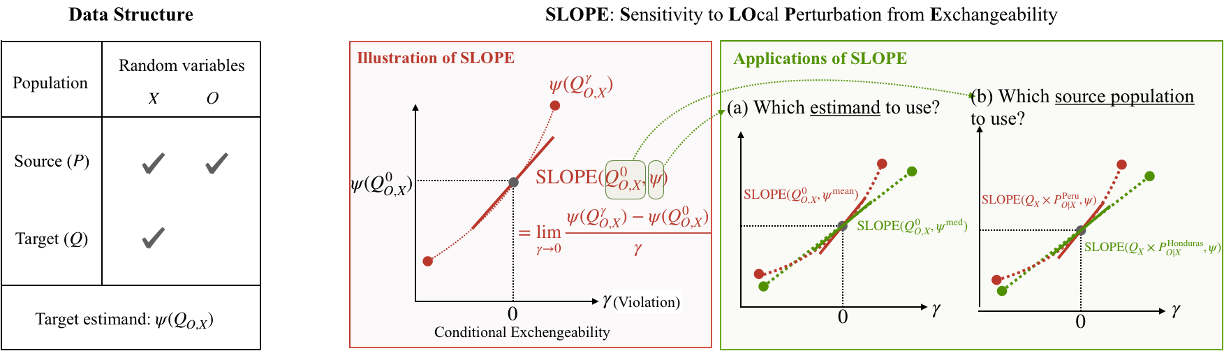}
  \caption{
   Left: the data structure of our setup. Right: an illustration of SLOPE. The $y$-axis plots the target estimand/functional $\psi(\cdot)$ and the x-axis plots $\gamma$, which represents the magnitude of violating conditional exchangeability. The point $\gamma = 0$ indicates no violation of conditional exchangeability.
    }
    \label{fig:illustration}
\end{figure}

To this end, the main contribution of the paper is to propose a simple and novel tool that helps gauge which study designs are robust to violations of conditional exchangeability
and we present a high-level summary of the tool.
Let $\QOmidXgamma$ be the distribution of $\QOmidX$ when conditional exchangeability is violated by a degree quantified by a sensitivity parameter $\gamma \in \R$ and let $\gamma = 0$ be the case where conditional exchangeability holds  (i.e., $\QOmidXzero = \POmidX$).
Then, we propose a metric called SLOPE, which stands for \textbf{S}ensitivity to \textbf{LO}cal \textbf{P}erturbations from \textbf{E}xchangeability:
\begin{equation*}
 {\rm SLOPE}(\QOXzero,\psi) = \lim_{\gamma \to 0} \frac{\psi(\QOXgamma) -\psi(\QOXzero)}{\gamma},
 \quad{} \text{ where }  \QOXgamma = \QOmidXgamma \times \QX.
\end{equation*}
As its name and definition imply, SLOPE is the slope of the target estimand $\psi(\QOXgamma)$ at $\gamma=0$ (see Figure \ref{fig:illustration} for a visual illustration). SLOPE measures how dramatically the target estimand changes when moving from a setting with no violation of conditional exchangeability (i.e., $\gamma = 0$) to a setting with a near-violation (i.e., $\gamma \to 0$).
Generally, a higher magnitude of SLOPE suggests that the target estimand is more sensitive/less robust to local violations, while a lower magnitude suggests the estimand is less sensitive/more robust; see Section \ref{subsec:slope.def} for further discussions on interpreting SLOPE.

SLOPE depends on two quantities: (a) the target estimand (i.e., $\psi$) and (b) the target distribution under conditional exchangeability (i.e., $\QOXzero = \POmidX \times \QX$). Importantly, SLOPE does not depend on the estimation procedure of $\psi$.
To put it differently,  SLOPE is an intrinsic, \emph{design-level characteristic} about (a) the \emph{target estimand} (i.e., $\psi$) and
(b) the \emph{source or target distributions} in the setup (i.e., $\POmidX$ and $\QX$) when there is a local violation of conditional exchangeability.

We briefly answer three common and important questions about SLOPE; see Sections \ref{sec:slope} and  \ref{sec:further} for detailed discussions. First, SLOPE is a local measure and for small deviations of $\gamma$ from $0$, SLOPE provides an accurate reflection about the change in $\psi$. For larger deviations of $\gamma$ from $0$, SLOPE may still provide valuable intuition, subject to the usual limitations of linear approximations based on the tangent line.
We remark that some well-known robustness measures are local, including \citet{hampel1974influence}'s celebrated influence function (IF), and these local measures yield valuable insights for designing robust estimators and tests  \citep{huber1981robust}. For more discussions behind the motivation for measuring local violations in robust statistics and how SLOPE yields valuable insights about robust study designs, see Sections \ref{sec:slope} and \ref{subsec:slope.if}. %
Second, by definition, the unit of SLOPE depends on the unit of the target estimand, thereby respecting the investigator's original choice of units for the target estimand.
In general, SLOPE is not unit-less and should be interpreted with caution; see Section \ref{subsec:slope.def} for more discussions.
Third, SLOPE depends on the parametrization of $\QOmidXgamma$ or equivalently, the sensitivity model for conditional exchangeability. Our main proposal of SLOPE uses the exponential tilting model \citep{robins2000sensitivity}, whose benefits and limitations are discussed in  Sections \ref{subsec:sensitivity} and \ref{subsec:slope.if}.
We discuss an extension of SLOPE to set-identified sensitivity models, such as \citet{tan2006distributional}'s marginal sensitivity model and \citet{dahabreh2023sensitivity,zeng2025efficient}'s conditional bias model, in Section \ref{subsec:slope.other.models}.
Regardless of the choice of the sensitivity model, we believe the high-level idea of SLOPE as a  derivative-based summary of a sensitivity analysis can provide new and important insights about designing robust studies for generalization.

\subsection{Prior Works} \label{sec:prior_works}
SLOPE fits into the large literature on robust statistics. Specifically, SLOPE and \citet{hampel1974influence}'s IF are related in that both SLOPE and IF use local derivatives to quantify robustness. But, \citet{hampel1974influence}'s IF measures a local change in the target estimand $\psi$ due to  contamination of a single data point whereas SLOPE measures a local change in $\psi$ due to violations of conditional exchangeability. Despite these differences, Section \ref{subsec:slope.if}
reveals an interesting analytic and geometric connection between SLOPE and IF.

Our work also fits into the  literature on sensitivity analysis of conditional exchangeability for generalizability and transportability \citep{nguyen2017sensitivity,
nie2021covariate, colnet2021generalizing,dahabreh2022global,dahabreh2023sensitivity,duong2023sensitivity,ek2023externally,huang2024sensitivity, jin2024beyond}. %
What differentiates our paper from most existing works on sensitivity analysis for generalizability is that existing works studied sensitivity of \emph{estimators} or \emph{tests} for a specific value of the sensitivity parameter $\gamma \neq 0$.
In contrast, our work studies sensitivity of \emph{study designs};
as mentioned in our summary above, SLOPE is agnostic to how the target estimand $\psi$ is estimated and is a population-level characteristic.

There have been some works on presenting simple, numerical summaries about the impact of violating key assumptions in different fields and we highlight some relevant examples.
\citet{gupta2023s} proposed the directional s-value, which quantifies the minimum amount of covariate shift that alters the sign of a target estimand, usually the mean.
\citet{rosenbaum2004design} proposed design sensitivity, which is a scalar, odds-ratio based summary of the power of a test statistic in a sensitivity analysis.
\cite{andrews2017measuring} considered local mis-specification of generalized method of moments (GMM) and their matrix-based measure, denoted as $\Lambda \in \mathbb{R}^{p \times p}$, measured the change in the estimator for GMM parameters.  In missing data, \citet{troxel2004index} proposed a measure which quantifies the change in an estimator when there is a local violation of ignorable missingness. %
\citet{ding2016sensitivity}, \citet{oster2019unobservable}, \citet{zhao_sensitivity_value}, and
\citet{cinelli2020making} proposed scalar measures, which summarize the impact of unmeasured confounding in observational studies. Specifically,  each proposed metrics that quantify the ``minimum unmeasured confounding bias'' necessary to alter the study's conclusion under no unmeasured confounding. %
Except for \citet{gupta2023s} and \citet{rosenbaum2004design} to some extent, all the above works measure robustness of estimators or tests rather than that of study designs. Also, except for \citet{troxel2004index}, all these works do not use derivative-based measures of robustness.

Finally, we briefly mention another important line of work on robustness and sensitivity analysis when conditional exchangeability holds, but the distributions of shared characteristics $X$ of the two populations differ dramatically, especially in the observed data; this setting referred to as limited overlap \citep{stuart2011use,tipton2014generalizable,chen2023generalizability,huang2025overlap}. Except for \citet{huang2025overlap}, a key distinction between these works and our work is that limited overlap can be, in principle, checked from the observed data since $X$ is observed in both populations. In contrast, conditional exchangeability cannot be checked from the observed data since $O$ is not observed in the target population.

\subsection{Organization of Paper and Notation}
The paper is organized as follows.
Section \ref{sec:setup} introduces the setup and the sensitivity model.
Section \ref{sec:slope} formally introduces SLOPE, its properties, and results on robust study designs.
Section \ref{sec:further} discusses more insights and results, including the relationship between SLOPE and IF and estimating SLOPE.
 Section \ref{sec:data} showcases an application of SLOPE through a re-analysis of a multi-site experiment by \citet{banerjee2015multifaceted}.
Section \ref{subsec:slope.other.models} discusses extensions of SLOPE to other sensitivity models.
Section \ref{sec:discussion} concludes with  a discussion on practical considerations.
Proofs and other results are in the Appendix.

We define the notations that we use in the paper. %
For a population $P$ %
and random vectors $X_1$, $X_2$,
we let $P_{X_j}$ be the marginal distribution of $X_j$ for $j=1,2$. Given $P_{X_j}$, we let  $\E_{P_{X_j}}(\cdot)$ and $F_{P_{X_j}}$ be the expectation and cumulative distribution function (c.d.f.), respectively, under $P_{X_j}$.
We let $P_{X_1\mid X_2}(\cdot\mid \cdot)$ be the conditional distribution of $X_1$ given $X_2$. Similarly, we let $\E_{X_1\mid X_2}(\cdot\mid\cdot)$, $F_{X_1\mid X_2}(\cdot\mid\cdot)$, $f_{X_1\mid X_2}(\cdot\mid\cdot)$ be the conditional distribution, conditional c.d.f., and conditional probability density function, respectively. %
We also let $\E_{X_1\mid X_2}(\cdot\mid x_2)$ be the conditional expectation given a specific $X_2=x_2$.
Throughout the paper, we assume sufficient regularity conditions for conditional distributions, conditional densities, and conditional expectations to exist; see \citet[Chapter 1]{shao2008mathematical} for the regularity conditions. Finally, for a probability distribution $Q$ defined on the same measurable space of $(X_1,X_2)$, we use
$P_{X_1\mid X_2}\times Q_{X_2}$
to denote the joint distribution where $P_{X_1\mid X_2}\times Q_{X_2}(A\times B) = \int_{B}P_{X_1\mid X_2}(A\mid x_2)dQ_{X_2}(x_2)$ for $A\in\B$ and $B\in\B^d$ and $\B$ is a Borel $\sigma$-field with respect to the reals.

\section{Setup}\label{sec:setup}
\subsection{Goal in Generalization and Key Assumptions} \label{subsec:review}

Let $P_{O,X}$ and $Q_{O,X}$ be the joint distributions of  random variables $(O,X)$ from a source population $P$ and a target population $Q$, respectively, where $O$ is a scalar and $X$ is a vector.
The goal in generalization is to learn a functional (i.e., $\psi$) of the target distribution $\QOX$, which we denote as $\psi(\QOX)$ and refer to as the target estimand or target functional,
from  (a) ``full'' data $(\O,X)$ from the source distribution $\POX$ and (b)  ``partial'' data $X$ from the target distribution $\QX$; %
see Figure \ref{fig:illustration} for a visual illustration of the data setup.

Before we go any further, we make two brief remarks about the setup. First, while our exposition below considers a scalar functional $\psi$ (e.g., means, medians, average potential outcomes),  all of our results extend to a low-dimensional, vector-valued $\psi$; see Section \ref{subsec:slope.def} and Section \ref{supp.subsec:slope.vector} of the Appendix where we discuss the setting when $\psi$ is the regression coefficient of $O$ regressed on $X$ in the target population. %
Second, almost all results below are agnostic to how the data were sampled within each population, for instance by simple random sampling, i.i.d. sampling, or even adaptive sampling. Specifically, our results remain at the population level until Section \ref{subsec:estimation}, where we propose estimators of SLOPE.

Under the setup, the two most popular assumptions for identifying the target estimand $\psi$ (e.g., \citet{cole2010generalizing,tipton2014generalizable,kern2016assessing,dahabreh2019generalizing,huang2023leveraging,zeng2025efficient,degtiar2023review}) are as follows.

\begin{assumption}[Overlap]\label{assump:overlap}
    $\QX$ is absolutely continuous with respect to $\PX$.
\end{assumption}

\begin{assumption}[Conditional Exchangeability]\label{assump:exchange}
Equation \eqref{eq:condExch} holds.
\end{assumption}

\begin{remark}[Alternative Formulation of Assumption \ref{assump:exchange}]\label{remark:exchange.in.ratios}
When the densities of $\POmidX$ and $\QOmidX$ exist with respect to a common measure (e.g., the Lebesgue measure or the counting measure), Assumption \ref{assump:exchange} can be re-formulated with respect to the corresponding density functions, i.e.,
       $f_{\QOmidX}(O,X)/ f_{\POmidX}(\O,X)=1$
       almost everywhere in $\POmidX\times \QX$.
\end{remark}

\noindent Assumption \ref{assump:overlap} states that the support of $\QX$ is within the support of $\PX$.
Assumption \ref{assump:exchange} enables replacing $\QOmidX$ with $\POmidX$, which can be identified from the ``full data'' $(O,X)$ in the source population $\POX$. Under Assumptions \ref{assump:overlap} and \ref{assump:exchange}, the target estimand  can be identified as $\psi(\QX\times \POmidX)$ and
some examples of target estimands are listed below.

\begin{example}[Mean]\label{ex:transfer.mean}
Suppose we are interested in the mean of $\O$ in the target distribution, denoted as $\psi^{\mean}(\QOX) = \E_{\QO}(\O)$. Under Assumptions \ref{assump:overlap} and \ref{assump:exchange}, the mean is identified via
\begin{align*}
   \psi^{\mean}(\QOX) = \E_{\QO}(\O) =& \E_{\QX}\left[\E_{\QOmidX}(\O\mid X)\right]
    = \E_{\QX}\left[\E_{\POmidX}(\O\mid X)\right].
\end{align*}
\end{example}

\begin{example}[Median]\label{ex:transfer.median}
Suppose we are interested in the median of $\O$ in the target distribution, denoted as $\psi^{\med}(\QOX) =F_{\QO}\inv(1/2)$, and $O$ is continuous. Under Assumptions \ref{assump:overlap} and \ref{assump:exchange},
the median is identified as the solution to the following equation:
\begin{align*}
\frac{1}{2} = \int\int_{-\infty}^{\psi^{\med}}  d \QOmidX d\QX
 =  \int\int_{-\infty}^{\psi^{\med}}  d \POmidX d\QX.
\end{align*}
\end{example}

\begin{example}[Z-Estimand]\label{ex:transfer.z.est}
Suppose $\psi(\QOX)$ is defined as  the solution to
\begin{align}\label{eq:z.est}
    \E_{\QOX}\{s\left(\O,X,\psi(\QOX)\right)\}=0,
\end{align}
where $s(\O,X,\cdot)$ is a user-specified function, usually a score function of the same dimension as $\psi$. Under Assumptions \ref{assump:overlap} and \ref{assump:exchange}, the target estimand is identified as the solution to
\begin{align*}
  0 %
   =\E_{\QX}\left[\E_{\QOmidX}\left\{s(\O,X,\psi(\QOmidX\times\QX))\mid X\right\}\right]
    = \E_{\QX}\left[\E_{\POmidX}\left\{s(\O,X,\psi(\QOXzero))\mid X\right\}\right].
\end{align*}
\end{example}

\begin{example}[Mean or Median of Potential Outcomes]\label{ex:mean.potential.outcome}
    Consider a randomized experiment in the source population to measure the average treatment effect (ATE). Let  $Y(a)$ be the potential outcome if, contrary to fact, a study unit was assigned to treatment $a \in \calA \subset \mathbb{R}$ where $\calA$ is a set of all possible treatment (e.g., $\calA=\{0,1\}$) and let
    $X$ be pre-treatment covariates. The goal is to learn about the ATE in a new, target population based on (a) the randomized experiment in the source population and (b) the distribution of $X$ in the target population. Under Assumptions \ref{assump:overlap} and \ref{assump:exchange}, the mean and the median of the potential outcome $Y(a)$ in the target population are identified using Examples \ref{ex:transfer.mean} and \ref{ex:transfer.median}:
    \begin{align*}
         \psi^{\mean}(Q_{Y(a),X}) = \mathbb{E}_{Q_{Y(a)}}\{Y(a)\} =  \E_{\QX}\left[\E_{P_{Y(a)\mid X}}\{Y(a)\mid X\}\right],\text{ and }\quad{}
    \frac{1}{2} = \int\int_{-\infty}^{\psi^{\med}}  d P_{Y(a)\mid X} d\QX.
    \end{align*}
    We remark that both equations  involve potential outcomes (i.e., $P_{Y(a)\mid X}$), which is identified under a randomized experiment in the source population; see Section \ref{supp.sec:causal.functional}  in the Appendix for details.
\end{example}

\subsection{Model for Sensitivity Analysis of Conditional Exchangeability}\label{subsec:sensitivity}

Suppose we suspect that conditional exchangeability (i.e., Assumption \ref{assump:exchange}) is implausible and we wish to assess how the conclusion of the study may change if conditional exchangeability is violated. A sensitivity analysis addresses this question by supposing that there is a ``$\gamma$ violation'' of conditional exchangeability and quantifying the downstream consequences of this violation. This section presents a model-based sensitivity analysis \citep{rosenbaum1983assessing,robins2000sensitivity,franks2019flexible} for quantifying violations of conditional exchangeability based on exponential tilting.

Formally, %
for each $\gamma \in \mathbb{R}$,  %
let $\QOmidXgamma(\cdot\mid X)$ be absolutely continuous with respect to $\POmidX(\cdot\mid X)$ almost everywhere $\QX$. Suppose the corresponding densities $f_{\QOmidXgamma}(\O,X)$ and $f_{\POmidX}(\O,X)$ satisfy the following relationship:
\begin{align}\label{eq:sensitivity}
  \dfrac{f_{\QOmidXgamma}(\O,X)}{f_{\POmidX}(\O,X)}\propto \exp(\gamma\cdot  \O) \text{, almost everywhere in } \POmidX\times\QX.
\end{align}
The notation ``$\propto$'' means ``proportional to'' and the normalizing constant satisfies
$ \int \exp(\gamma \O) d \POmidX(\O\mid X) <\infty$ almost everywhere $\QX$.
The term $\gamma$ is often referred to as the sensitivity parameter and it measures the difference between $\QOmidX$ and $\POmidX$. If $\gamma=0$, the sensitivity model \eqref{eq:sensitivity} reduces to Assumption \ref{assump:exchange} where the two distributions are identical, i.e., $\QOmidXzero=\POmidX$ almost everywhere in $\QX$. As $\gamma$ moves away from zero, the difference between $\QOmidX$ and $\POmidX$ becomes larger and conditional exchangeability is violated by a larger amount.

We make some brief remarks about the sensitivity model \eqref{eq:sensitivity}. Model \eqref{eq:sensitivity}  was proposed by \cite{scharfstein1999adjusting} and \citet{robins2000sensitivity} as a non-parametric (just) identified model for describing selection bias in missing data and has been used by several others, notably \citet{dahabreh2022global} in the context of generalization.
Model \eqref{eq:sensitivity} can be extended so that $\gamma$ depends on $X$ and $\O$ or the exponential tilting term can be replaced with a non-negative tilting term (see Section \ref{subsec:slope.other.models}). Also, model \eqref{eq:sensitivity} (i) posits no testable implications on the data \citep{franks2019flexible}, (ii) leads to an analytic connection to the influence function (see Section \ref{subsec:slope.if}), and (iii) provides simple and empirically validated principles about designing robust study designs for generalizations (see Sections \ref{subsec:slope_mean}, \ref{subsec:slope_median} and \ref{sec:discussion}).

An important caveat of model \eqref{eq:sensitivity} is the non-collapsibility of the model with respect to $X$; %
see Section 7 of \cite{scharfstein1999adjusting}. Also, model \eqref{eq:sensitivity} induces a point identification of the target estimand, which differs from a ``set-identified'' sensitivity analysis  that induces a set identification (e.g., \citet{rosenbaum1987sensitivity,tan2006distributional,jin2026sensitivity}). Section \ref{subsubsec:slope.bound} discusses an extension to such models.

\section{SLOPE: Sensitivity to Local Perturbations from Exchangeability}
\label{sec:slope}

\subsection{Definition and Basic Properties}\label{subsec:slope.def}

Under Assumption \ref{assump:overlap} and the sensitivity model \eqref{eq:sensitivity}, with a chosen value of the sensitivity parameter $\gamma$, the target estimand can be identified via $\psi(\QOXgamma)$ where $\QOXgamma = \QOmidXgamma\times \QX$ represents the joint distribution of $\QOmidXgamma$ induced from \eqref{eq:sensitivity} and $\QX$. But, $\gamma$ is never known in practice because it represents the magnitude of violating conditional exchangeability (i.e., Assumption \ref{assump:exchange}). Instead,  sensitivity analysis seeks to understand how $\psi(\QOXgamma)$ changes from $\gamma = 0$ (i.e., when Assumption \ref{assump:exchange} holds)
to $\gamma \neq 0$ (i.e., when Assumption \ref{assump:exchange} doesn't hold). This is typically done by presenting a table or a plot of $\psi(Q_{O,X}^\gamma)$ for a plausible range of $\gamma$ with the range determined by domain knowledge \citep{scharfstein1999adjusting,rotnitzky2001methods,nabi2024semiparametric},  benchmarking \citep{huang2024sensitivity} or calibration \citep{miao2024transfer}.

Instead of a table or plot of $\psi(Q_{O,X}^\gamma)$ with benchmarked/calibrated $\gamma$s, our approach to studying violations of Assumption \ref{assump:exchange} is inspired by a general principle from robust statistics  that ``robustness signifies insensitivity to \emph{small deviations} from the assumptions'' \citep[Chapter 1]{huber1981robust} where we added the emphasis on ``small deviations.'' Specifically, it would not be surprising if a large departure from Assumption \ref{assump:exchange} (i.e., a large $\gamma$) corresponds to a large change in the target estimand $\psi(\QOXgamma)$. But, it would be surprising and worrisome if a small departure from Assumption \ref{assump:exchange} (i.e., a small $\gamma$) corresponds to a large change in $\psi(\QOXgamma)$. Our proposed metric SLOPE formalizes this idea by measuring the ``instantaneous change'' (i.e.,  the slope) of $\psi(\QOXgamma)$ at $\gamma = 0$; see Figure \ref{fig:illustration} for a visual illustration.
\begin{definition}[SLOPE]\label{def:proposal}
The  sensitivity to local perturbations from exchangeability (SLOPE\footnote{We note that this acronym ``SLOPE'' was used in short for ``Sorted L-One Penalized Estimation'' \citep{bogdan2015slope}.
}) of a target functional/estimand $\psi$ with respect to the sensitivity model in \eqref{eq:sensitivity} is defined as
   \begin{align}\label{eq:slope.def}
      \SI(\QOXzero, \psi)
      &= \lim_{\gamma \to 0}
      \dfrac{ \psi(\QOXgamma)
      -\psi(\QOXzero)
      }
      {\gamma},
   \end{align}
   provided the limit exists.
\end{definition}
A large magnitude of SLOPE means that the target estimand $\psi$ will change more drastically if conditional exchangeability is slightly violated (i.e., $\gamma \to 0$).
In contrast, a small magnitude of $\SI$ means that the target estimand will change less drastically. %
Note that the magnitude of $\SI$ is the absolute value of SLOPE when the target estimand $\psi$ is a scalar. When $\psi$ is a vector, the magnitude of $\SI$ corresponds to the researcher's choice of measuring the magnitude of vectors (e.g., $\ell_2$ norm). As remarked in Section \ref{subsec:review}, for expositional purposes, we focus on a scalar $\psi$, and thereby a scalar SLOPE, but our results hold for low-dimensional, vector-valued $\psi$.

When comparing SLOPE, investigators should note that SLOPE is not unit-less. Instead, its unit depends on the target estimand as well as the sensitivity parameter. This is analogous to a slope parameter in simple linear regression, which depends on both the response variable and the explanatory variable. In practice, we recommend comparing SLOPE between distributions, or between estimands with the same unit (e.g., between mean and median).
One may also benchmark the magnitude of SLOPE using a procedure described in Section \ref{supp:subsec:benchmark} of the Appendix. More broadly, although SLOPE can be defined in different sensitivity models (see Section \ref{subsubsec:slope.bound}), we do not recommend comparing SLOPE across sensitivity models with different parameterization due to the difference in the units of $\gamma$. Defining SLOPE in a unit-less way is an important direction for future work.

A key property of SLOPE is that it does not depend on any particular estimation procedure. Instead, SLOPE measures an intrinsic property about the robustness of a \emph{study design}
and is determined by two design quantities: (a) the target estimand $\psi$, and (b) the target distribution under conditional exchangeability ($\QOXzero=\POmidX\times \QX$).
For researchers, these choices roughly correspond to answering two questions: (a) ``what quantity do I want to study?'' and (b) ``which dataset should I use to study the quantity?''.
Changing the answers to either question can alter the value of SLOPE. %
Consequently, SLOPE can help researchers pick a robust study design by selecting an estimand
(e.g., the mean or the median of $O$ in the target distribution), target distribution (e.g., $\QX$), or the source distribution (e.g., $\POX$) that leads to a lower magnitude of SLOPE; %
see  Sections \ref{subsec:slope_mean} and \ref{sec:data} for illustrations.

When communicating the meaning of SLOPE, some researchers may find it useful to interpret SLOPE as a measure of the ``first-order change'' of the estimand when conditional exchangeability is violated. Specifically, a first-order Taylor expansion of $\psi$ yields
\begin{align}\label{eq:first.order.approx}
     \psi(\QOXgamma)- \psi(\QOXzero) \approx  \gamma\cdot\SI(\QOXzero, \psi).
 \end{align}
The Taylor expansion suggests that for a small $\gamma$ that is near zero, SLOPE provides an accurate measure of the change in $\psi$ when conditional exchangeability is violated. When $\gamma$ is large in magnitude, SLOPE may still provide some intuition about the change in $\psi$, with the usual limitations of first-order linear approximations. We remark, however, that SLOPE cannot identify the bias of an estimator of $\psi$ from violating conditional exchangeability since, as mentioned in Section \ref{subsec:sensitivity}, $\gamma$ and $\QOXgamma$ are not identifiable.

Finally, we remark that SLOPE  does not always exist for every target estimand. For example, if $\psi$ is the sign of the mean of $\O$ in the target population, the limit in Definition \ref{def:proposal} may not exist when the sign changes near $\gamma = 0$.
One general condition for SLOPE to exist is to satisfy the conditions for the chain rule under Hadamard differentiability; %
see Section \ref{supp.sec:slope.exist} of the Appendix for details. For Z-estimands in Example \ref{ex:transfer.z.est}, a sufficient condition for the existence of SLOPE is to impose smoothness and boundedness conditions on $s$; note that these conditions are common to establish consistency of Z-estimators.

  \begin{condition}[Existence of SLOPE for Z-Estimands]\label{condition:regular.z.slope}
  (i) $\E_{Q_{O\mid X}^{\gamma}}[s(O,X,\psi(Q_{O,X}^0))]$ is bounded %
  for $\gamma$ in a neighborhood of zero, and $\E_{\QOXzero}\left[ s(O,X,\psi(Q_{O,X}^0))  \left\{\O - \mu(X)\right\} \right]$ exists where $\mu(X) = \E_{\POmidX}[O \mid X]$; (ii)  $s(O,X,\cdot)$ is differentiable almost everywhere with the derivative $\dot{s}(O,X,\cdot)$; and (iii)  $\E_{\QOXzero}
       \{ \dot{s}(\O,X,\psi(\QOXzero))\}$ exists and is non-singular.
  \end{condition}
\subsection{Example: SLOPE for the Mean and Robust Study Designs} \label{subsec:slope_mean}

This section uses the mean functional  to make two illustrations. The first is to show that how researchers can derive SLOPE for a given $\psi(\QOX^\gamma)$ using basic calculus. The second is to illustrate how SLOPE can yield useful insights about robust study designs for generalization.

To begin, consider the mean of $O$ in the target population, i.e., $\psi^{\mean}$ in Example \ref{ex:transfer.mean} where  $\psi^\mean(\QOX) = \E_{\QO}(O)$. From Section \ref{subsec:slope.def}, for a given $\gamma$, the sensitivity model \eqref{eq:sensitivity}  implies the following equality:
\begin{align*}
    \psi^{\mean}(\QOXgamma) =
    \E_{\QOXgamma}(O)=
    \E_{\QX}
    \left\{
    \E_{\QOmidXgamma}\left(O\mid X\right)
    \right\}=
    \E_{\QX}\left[
    \dfrac{
    \E_{\POmidX}\left\{\O\exp(\gamma\O)\mid X\right\}
    }{ \E_{\POmidX}\left\{\exp(\gamma\O)\mid X\right\}}
    \right].
\end{align*}
Then SLOPE for the mean is the derivative of $\psi^{\mean}(\QOXgamma)$ with respect to $\gamma$, evaluated at $\gamma=0$. In principle, researchers can compute this derivative using single-variable calculus and
Theorem \ref{thm:slope.mean} states the regularity conditions to ensure the existence of this derivative.

\begin{theorem}[SLOPE of Mean]\label{thm:slope.mean}
    Suppose Condition \ref{condition:regular.z.slope} holds with $s(O,\psi^{\mean})=O-\psi^{\mean}$.
    Then the SLOPE of the mean $\psi^{\rm mean}$ from Example \ref{ex:transfer.mean} is
\begin{align}\label{eq:slope.mean}
       \SI(\QOXzero,\psi^{\mean})
    = \E_{Q_X}\{\sigma^2(X)\}, \text{ where } \sigma^2(X) = \var_{\POmidX}(\O\mid X).
    \end{align}
\end{theorem}
In words, the SLOPE of the mean is the average variability of $O$ after adjusting for $X$ in the source population and the average is taken over the target's $\QX$.
When this variation is homoskedastic/constant across $X$ (i.e., $\sigma^2(X) = \sigma^2$), the mean's SLOPE simplifies to $\SI(\QOXzero,\psi^{\mean}) = \sigma^2$. In this case,  SLOPE is only determined by the source distribution, specifically $\POmidX$; the target distribution $\QX$ does not change SLOPE.

Two immediate implications follow from Theorem \ref{thm:slope.mean} about robust study designs. First, suppose  $X$ explains almost all the variation in $O$ in the source population, then $\sigma^2(X)$ will be close to zero. Thus, SLOPE will be close to zero, meaning that the target mean will not change dramatically even if conditional exchangeability is slightly violated.

As a concrete example, consider Example \ref{ex:mean.potential.outcome} where the goal is to generalize the ATE by letting $O = Y(1) - Y(0)$. If the randomized experiment in the source population suggests that the individual treatment effect is nearly constant (i.e., $Y(1) - Y(0) \approx c$ for some constant $c$), then $\sigma^2(X)$ will be close to zero and according to SLOPE, the ATE will not be sensitive even if conditional exchangeability is violated. In short, a near-constant treatment effect is robust in generalization.
We remark that \citet{tipton2018review} made a similar comment in the context of generalizing ATEs that ``generalizability is straightforward under the strong -- and often implausible -- assumption that treatment impacts are constant.''
The main  difference between \citet{tipton2018review} and our work is that we provide a general and quantitative metric to this comment where this metric encompasses constant $O$ (i.e., ``treatment impacts are constant'') as a special case.

\begin{figure}[!h]
    \centering
    \includegraphics[width=1\linewidth]{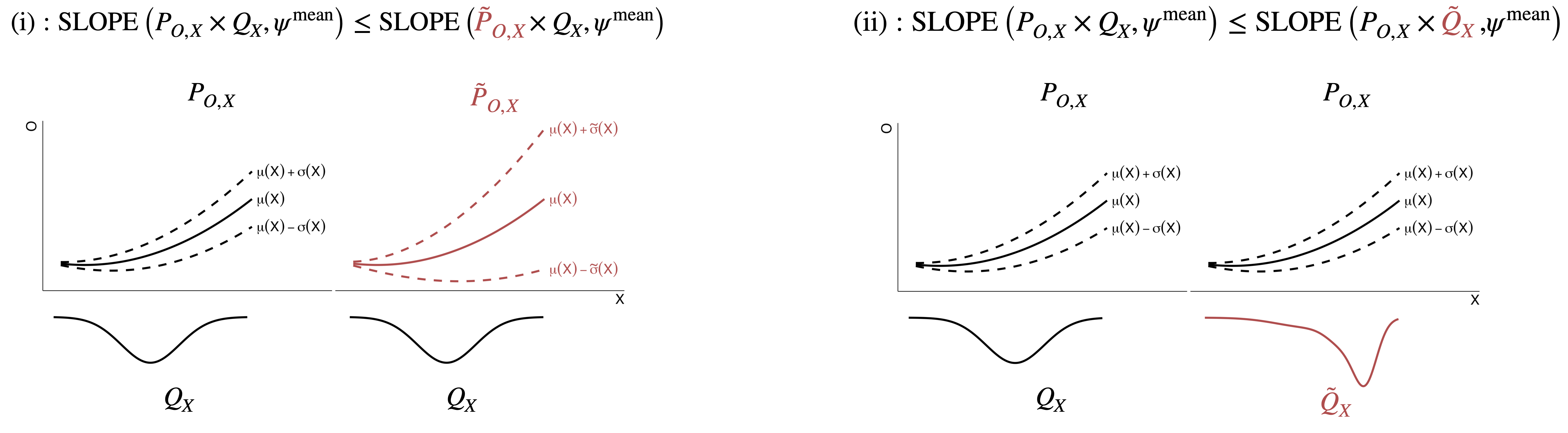}
    \caption{Illustration of designing robust studies for generalization with $\psi^{\mean}$. The x-axis represents $X$ and the y-axis represents $O$.
    For a formal result behind the illustrations, see Corollary \ref{cor:slope.mean.source}.}
    \label{fig:cor1}
\end{figure}

Second, suppose the variation in $O$ is not well-explained by $X$ in some regions of $X$ from the source population. Then, the mean's SLOPE can be made small by choosing a target distribution $\QX$ that concentrates around a region of $X$ that shows the smallest variation in $O$ in the source population. Figure \ref{fig:cor1} provides a visual illustration of these examples and Corollary \ref{cor:slope.mean.source} formalizes these observations.

\begin{corollary}[Robust Study Design for Learning $\psi^{\rm mean}$]\label{cor:slope.mean.source}
(i) Consider %
two source populations $P$ and $\widetilde{P}$ that satisfy  $\sigma^2(X) \leq \widetilde\sigma^2(X)$ almost surely for a target distribution $\QX$, where $\sigma^2(X)=\var_{P_{O\mid X}}(O\mid X)$ and $\widetilde{\sigma}^2(X)=\var_{\widetilde{P}_{O\mid X}}(O\mid X)$.
Then
\begin{align*}
    \SI(\POmidX\times\QX,\psi^{\mean})\leq \SI(\widetilde{P}_{O\mid X}\times \QX, \psi^{\mean}).
\end{align*}
(ii)
Next, consider two target distributions $\QX$ and $\widetilde{Q}_X$ over a common support $\SX$ such that there exists a subset $\S_{X,\textnormal{low}}\subseteq\SX$ that satisfies $Q_X(\S_{X,\textnormal{low}})\geq\widetilde{Q}_X(\S_{X,\textnormal{low}})$. If there exists a constant $c$ such that $\sigma^2(X) < c$ for $X\in\S_{X,\textnormal{low}}$ and $\sigma^2(x)>c$ for $x\not\in\S_{X,\textnormal{low}}$, then
\begin{align*}
     \SI(\POmidX\times\QX,\psi^{\mean})\leq \SI(P_{O\mid X}\times \widetilde{Q}_X, \psi^{\mean}).
\end{align*}
\end{corollary}

\subsection{Example: SLOPE for Median} \label{subsec:slope_median}
Similar to the SLOPE of the mean, the SLOPE of the median also depends on the dispersion of the underlying distribution $\QOXzero$. Theorem \ref{thm:slope.median} provides a general formula for the SLOPE of the median along with two special cases.

\begin{theorem}[SLOPE of   Median]\label{thm:slope.median}
    Suppose Condition \ref{condition:regular.z.slope} holds with $s(O,X,\psi^{\med}) = \ind(O\leq \psi^{\med})-\ind(O>\psi^{\med})$ and $\psi^{\rm med}$ is unique, where $\ind(\cdot)$ is the indicator function which is one if the event holds and zero otherwise. Then, the SLOPE of the $\psi^{\rm med}$ from Example \ref{ex:transfer.median}
is    \begin{align}\label{eq:slope.median}
   \SI(\QOXzero,\psi^{\med}) =& \dfrac{
        \E_{\QX}\left[F_{\POmidX}(m_{1/2}\mid X)\mu( X)\right] - \E_{\QOXzero}\left[\O\ind(\O\leq m_{1/2})\right]
        }{f_{\QOzero}(m_{1/2})},
    \end{align}
    where $m_{1/2}=F_{\QOzero}\inv(1/2)$ is the median of $O$ on the target population under conditional exchangeability and we recall  $\mu(X)=\E_{\POmidX}(O\mid X)$ is the conditional expectation of $O$ given $X$ in the source population.\\
(i) If $\POmidX$ is symmetric with respect to O and $\mu(X)=m_{1/2}$  almost surely $\QX$, then \eqref{eq:slope.median} simplifies to
    \begin{align}\label{eq:slope.median.symmetry}
        \SI(\QOXzero,\psi^\med) = \dfrac{m_{1/2}-\E_{\QOzero}\left(\O\mid\O\leq m_{1/2}\right)}
        {2f_{\QOzero}(m_{1/2})}.
    \end{align}
(ii) If $\POmidX$ is Gaussian, i.e., $\POmidX  \sim N\left(\mu(X),\sigma^2(X)\right)$,
then \eqref{eq:slope.median} simplifies to
\begin{align*}
       \SI(\QOXzero,\psi^{\med}) = \E_{\QX}\left[
   \sigma^2(X)\cdot \dfrac{f_{\POmidX}(m_{1/2}\mid X)}{\E_{\QX}\left\{f_{\POmidX}(m_{1/2}\mid X)\right\}}
   \right].
\end{align*}
\end{theorem}

Compared with the SLOPE of the mean, the SLOPE of the median depends on the spread of $Q_{O,X}^0$ in a way that is more complicated than $\sigma^2(X)$.
For example, in part (i) which induces symmetry,
the SLOPE of the median in \eqref{eq:slope.median.symmetry} depends on two quantities:
(a) the difference between the mean and the truncated mean that is lower than the median $\left\{m_{1/2}-\E_{\QOzero} \left(\O\mid\O\leq m_{1/2}\right)\right\}$,
and (b) the inverse of the marginal density at the median $1/f_{\QOzero}(m_{1/2})$. In essence, (a) measures the spread of $\QOzero$ and (b) is called Tukey's sparsity \citep{tukey1965part}, which is designed to measure the inverse of the concentration of $\QOzero$.
In part (ii) when $\POmidX$ is Gaussian, the SLOPE of the median becomes a weighted average of the conditional variance $\sigma^2(X)$ where the weight is determined by the conditional density $f_{\O\mid X}$ evaluated at the  median $m_{1/2}$. This form of the median's SLOPE resembles the mean's SLOPE \eqref{eq:slope.mean}, which is an (unweighted) average of $\sigma^2(X)$. In general, whether the researcher should study the median or the mean as a measure of centrality of $O$ in their study will depend on the conditional variance $\sigma^2(X)$ and the shape of the tail of $\POmidX$.

\section{Further Insights and Results}\label{sec:further}

\subsection{Relationship Between SLOPE and Influence Function}\label{subsec:slope.if}

Another local, derivative-based measure of robustness that precedes and complements our work is \citet{hampel1974influence}'s celebrated influence function (IF)\footnote{\citet{hampel1974influence} originally called IF the ``influence curve.'' In later works, others, including Hampel, referred to the influence curve as the IF due to its generalization to higher dimensions \citep{hampel2011robust}.}. In this section, we show how SLOPE is related to IF. Briefly, the IF of $\psi(\cdot)$ under $\QOXzero$ is a derivative-based local measure of robustness that quantifies the effect of an infinitesimal contamination at the point $(o,x)$ on the estimand $\psi(\QOXzero)$,
\begin{align}
    \IF(o,x, \psi(Q_{O,X}^0))
    =& \lim_{t\downarrow0} \dfrac{
    \psi\left((1-t)Q_{O,X}^0 + t\delta_{o,x}\right) - \psi(Q_{O,X}^0)
    }{t}.
    \label{eq:if.direction}
\end{align}
The term $\delta_{o,x}$ represents the dirac delta function at $(o,x)$. From the definitions, both SLOPE and IFs are local measures of robustness. The main difference is that IFs measure the local change of the target estimand due to contamination of a single data point whereas SLOPE measures the local change of the target estimand due to violation of conditional exchangeability from the entire distribution.

In addition to their definitions, we can also compare the IF and SLOPE from a geometrical perspective. Specifically, both are directional derivatives at the ``origin'' (i.e., $t=0$ or $\gamma=0$), but they differ in their direction.  Theorem \ref{thm:slope.if} shows that SLOPE is equal to IF if the IF it ``tilted'' towards a particular direction.

\begin{theorem}[Connection between SLOPE and IF]\label{thm:slope.if}
Suppose either (i) $\psi$ is a Z-estimand defined in \eqref{eq:z.est} and Condition \ref{condition:regular.z.slope} hold, or (ii) Conditions \ref{condition:regular.hadamard}-\ref{condition:regular.if.hadamard} in the Appendix hold.
Then SLOPE can be written as
\begin{align}\label{eq:slope.if}
       \SI (\QOXzero,\psi)
       &= \E_{\QX}\left(\E_{\POmidX}\left[\IF(\O,X,\psi(\QOXzero))  \left\{\O - \mu(X)\right\} \mid X\right]\right).
    \end{align}
\end{theorem}

Similar to Condition \ref{condition:regular.z.slope}, Conditions \ref{condition:regular.hadamard}-\ref{condition:regular.if.hadamard} are regularity conditions that ensure the chain rule under Hadamard differentiability holds.
In words,  \eqref{eq:slope.if} states that SLOPE is the projection of the IF onto the residual variation $O-\mu(X)$.
The residual variation term ($O-\mu(X)$) is due to the exponential tilting term induced by the sensitivity model \eqref{eq:sensitivity}; see Theorem \ref{thm:slope.if.rho} for a more quantitative explanation.
If the IF is nearly orthogonal to the residual subspace $\O-\mu(X)$ based on violating conditional exchangeability, then
SLOPE will be close to zero. For a Z-estimand in \eqref{eq:z.est}, its SLOPE will be smaller if the score function $s$ is chosen to be nearly orthogonal to the subspace spanned by  $\O-\mu(X)$.

In addition to the geometric interpretation of SLOPE, Theorem \ref{thm:slope.if} provides a general formula to derive the SLOPE given an IF. Sections \ref{subsec:slope_mean} and \ref{subsec:slope_median} derived SLOPEs for the mean and the median using their respective IFs. Also, Section \ref{supp.sec:slope.other.estimands} of the Appendix derives SLOPEs for scale parameters,
ordinary least squares (OLS) coefficients, Pearson correlation, L-estimands and other Z-estimands using Theorem \ref{thm:slope.if}.

\subsection{Estimation}\label{subsec:estimation}
We briefly discuss two estimators of SLOPE, the weighting estimator and the regression estimator, for Z-estimands in Example \ref{ex:transfer.z.est}. In short, estimation of SLOPE follows from existing theory on M-estimation (e.g., Chapter 5 of \citet{van2000asymptotic}). %
Details behind the estimators, including implementation, asymptotic properties, and simulations to assess their finite-sample performance, as well as alternative estimators based on efficient influence functions (when they exist), are in Sections \ref{supp.sec:est} -- \ref{supp.sec:simulation} of the Appendix. Our data analysis in Section \ref{sec:data} uses the regression estimator below due to better finite-sample performance.

Suppose we have $\nq$ independent and identically distributed (i.i.d.) samples $X_i\sim \QX$  and another independent  $\np$ i.i.d. samples $(X_i,\O_i)\sim \POX$ with $i=1,\ldots,\nq,\nq+1,\cdots,\np+\nq=n$.
Under Condition \ref{condition:regular.z.slope}, the SLOPE of a Z-estimand is
\begin{align}\label{eq:slope.z.scalar}
   \SI(\QOXzero,\psi)=& -\left\{\E_{\QX}\left(\E_{\POmidX}
       \left[ \dot{s}(\O,X,\psi)\right]\right)\right\}\inv
       \E_{\QX}\left(\E_{\POmidX}\left[\{\O-\mu(X)\} s\left(\O,X,\psi\right)\right]\right).
\end{align}

Let $\omega(X) = f_{\QX}(X) / f_{\PX}(X)$ be the density ratio of $X$ between the two populations. The weighting estimator of \eqref{eq:slope.z.scalar} re-weighs the source samples to match the target distribution $\QX$:
\begin{align*}
\widehat{\SI}^{\wt} (\wh{Q}_{O,X}^0, \wh\psi)
&=- \left\{ \sum_{i=\nq+1}^{n}\wh{\omega}(X_i)\dot{s}(O_i,X_i,\wh{\psi})\right\}
\inv  \sum_{i=\nq+1}^{n}\wh{\omega}(X_i)\{O_i-\wh{\mu}(X_i)\}s(O_i,X_i,\wh{\psi})
\end{align*}
The term $\wh{Q}_{O,X}^0$ represents samples from the source and target distributions, and $\wh\omega$, $\wh\mu$, and $\wh\psi$ are estimates of $\omega$, $\mu$, and $\psi(\QOXzero)$, respectively, say by parametric or semi-parametric methods. For example, $\wh\mu$ can be from OLS,  $\wh\omega$ can be estimated from balancing methods or selection models (see Section \ref{supp.subsec:est.nuisance} of the Appendix), and $\wh\psi$ can be estimated by a Z-estimator with data from $\wh{Q}_{O,X}^0$.

The regression estimator of \eqref{eq:slope.z.scalar} works  by estimating the corresponding conditional expectations $\E_{\POmidX}$, denoted as $\wh\E_{\POmidX}$, using parametric or semi-parametric methods:
\begin{align*}%
\widehat{\SI}^{\reg} (\wh{Q}_{O,X}^0, \wh\psi)
= &-\left[\sum_{i=1}^{\nq}\wh\E_{\POmidX}\{\dot{s}(O_i,X_i,\psi)\mid X_i\}\right]\inv
\sum_{i=1}^{\nq}\wh\E_{\POmidX}\left[\{O_i-\wh{\mu}(X_i)\}s(O_i,X_i,\psi)\mid X_i\right].
\end{align*}
As a concrete example, consider the target mean $\psi^{\mean}$ with $s(O,X,\psi^{\mean})=O-\psi^{\mean}$. The weighting estimator and the regression estimator of the mean's SLOPE are
\begin{align*}
\widehat{\SI}^{\wt}(\wh{Q}_{O,X}^0,\wh\psi^{\mean}) &= \dfrac{1}{\np}\sum_{i=\nq+1}^{n}\wh{\omega}(X_i)\{O_i-\wh{\mu}(X_i)\}(O_i-\wh\psi^{\mean}), \\
\widehat{\SI}^{\reg}(\wh{Q}_{O,X}^0,\wh\psi^{\mean})  &=
\dfrac{1}{\nq}\sum_{i=1}^{\nq}\wh\sigma^2(X_i).
\end{align*}
Here, $\hat{\psi}^{\mean}$ is the estimate of the target mean under conditional exchangeability and $\wh\sigma^2(x)$ is an estimate of $\sigma^2(x)$.
A simple way is to estimate $\widehat{\mu}(\X)$ is via OLS and $\widehat{\sigma}^2(\X)$ is with a log-variance linear model from weighted least squares (e.g., \citet{harvey1976estimating,carroll1982robust,davidian1987variance}).

\section{Application}\label{sec:data}

\subsection{Data Background}\label{subsec:ex.data.data}

We illustrate how to use SLOPE to inform robust study designs by re-analyzing \citet{banerjee2015multifaceted}'s multi-national experiment. The goal of the experiment was to evaluate the Graduation program in six countries (Ethiopia, Ghana, Honduras, India, Pakistan, and Peru).
The Graduation program provides a holistic set of services, including asset transfers, consumption support and other career and health services, to poor households. %
Between 2007 and 2014, eligible households in each village were randomized to the intervention or the control group and the experiment lasted for 24 months. %

We adopt the potential outcome notation stated in Example \ref{ex:mean.potential.outcome} where $Y(a)$ denotes the potential outcome under treatment $a$, with $a=1$ denoting participation in the program and $a=0$ denoting otherwise.
In subsections below, we use  SLOPE to study the violation of conditional exchangeability in transporting the potential outcome of treatment, i.e., $O=Y(1)$, from one country (i.e., the source population $P$) to another country (i.e., the target population $Q$). We remark that identification of the SLOPE requires additional causal assumptions in the source population (i.e., stable unit treatment variable assumption [SUTVA] and strong ignorability), on the source population and these assumptions are satisfied because a randomized experiment was conducted; see Sections \ref{supp.sec:causal.functional} and \ref{supp.subsec:data.identify.slope} in the Appendix for details. %

We focus on two types of outcome variables, the per capita consumption and the physical health index, respectively, in Sections \ref{subsec:data.choice.source} and \ref{subsec:data.weight}.
For each outcome, the baseline covariate corresponds to the same variable measured prior to intervention.
Our analyses are based on  complete data with overlapped baseline measurement (i.e., Assumption \ref{assump:overlap} holds).
To harmonize $X$ across countries for generalizability, we discretize $X$ into coarse categorical variables; %
note that due to the randomization of $A$, these transformations of $X$ will not affect the plausibility of strong ignorability of $A$.
Finally, for estimation, we have assumed that the conditional expectation of the outcome is linear in the baseline $X$ and the village where the household locates. Moreover, for SLOPE for the median in Section \ref{subsec:data.choice.source}, the residual of the linear model is assumed to be normal for simplicity and diagnostics in Section \ref{supp.subsec:data.choice.source} of the Appendix suggest that this is a reasonable assumption.
For details on data-preprocessing, estimation, and auxiliary results, %
see Sections \ref{supp.subsec:data.choice.source}  and \ref{supp.subsec:data.weight} in the Appendix.

\subsection{Which Source Country Is More Robust?}\label{subsec:data.choice.source}

In this section, we study the SLOPE of transporting the per capita consumption across countries.
The outcome variable is the log of the average of per capita consumption at two time points and ranges from 1.4 to 6.4. %
One country (Ethiopia) was excluded from the analysis because their consumption support %
was substantially different from other countries.
Sample sizes ranged from 771 to 2379 households across countries, with an average treatment rate of $35\%$.
For each pair of countries, we treat one  as the target population, the other as the source population, and estimate the SLOPE of the mean and median (i.e., $\psi^{\rm \mean}$ and $\psi^{\med}$).

Results are shown in Table \ref{tab:data.appI}.
Given a target country, the SLOPE is primarily determined  by the data distribution in the source country and does not vary much between the mean and the median. Using India or Peru as the source population  yields a lower SLOPE (i.e., a lower sensitivity) compared with using other source countries. %
Also, there is minimal difference in the SLOPEs of the mean and median, although the median's SLOPEs are consistently, but slightly, lower than the mean's SLOPEs.
Section \ref{supp.subsubsec:data.choice.source.result} in the Appendix further shows that the estimated median and mean themselves (not their corresponding SLOPEs) have comparable estimated variances and thus, there is no (empirical) loss in efficiency when we estimate the median. Therefore, from this analysis based on SLOPE, we generally recommend using the median in transporting the average of per capita consumption.

\begin{figure}[!h]
    \centering
    \includegraphics[width=1\linewidth]{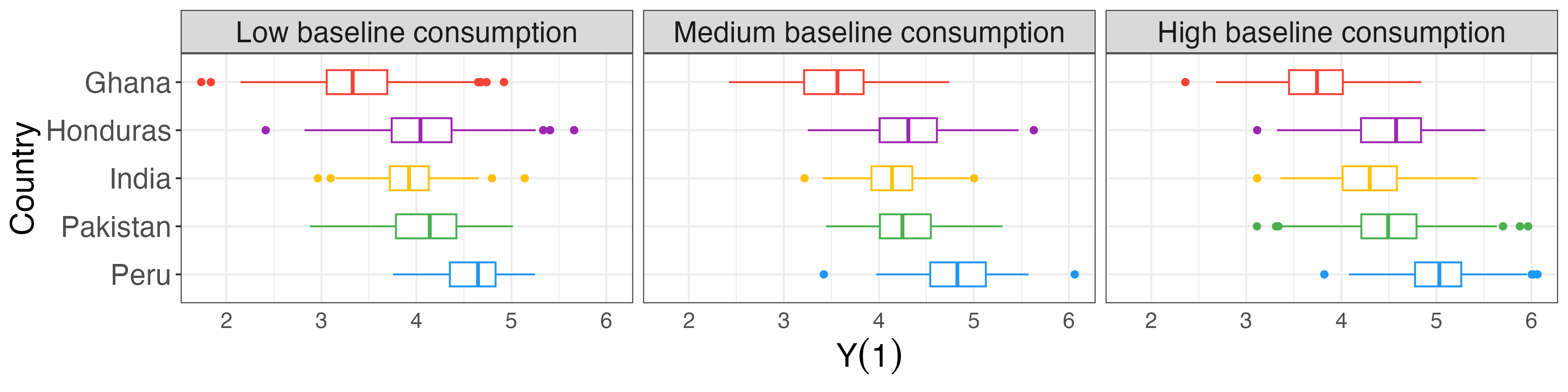}
    \caption{Boxplots of $Y(1)$ across countries (in y-axis) and categories of $X$ (in panels). }
    \label{fig:consumption.box}
\end{figure}

To explain why India and Peru generally have smaller SLOPEs, Figure \ref{fig:consumption.box} plots the
the distribution of $Y(1)$ given  some discrete values of $X$ across  countries. We see that Ghana, which is a source country that almost always has the highest SLOPE,  has higher spread of $Y(1)$ conditional on each category of $X$.
Conversely, India and Peru, which are the source countries with lower SLOPEs, have more concentrated values of $Y(1)$ relative to others.

We conducted some sanity checks of our recommendation and we provide a summary of them. First, because the outcome data is actually measured in the target country, we are in a unique situation where we can assess whether the sensitivity model \eqref{eq:sensitivity} is reasonable for our analysis. Specifically, we can assess the bias from violating conditional exchangeability %
empirically and check whether SLOPE can approximate this bias using the first-order approximation in \eqref{eq:first.order.approx}. Figure \ref{fig:data.bias.slope.consumption}  in Section \ref{supp.subsec:data.approx.bias} of the Appendix shows that the first-order approximation works well where the bias can be well-approximated by SLOPE. Second, a deeper subject-matter expertise explanation for why India and Peru are robust compared to other countries with respect to violation of conditional exchangeability is beyond the scope of this work. Nevertheless, we present one hypothesis in Section \ref{supp.subsec:data.choice.source} of the Appendix  based on our understanding of \cite{banerjee2015multifaceted}.

Finally, note that in comparing SLOPE across designs, sensitivity model \eqref{eq:sensitivity} is implicitly assumed to hold across designs. This can be restrictive if the sensitivity model is not plausible for some designs. However, it's important to keep the sensitivity model fixed when comparing SLOPE across designs as different sensitivity models can yield different conclusions about robustness \citep{huang2026design}; see Section \ref{subsubsec:slope.bound}  for more discussions.

\begin{table}[!h]
 \caption{
 The estimated SLOPEs for transporting the counterfactual log-transformed per capita consumption under treatment (i.e., $O=Y(1)$) from a source country (by rows) to a target country (by columns). Bootstrap standard errors are in the parentheses.
 For each target country and estimand, the lowest SLOPEs among the source countries are bold faced.
 }
    \label{tab:data.appI}
    \centering
   {\footnotesize \begin{tabular}{c | c | cccccc }
    \hline
    \multirow{2}{*}{Estimand ($\psi$)} & \multirow{2}{*}{Source ($\POmidX$)} & \multicolumn{5}{c}{Target ($\QX$)} \\
    &  & Ghana & Honduras & India & Pakistan & Peru \\
    \hline
    \multirow{5}{*}{Mean}
 & Ghana    &                 & 0.24 (0.01) & 0.24 (0.01) & 0.24 (0.02) & 0.24 (0.02) \\
 & Honduras & 0.24 (0.01)     &             & 0.24 (0.01) & 0.25 (0.02) & 0.25 (0.02) \\
 & India    & \bf{0.14} (0.01)     & \bf{0.13} (0.01) &             & 0.20 (0.04) & \bf{0.20} (0.04) \\
 & Pakistan & 0.21 (0.03)     & 0.21 (0.03) & 0.21 (0.04) &             & 0.20 (0.01) \\
 & Peru     & 0.15 (0.01)     & 0.15 (0.02) & \bf{0.15} (0.02) & \bf{0.15} (0.01) &               \\
\hline
     \multirow{5}{*}{Median}
    & Ghana    &                 & 0.24 (0.01) & 0.24 (0.01) & 0.24 (0.02) & 0.24 (0.02) \\
 & Honduras & 0.24 (0.01)     &             & 0.24 (0.01) & 0.25 (0.02) & 0.25 (0.02) \\
 & India    & \bf{0.13} (0.01)     & \bf{0.12} (0.01) &             & 0.19 (0.03) & \bf{0.19} (0.03) \\
 & Pakistan & 0.20 (0.03)     & 0.21 (0.03) & 0.21 (0.04) &             & 0.20 (0.01) \\
 & Peru     & 0.15 (0.02)     & 0.15 (0.02) & \bf{0.14} (0.02) & \bf{0.15} (0.01) &               \\
    \hline
    \end{tabular}}
\end{table}

\subsection{Which Health Index Is More Robust for Generalization?}\label{subsec:data.weight}
\citet{banerjee2015multifaceted} constructed a  physical health index to capture the overall physical health of individuals in a household. Specifically, the index is an (equally weighted) average of three standardized variables (z-scores):
did not miss work due to illness ($Y_{\text{notMiss}}(1)$), activities of daily living score ($Y_{\text{act}}(1)$), and perception of health status ($Y_{\text{perc}}(1)$).
In this section, we ask whether there is another way to define the physical health index so that it's less sensitive for generalization.
Formally, suppose we rewrite the physical health index as a weighted average of three z-scores:
\begin{align*}
    O=\alpha_{\text{notMiss}}Y_{\text{notMiss}}(1) + \alpha_{\text{act}}Y_{\text{act}} (1)+\alpha_{\text{perc}}Y_{\text{perc}}(1).
\end{align*}
In the original analysis, the weights $\alpha_{\text{notMiss}}$, $\alpha_{\text{act}}$, and $\alpha_{\text{perc}}$ were set to $1/3$ (i.e., equally weighted).
Our goal is to find a new vector of weights $\alpha = [\alpha_{\text{notMiss}},\alpha_{\text{act}},\alpha_{\text{perc}}]\trans$ in a simplex that minimizes the SLOPE of the mean, i.e., $\psi^{\mean}=\E_{\QO}(O)$.

We focus on households in three countries (India, Ethiopia, Peru) where all three variables that make up the index were measured. %
Also, like before, we filtered households so that the overlap was plausible.

\begin{figure}[!h]
    \centering
    \includegraphics[width=0.7\linewidth]{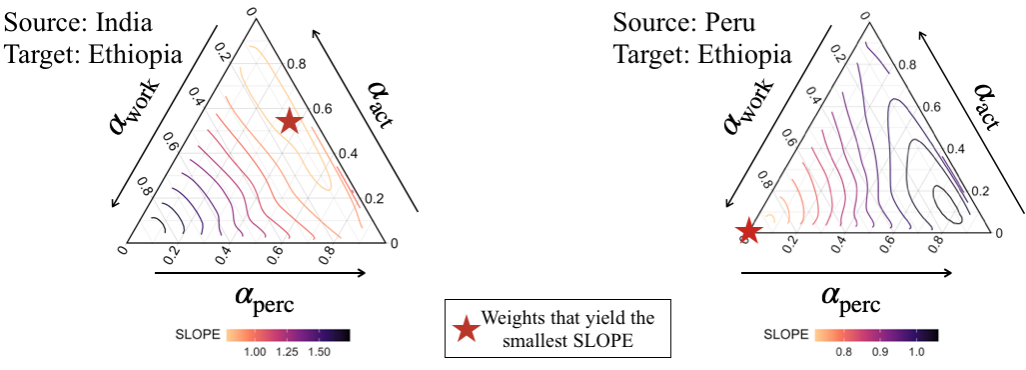}
    \caption{SLOPE for the mean of the physical health index across different weights. In each panel,  edges of the triangle represent the weight in percentage scale of the three z-scores that make up the index. The contours represent the SLOPE where a lighter color means a lower SLOPE. The  star represents the point in the simplex where SLOPE is minimized.}
    \label{fig:data.weights}
\end{figure}

Figure \ref{fig:data.weights} presents the SLOPE across different weight combinations where the target country is Ethiopia and the source country is (left) India or (right) Peru.
When the source country is India, the  SLOPE is minimized at $\alpha_{\text{notMiss}}=0.10$, $\alpha_{\text{act}}=0.55$, $\alpha_{\text{perc}}=0.35$. Upon closer examination, the variable \texttt{notMiss} has the highest variance compared to the other two variables and thus, putting a low weight on it minimizes the overall variance of $O$.
Also, when the source country is Peru,
the SLOPE is minimized with the weight representing the left bottom vertex where $\alpha_{\text{notMiss}}=1$. Again, upon closer examination, this was because the variance of $Z_{\text{notMiss}}$ in Peru is substantially lower than the other two variables.
More broadly, similar to the theoretical discussion from Section \ref{subsec:slope_mean} and Corollary \ref{cor:slope.mean.source}, SLOPE prefers distributions of the physical index that are less variable in order to be less sensitive to generalization.

\section{Extending SLOPE to Other Types of Sensitivity Analysis}\label{subsec:slope.other.models}
Our development of SLOPE has been based on the sensitivity model \eqref{eq:sensitivity}. While our reasons of choosing this model have been discussed in Section \ref{subsec:sensitivity}, we discuss two potential extensions of SLOPE to other sensitivity models.

\subsection{Extensions of the Exponential Tilting Term}
Consider an extension of \eqref{eq:sensitivity} where we replace the exponential tilting term $\exp(\gamma O)$ with a more general, non-negative function $\rho(O,X,\gamma)$. As we show below, SLOPE can still be well-defined with respect to $\rho$ and it maintains its analytic connection to the IF.
\begin{theorem}[SLOPE and IF for $\rho$-Based Sensitivity Model]\label{thm:slope.if.rho}
Consider a broader class of sensitivity models that tilts the density ratio by a non-negative function $\rho(\O,X,\gamma)$:
 \begin{align} \label{eq:sens_model_rho}
         \dfrac{ f_{\QOmidXgamma}(\O,X)}
   { f_{\POmidX}(\O,X)}&\propto \rho(\O,X,\gamma),
   \end{align}
   where $\rho(\O,X,0)=1$ and $\int\rho(\O,X,\gamma) O d \POmidX <\infty$.
  Suppose $\rho$ is differentiable at $\gamma = 0$
   with its derivative denoted as $\dot{\rho}(O,X,0)$
   and  Condition \ref{condiiton:regular.rho.slope}  in the Appendix holds. Then
     \begin{align*}
       \SI (\QOXzero,\psi)
        &= \E_{\QX}
\left\{\E_{\POmidX}\left(\IF(\O,X,\psi(\QOXzero))  \left[\dot{\rho}(\O,X,0) - \E_{\POmidX}\{\dot{\rho}(\O,X,0)\mid X\}\right]
       \right)\right\}.
    \end{align*}
 \end{theorem}
   Compared to Theorem \ref{thm:slope.if}, the relationship between IF and SLOPE defined under the sensitivity model \eqref{eq:sens_model_rho} is driven by %
   $\dot{\rho}$ at the point where conditional exchangeability holds (i.e., $\gamma = 0$). Specifically, the residual variation in $\dot{\rho}$, as measured by $\dot{\rho}(\O,X,0) - \E_{\POmidX}\{\dot{\rho}(\O,X,0)\mid X\}$ can be viewed as the subspace defined in the sensitivity analysis that is not explained by $X$. When $\rho(\O,X,\gamma)=\exp(\gamma\O)$ as in our original sensitivity model \eqref{eq:sensitivity},  the derivative at $\gamma = 0$ is $\dot{\rho}(\O,X,0)=\O$ and the subspace is defined by $O - \mu(X)$.

\subsection{Extensions to Set-Identified Sensitivity Models}\label{subsubsec:slope.bound}
Given a sensitivity parameter $\gamma$, either the preceding sensitivity model \eqref{eq:sensitivity} or the extension in \eqref{eq:sens_model_rho}   induces a \emph{point} identification of the target estimand. In contrast, another popular type of sensitivity models encompasses a \emph{set} of distributions for each $\gamma$, say $\calQ(\gamma)$, and thus  a \emph{set} of values for the target estimand, $\{\psi(\QOX)\text{ for }\QOX\in\calQ(\gamma)\}$; see, for example, \citet{rosenbaum1987sensitivity} and \citet{tan2006distributional}. We call such models ``set-identified'' sensitivity models and this section proposes an extension of SLOPE to such models. We then illustrate the extension  with  \citet{tan2006distributional}'s marginal sensitivity model and \citet{dahabreh2023sensitivity,zeng2025efficient}'s conditional bias model.

For an estimand $\psi(\QOX)$, consider a set-identified sensitivity model that induces the largest and smallest possible values of $\psi(\QOX)$ for a given sensitivity parameter $\gamma \geq 0$,
\begin{align*}
        \psi_+(\calQ(\gamma))=\sup_{\QOX\in\calQ(\gamma)}\psi(\QOX),\quad \psi_-(\calQ(\gamma))=\inf_{\QOX\in\calQ(\gamma)}\psi(\QOX).
    \end{align*}
We assume that $\psi_+(\calQ(\gamma))$ and $\psi_-(\calQ(\gamma))$ are finite and sharp for every $\gamma > 0$ and
$\psi_+(\calQ(0)) = \psi_-(\calQ(0)) = \psi(\QOX)$ for $\gamma = 0$ (i.e., under conditional exchangeability, the target estimand $\psi$ is point-identified). We briefly remark that sharpness is not strictly necessary to define our proposed extension of SLOPE, but can aid interpretation.

 The following definition extends SLOPE for set-identified sensitivity models. \begin{definition}[Extension of SLOPE to Set-Identified Sensitivity Models]\label{def:slope.set}
   Suppose we have a set-identified sensitivity model  $\calQ(\gamma)$ that is parametrized by $\gamma \geq 0$ and induces upper and lower bounds $ \psi_+(\calQ(\gamma))$ and $ \psi_-(\calQ(\gamma))$.
We define the upper and lower SLOPEs as
   \begin{align*}
	    \SI_+(\QOXzero,\psi,\psi_+) =& \lim_{\gamma\downarrow 0}\dfrac{\psi_+(\calQ(\gamma))-\psi(\QOXzero)}{\gamma},\\
	        \SI_-(\QOXzero,\psi,\psi_-) =& \lim_{\gamma\downarrow 0}\dfrac{\psi_-(\calQ(\gamma))-\psi(\QOXzero)}{\gamma},
    \end{align*}
    provided the limits exist.
\end{definition}
  Much like the original proposal, $\SI_-$ and $\SI_+$ are population-level quantities that measure the robustness of a design by measuring the change in the target estimand under a local violation of the conditional exchangeability. For this reason,  $\SI_-$ and $\SI_+$ carry the same interpretations as the original SLOPE in Definition \ref{def:proposal}. When  $\SI_+=\SI_-$,  Definition \ref{def:slope.set} reduces to the original Definition \ref{def:proposal}. When  $\SI_+$ and $\SI_-$ differ in magnitude, one side of the bound may be more sensitive than the other.

To illustrate Definition \ref{def:slope.set}, we consider the mean functional $\psi^{\mean}$ and derive $\SI_-$ and $\SI_+$ under two sensitivity models: \citet{tan2006distributional}'s marginal sensitivity model and \citet{dahabreh2023sensitivity,zeng2025efficient}'s conditional bias model. Notably, the generic representation of SLOPE in terms of IF (Theorem \ref{thm:slope.if} and Theorem \ref{thm:slope.if.rho}) is not guaranteed to hold for sensitivity models other than \eqref{eq:sensitivity} or \eqref{eq:sens_model_rho}. Therefore,  deriving  $\SI_-$ and $\SI_+$ for  estimands  other than $\psi^{\mean}$ will require more knowledge on their corresponding bounds  $\psi_-$ and $\psi_+$; we leave that as an important future direction.

\begin{example}[Marginal Sensitivity Model \citep{tan2006distributional}]\label{ex:msm}
Consider bounding the density ratio between the two conditional distributions with a range characterized by $\gamma$:
\begin{align}
    \exp(-\gamma)\leq \dfrac{f_{\QOmidX}(O,X)}{f_{\POmidX}(O,X)}\leq\exp(\gamma).\label{eq:sensitivity.tan}
\end{align}
The sensitivity model is the set $\calQ^{\text{MSM}}(\gamma) = \left\{
    \QOmidX\times\QX, \text{ where }\QOmidX\text{ satisfies }\eqref{eq:sensitivity.tan}
    \right\}$. Following \citet{dorn2023sharp}, under Condition \ref{condition:regular.msm} of the Appendix, sharp bounds of the target estimands are
    \begin{align*}
        \psi_+^{\mean}(\calQ(\gamma)) =& \E_{\QOXzero}
	        \left[O\cdot \exp\left(\gamma\cdot \sign\left\{O-F_{\POmidX}\inv\left(\dfrac{\exp\{\gamma\}}{1+\exp\{\gamma\}}\,\mid X\right)\right\}\right)\right],\\
	       \psi_-^{\mean}(\calQ(\gamma)) =& \E_{\QOXzero}
	        \left[O\cdot \exp\left(\gamma\cdot \sign\left\{F_{\POmidX}\inv\left(\dfrac{1}{1+\exp\{\gamma\}}\,\middle|\,X\right)-O\right\}\right)\right].
    \end{align*}
   Further suppose Condition \ref{condition:regular.msm.slope} of the Appendix holds. Then, the upper and lower SLOPEs are
    \begin{align*}
        \SI_+^{\msm}(\QOXzero, \psi^{\mean},\psi^{\mean}_+) &= \E_{\POmidX\times \QX}\left(\left[\ind\{O>F\inv_{\POmidX}(1/2)\} -  \ind\{O\leq F\inv_{\POmidX}(1/2)\}\right]O\right), \\
        \SI_-^{\msm}(\QOXzero, \psi^{\mean},\psi^{\mean}_-)&=- \SI_+^{\msm}(\QOXzero, \psi^{\mean},\psi^{\mean}_+).
    \end{align*}
\end{example}

\begin{example}[Conditional Bias Model \citep{dahabreh2023sensitivity,zeng2025efficient}]\label{ex:conditional.mean}
    Consider bounding the absolute difference in conditional means between the target and the source distributions by $\gamma$:
    \begin{align}
      -\gamma\leq   \E_{\QOmidX}(O\mid X)-\E_{\POmidX}(O\mid X)\leq \gamma.\label{eq:sensitivity.cmm}
    \end{align}
    The corresponding sensitivity model is $\calQ^{\cbm}(\gamma) = \left\{
    \QOmidX\times\QX, \text{ where }\QOmidX\text{ satisfies }\eqref{eq:sensitivity.cmm}
    \right\}$. The sharp bounds of the mean functional are $\psi_+^{\mean}=\E_{\QOXzero}(O)+\gamma$ and $\psi_-^{\mean}=\E_{\QOXzero}(O)-\gamma$.  The upper and lower SLOPEs are
    \begin{align*}
    \SI_+^{\cbm}(\QOXzero, \psi^{\mean},\psi^{\mean}_+) &= 1, \\
    \SI_-^{\cbm}(\QOXzero, \psi^{\mean},\psi^{\mean}_-) &= -\SI_+^{\cbm}(\QOXzero, \psi^{\mean},\psi^{\mean}_+) = -1.
    \end{align*}
\end{example}

From Example \ref{ex:msm}, the upper and lower SLOPEs of the mean functional concern the dispersion of $O$ around the conditional median of $\POmidX$, echoing a similar phenomenon under the original sensitivity model \eqref{eq:sensitivity} in Section \ref{sec:slope}. %
In particular, when  $\POmidX\sim N(\mu(X),\sigma^2(X))$, suppressing the arguments of SLOPE for notational simplicity, $\SI_+^{\msm}=-\SI^{\msm}_-=\sqrt{2/\pi}\E_{\QX}\{\sigma(X)\}$. This is a re-scaled version of SLOPE under the original sensitivity model \eqref{eq:sensitivity}, i.e.,  $\SI = \E_{\QX}\{\sigma^2(X)\}$, indicating both the original sensitivity model and the marginal sensitivity model offer similar insights about robust study designs for the mean functional. This observation also matches with \citet{huang2026design} in which a higher variance leads to lower robustness.
We also repeat a remark in Section \ref{subsec:slope.def} that the magnitude of SLOPE cannot be directly compared between two different sensitivity models \eqref{eq:sensitivity} and \eqref{eq:sensitivity.tan}.
In contrast, in Example \ref{ex:conditional.mean}, the upper and lower SLOPEs  are $1$ and $-1$, respectively. They do not change with respect to $\POmidX$ or $\QX$ and thus, offer little insight about robust study designs.

More broadly, studying SLOPE across sensitivity models reproduces the observation in \citet{huang2026design} that different sensitivity models may lead to different design recommendations. From this perspective, SLOPE offers another unintended feature that may be important in practice: an approach to compare sensitivity models and their usefulness to inform study designs. Comparing the three sensitivity models and their SLOPEs for the mean functional, it's clear that the conditional bias model in Example \ref{ex:conditional.mean} provides little insight about robust study designs compared to the marginal sensitivity model in Example \ref{ex:msm} or the original sensitivity model \eqref{eq:sensitivity}.

\section{Discussion}\label{sec:discussion}
This main contribution of this paper is a simple and novel measure, SLOPE, to design robust studies for generalization. Specifically, SLOPE is inspired by principles from robust statistics and is a derivative-based metric that measures the change in the target estimand when there is a local violation of the conditional exchangeability assumption. SLOPE depends on two design-level quantities, (i) the target estimand and (ii) the source and target  distributions $\POmidX$ and $Q_X$, respectively. Changing either of these quantities can result in a different SLOPE and thus, researchers can assess the robustness of competing study designs for generalization.

Inspired by \citet{tipton2018review}'s recommendations for robust generalization, we summarize some advice for designing robust studies based on SLOPE.
\begin{enumerate}
    \item If the target population is fixed and the target estimand is the mean of $O$,  a source population will be less sensitive to local violations of conditional exchangeability if the spread of $O$ given $X$ is small. For instance, the mean's SLOPE becomes smaller when $\sigma^2(X)$ becomes smaller (Corollary \ref{cor:slope.mean.source}). From our data example in Section \ref{subsec:data.choice.source}, countries with a smaller SLOPE (e.g., India and Peru as shown in Table \ref{tab:data.appI}) also have a less variable distribution of $O$ given $X$ (Figure \ref{fig:consumption.box}).
    \item If the source population is fixed and the target estimand is the mean of $O$, it's essential to understand which region of $X$ leads to the least amount of variation in $O$ in the source population. Once this region of $X$ is identified, a target population will be robust for generalization if it  is homogeneous with respect to its $X$ and its $X$s focus around this region. More concretely, Section \ref{subsec:slope_mean} and Corollary \ref{cor:slope.mean.source} discuss an example where $Q_X$ is selected to focus on regions with the least variability in $O$.
    \item If both the source and the target populations are fixed, it's less sensitive to choose a target estimand whose influence function projects more (in proportion) onto the space of shared variable $X$.
    In a trivial example inspired by Theorem \ref{thm:slope.if}, a target functional that concerns $X$ only, i.e., $\psi(\QOX)=\psi(\QX)$, has zero SLOPE. Additionally, as shown in Section \ref{subsec:data.weight}, for a weighted average of several physical health variables, increasing weights to variables that are better explained by $X$ will also reduce the magnitude of SLOPE, thereby improving robustness.
\end{enumerate}

We re-emphasize that SLOPE is developed under the validity of the overlap assumption (i.e., Assumption \ref{assump:overlap}), which means that the source population is already sufficiently large compared with the target population,  in terms of $\PX$ and $\QX$. This assumption should always be empirically assessed.
Moreover, the absolute continuity condition in sensitivity model \eqref{eq:sensitivity} suggests an overlap condition in $O\mid X$, in that the support of $O\mid X$ in the target population cannot exceed that in the source population, and our design recommendations are based on this assumption.
Settings where overlap and conditional exchangeability are simultaneously violated are an important topic, especially in terms of their implications for study design. Recent work by \citet{yu2025using} offers a promising path to extend SLOPE to such settings.
In general, we echo \citet{tipton2018review} and \citet{degtiar2023review} on the importance of guaranteeing conditional exchangeability through discussion with domain experts and careful data collection processes (e.g., by collecting a rich $X$). However,  when it is infeasible or impractical to plausibly satisfy conditional exchangeability with the observed set of $X$, we believe SLOPE is a useful tool to assess the sensitivity/robustness of the underlying study design, and to guide future designs for generalization.

Finally, we highlight extensions and future directions.
First, while this paper focuses on the context of generalizability, by defining $P$ and $Q$ differently, SLOPE  extends to measuring the sensitivity of violating conditional exchangeability in causal inference and missing data problems;
see Section \ref{supp.sec:slope.other.contexts} of the Appendix.
Second, Section \ref{supp.subsec:MIE} of the Appendix connects SLOPE to the marginal interventional effect proposed by \citet{zhou2022marginal} with the incremental propensity score intervention \citep{kennedy2019nonparametric}, highlighting some potential mathematical connections between SLOPE and incremental treatment effects.
Third, as SLOPE is a first-order derivative, another  future direction  is towards higher-order derivatives. For example, for the mean functional $\psi^{\mean}$, the second-order SLOPE is $\E_{\QOXzero}\left[\{O-\mu(X)\}^3\right]$, the expectation of an un-standardized conditional skewness. For other functionals, the second-order SLOPE can be rather complicated; open questions remain on how to use second-order SLOPEs to guide practice.
Fourth, in addition to choosing one source population for a fixed target population in Section \ref{subsec:data.choice.source}, SLOPE can be used to find multiple (instead of single) source populations; see Section \ref{supp.sec:extend.app} of the Appendix for more discussions. Or, given a fixed source population, SLOPE can be used to select the best target population among a set of target populations.

% \section{Data Availability Statement}
% The dataset used in Section \ref{sec:data}  is available at \href{https://doi.org/10.7910/DVN/NHIXNT}{https://doi.org/10.7910/DVN/NHIXNT}.

% \section{Disclosure Statement}
% No potential competing interest was reported by the author(s).

\section*{Acknowledgment}
The authors gratefully acknowledge Guanhua Chen, Melody Huang, Edward Kennedy, Jae-Kwang Kim, Ed Vytlacil, Qingyuan Zhao, participants of the 2024 International Conference on Statistics and Data Science, 2024 Joint Statistical Meetings, 2025 American Causal Inference Conference, and statistics seminar at Seoul National University for their valuable feedback. The work of Xinran Miao is supported in part by funding from American Family Funding Initiative and the Morgridge Summer Research Fellowship.

\bibliographystyle{apalike}
\bibliography{ref}

\clearpage
\appendix
\addcontentsline{toc}{section}{Appendix}
\part*{Appendix}

The Appendix is organized as follows. Section \ref{supp.sec:slope.exist} discusses the existence of SLOPE, including the definition of SLOPE through Hadamard differentiability and some regularity conditions deferred from the main text. Section \ref{supp.sec:slope.choice.estimand} exemplifies the use of SLOPE in choosing a robust location estimand. Section \ref{supp.sec:slope.other.estimands} discusses SLOPE for other estimands, including risks, quantiles, trimmed mean, OLS coefficients, some scale parameters, Pearson correlation coefficient, and general formulas of SLOPEs for L-estimands and Z-estimands.
Section \ref{supp.sec:causal.functional} discusses transporting  functionals of potential outcomes.
Next,  Section \ref{supp.sec:est}  and Section \ref{supp.sec:data} supplement  Section \ref{subsec:estimation} (estimation)  and Section \ref{sec:data} (data application)  in the main text, respectively, by providing details, auxiliary results, and some deferred discussions.
Section \ref{supp.sec:simulation} includes a simulation study that verifies the asymptotic properties of proposed estimators.
Afterwards, Sections \ref{supp.sec:proof.slope} and \ref{supp.sec:proof.estimation} provide proof for the derivation of the SLOPE (at  population level) and the estimation of the SLOPE, respectively.
Section \ref{supp.sec:extended.remarks} details some extended remarks, including SLOPE with a vector valued $\psi(\cdot)$, SLOPE for other types of conditional exchangeability assumptions, and the mathematical connection between SLOPE and the marginal interventional effect.
Finally, Section \ref{supp.sec:slope.bound} provides supplementary materials for SLOPE defined under set-identified sensitivity models.

We note that except for Section \ref{supp.sec:slope.bound}, SLOPE is defined with respect to the sensitivity model \eqref{eq:sensitivity}.

\section{Existence of SLOPE}\label{supp.sec:slope.exist}
\subsection{Notation}\label{supp.subsec:exist.notation}
We introduce some notation for normed spaces and functional analysis \citep{van1996weak}.
For an arbitrary set $T$,  let $l^\infty(T)$ be the set of all uniformly bounded, real functions on $T$: the set with elements satisfying $z:T\to\mathbb{R}$ such that $\lVert z\rVert_T:=\sup_{t\in T}|z(t)|<\infty$.
For two normed spaces $\mathbb{D}$ and $\mathbb{E}$, a map $\phi:\mathbb{D}_{\phi}\subset\mathbb{D}\to \mathbb{E}$ is Hadamard differentiable at $\theta\in\mathbb{D}_{\phi}$ tangentially to $\mathbb{D}_{0}$ if there exists a continuous linear map $\phi_{\theta}':\mathbb{D}\to\mathbb{E}$ such that $\{\phi(\theta+t_nh_n)-\phi(\theta)\}/t_n\to\phi_{\theta}'(h)$ as $n\to\infty$, for all converging sequences $t_n\to0$ and $h_n\to h$ such that $\theta+t_nh_n\in\mathbb{D}_{\phi}$ and $h\in\mathbb{D}_0$, where $\mathbb{D}_{\phi}$ and $\mathbb{D}_0$ are two subsets of $\mathbb{D}$.

We introduce notations for the supports under $\QOXzero=\POmidX\times\QX$. Let $\SOX$ be the support under $\QOXzero$ and $\SX$ and $\SO$ be the supports for the marginals $\QX$ and $\QOzero$, respectively.

\subsection{SLOPE  Through Hadamard Differentiability.}
We define SLOPE through the derivative of a composite of two functionals. First,   define the map from $\gamma$ to $\QOX ^{\gamma}$ as $\phi:\left(\mathbb{R},l^{\infty}(\SOX)\right)\to l^{\infty}(\SOX)$ such that $(\gamma,\QOXzero )\mapsto \QOX ^{\gamma}$, with domain $\mathbb{D}_\phi = \left([-\varepsilon,\varepsilon], \QOXzero \right)\subset \left(\mathbb{R},l^{\infty}(\SOX)\right)$  for some $\varepsilon>0$.
    Next, consider the functional $\psi$ as $l^\infty(\SOX) \to \mathbb{R}$ with domain $\mathbb{D}_{\psi}\subset l^\infty(\SOX)$ which contains probability distributions on $\SOX$. Then holding $\QOXzero $ fixed, SLOPE as defined in \eqref{eq:slope.def} is  the Hadamard derivative of the composite function $\psi\circ\phi$ with respect to $\gamma$ at zero. It exists under Condition \ref{condition:regular.hadamard}, the standard condition that enables the chain rule of Hadamard differentiability. For Z-estimands, SLOPE exists under Condition \ref{condition:regular.z.slope} of the main text. Part (i)  ensures exchanging integration and differentiation and the existence of SLOPE as an integral; parts (ii) and (iii) are analogous to regularity conditions that ensures the existence of the IF of $\psi$ (see Section \ref{subsec:slope.if}).

\begin{condition}[Existence of SLOPE]\label{condition:regular.hadamard}
Suppose $\phi$ as a function of $\gamma$ is Hadamard differentiable at $0$
    and $\psi$ is Hadamard differentiable at $\phi(0,\QOXzero )=\QOXzero $ tangentially to $\phi'_{0}(\{0\})$.
\end{condition}

 Condition \ref{condition:regular.if.hadamard} is a regularity condition that  ensures IF is an evaluation of the Hadamard derivative and therefore its connection to SLOPE follows from the linearity of Hadamard differentiability.
We note that Hadamard differentiability
(i.e., Condition  \ref{condition:regular.if.hadamard}) is stronger than the directional differentiability
 (i.e., \eqref{eq:if.direction} in the main text) since the former requires the limit to exist for every sequence of directions that converges to  $\delta_{o,x}-\QOXzero $ whereas the latter only requires the limit to exist in this single direction.
\begin{condition}[IF as an Evaluation of a Hadamard Derivative]\label{condition:regular.if.hadamard}
    Suppose $\psi$ is Hadamard differentiable at $\QOXzero $ tangentially to $\delta_{o,x}-\QOXzero $.
\end{condition}

\subsection{Regularity Conditions}
\begin{condition}[Existence of SLOPE for the Mean]\label{condition:regular.mean}
 Suppose $\E_{\POmidX}\{\O\exp(\gamma\O)\mid X\}/\E_{\POmidX}\{\exp(\gamma\O)\mid X\}$ is uniformly bounded by an integrable function under $\QX$ for $\gamma$ in a neighborhood of zero, and $\sigma^2(X)<\infty$ almost surely on $\QX$.
\end{condition}

\begin{condition}[Existence of SLOPE for the Median]\label{condition:regular.median}
   (i) Suppose $F_{\QOzero}$ is differentiable at $m_{1/2}$ with a positive derivative.\\
   (ii) Suppose $\E_{\POmidX}\left[
 \O\ind(\O\leq m_{1/2})\exp(\gamma\O)\mid X
   \right]/\E_{\POmidX}(\O\mid X)$ is uniformly bounded by an integrable function under $\QX$ for $\gamma$ in a neighborhood of zero, and $ \E_{\QX}\left[F_{\POmidX}(m_{1/2}\mid X)\mu( X)\right]$ and $ \E_{\QOXzero}\left[\O\ind(\O\leq m_{1/2})\right]$ exist.
\end{condition}

\begin{condition}[Existence of SLOPE for the $q$-th Quantile]\label{condition:regular.quantile}
   (i) Suppose $F_{\QOzero}$ is differentiable at $m_{q}$ with a positive derivative.\\
   (ii) Suppose $\E_{\POmidX}\left[
 \ind(\O\leq m_{q})\O\exp(\gamma\O)\mid  X
   \right]/\E_{\POmidX}(\O\mid X)$ is uniformly bounded by an integrable function under $\QX$ for $\gamma$ in a neighborhood of zero, and $ \E_{\QX}\left[F_{\POmidX}(m_{q}\mid X)\mu( X)\right]$ and $ \E_{\QOXzero}\left[\O\ind(\O\leq m_{q})\right]$ exist.
\end{condition}

\begin{condition}[Existence of SLOPE for Sensitivity Model Defined Through $\rho(O,X,\gamma)$]\label{condiiton:regular.rho.slope}
  (i) $\E_{Q_{O\mid X}^{\gamma}}[s(O,X,\psi(Q_{O,X}^0))]$ is bounded %
  for $\gamma$ in a neighborhood of zero and $$\E_{\QOXzero}\left( s(O,X,\psi(Q_{O,X}^0))  \left[\dot{\rho}(O,X,0) - \E_{\POmidX}\{\dot{\rho}(O,X,0)\mid X\}\right] \right)$$ exists; (ii)  $s(O,X,\cdot)$ is differentiable almost everywhere with the derivative $\dot{s}(O,X,\cdot)$;  (iii)  $\E_{\QOXzero}
       \{ \dot{s}(\O,X,\psi(\QOXzero))\}$ exists and is non-singular;
       (iv) $\rho(O,X,\gamma)$ is twice differentiable with respect to $\gamma$  at a neighborhood of zero with derivative $\ddot{\rho}(O,X,\gamma)$.
\end{condition}

\clearpage
\section{Using SLOPE to Choose A Robust Location Parameter}\label{supp.sec:slope.choice.estimand}

In this section, we illustrate how SLOPE can guide the choice of a robust location parameter. We focus on comparing two estimands, the mean with functional $\psi^\mean$ and the median with functional $\psi^\med$. Their SLOPEs have been stated in Theorems \ref{thm:slope.mean} and \ref{thm:slope.median} in the main text.
When $\POmidX$ is normal, we recall that in this case the both SLOPEs are weighted average of the conditional variance $\sigma^2(X)$.
In this case, Theorem \ref{thm:mean.median.cov} lists some sufficient conditions where the SLOPE for the mean is larger than (or equal to) the SLOPE for the median.

\begin{theorem}[Comparison in SLOPEs of Mean and Median]\label{thm:mean.median.cov}
Suppose $\POmidX  \sim N\left(\mu(X),\sigma^2(X)\right)$.\\
(a) If $\sigma^2(X)=\sigma^2$ almost surely $\QX$, then $\SI(\QOXzero,\psi^{\mean})=\SI(\QOXzero,\psi^{\med})=\sigma^2$.\\
(b) If $\mu(X)=\mu$ almost surely $\QX$, then $\SI(\QOXzero,\psi^{\mean})\geq \SI(\QOXzero,\psi^{\med})$.  \\
(c) More generally, $\SI(\QOXzero ,\psi^\mean)\geq \SI(\QOXzero ,\psi^\med)$ if and only if
  \begin{align*}%
     \cor_{\QX}[
            \sigma^2(X), f_{\POmidX}(m_{1/2}\mid X)
            ]\leq 0,\text{ where }\cor\text{ represents correlation}.
 \end{align*}
\end{theorem}
In part (a) of Theorem \ref{thm:mean.median.cov} where the conditional variance is constant, the SLOPEs for the median and the mean are identical.
From part (b) of Theorem \ref{thm:mean.median.cov},  roughly speaking, when the conditional means of $\POmidX$ are sufficiently uniform, the median is more robust than or equally robust with the mean in terms of violation of conditional exchangeability.
Intuitively, this is because the conditional density at the median, i.e., $f_{P_{O\mid X=x}}(m_{1/2})$, should be higher for $x$'s with a lower conditional variance ($\sigma^2(x)$).
Then $\SI(\QOXzero,\psi^{\med})$, as a weighted average of $\sigma^2(x)$, will assign a higher weight  to $x$'s with a lower $\sigma^2(x)$. This makes the SLOPE for the median no larger than the SLOPE for the mean, which is an (unweighted) average of $\sigma^2(x)$.

\begin{figure}[!h]
    \centering
    \includegraphics[width=1\linewidth]{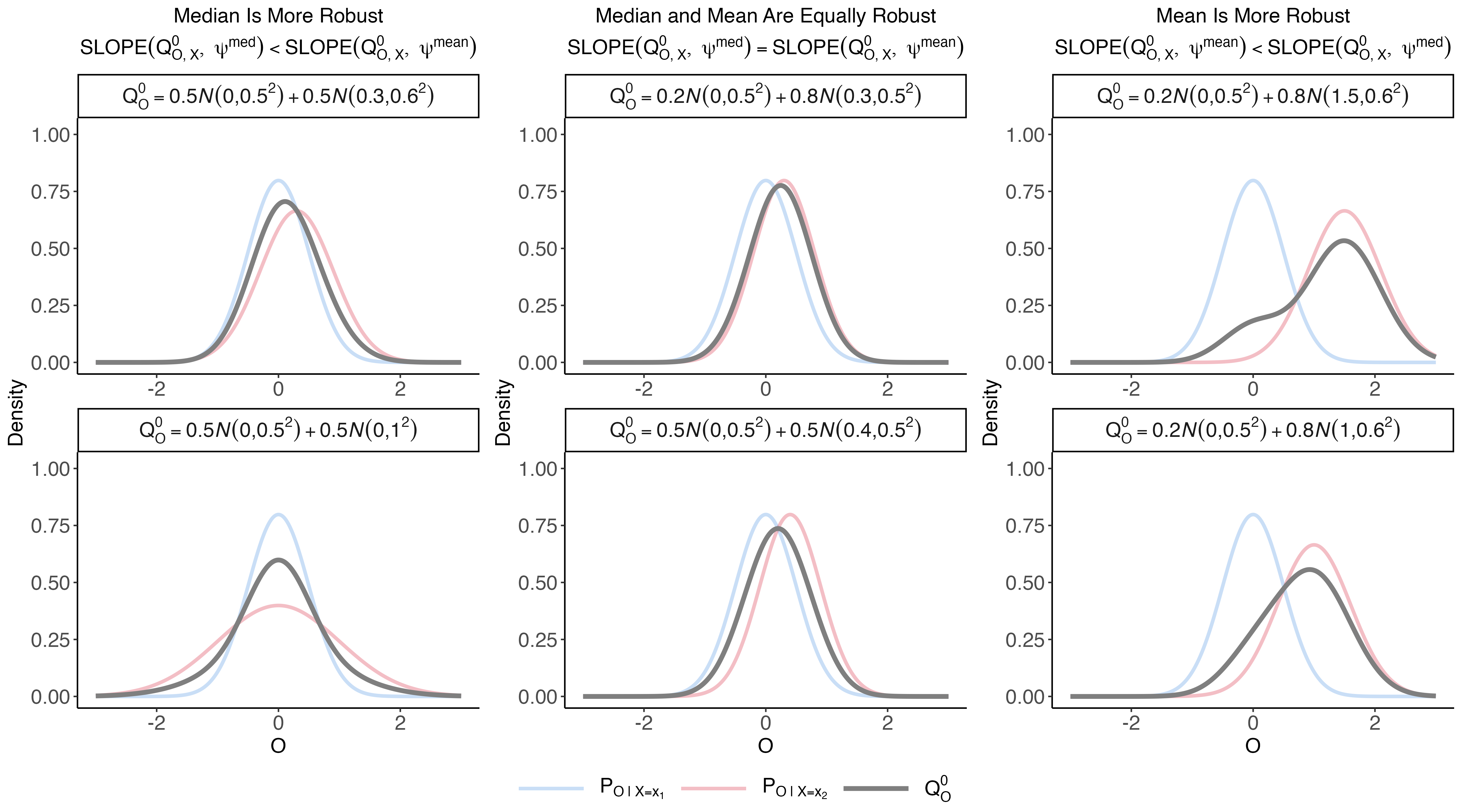}
    \caption{Some toy examples where the SLOPE of the median is more, equal, or less than the SLOPE of the mean.}
    \label{fig:toy.density.compare}
\end{figure}

While Theorem \ref{thm:mean.median.cov} lists some sufficient conditions for the median to be more robust than the mean, in general, the relationship can be reversed.
For example, suppose the target population contains two subgroups  indicated by $X\in\{x_1,x_2\}$. In each subgroup, the source distribution is normal with $P_{O\mid X=x_1}\sim N(\mu_1,\sigma_1^2)$ and $P_{O\mid X=x_2}\sim N(\mu_2,\sigma_2^2)$, where we set $\mu_1=0$, $\sigma_1=0.5$, and $\sigma_2=0.6$.
Then SLOPE of the median becomes
\begin{align*}
\SI(\QOXzero ,\psi^\med) =    (1-w_2)\sigma_1^2 + w_2\sigma_2^2,
\end{align*}
where $w_2$ is the weight for the more heterogeneous subgroup ($X=x_2$) with the higher  variance $\sigma_2^2$.
As shown in  Lemma \ref{lemma:slope.median.gaussian.two}, once $\mu_1$ and $\sigma_2>\sigma_1$ are fixed,  $w_2$ (as a function of $\mu_2-\mu_1$) is monotonically increasing with  $\mu_2-\mu_1$, the mean difference between the two subgroups.
Therefore, as $w_2$ increases, the SLOPE for the median assigns an increasing weight for the heterogeneous subgroup, which will eventually exceed the SLOPE for the mean. %
Intuitively, as $\mu_2$ increases, the marginal target distribution under exchangeability $\QOzero \sim 0.2N(0,0.5^2)+0.8N(\mu_2,0.6^2)$ becomes more asymmetric and less uni-modal, and thus the usual understanding on median's robustness no longer holds. Figure \ref{fig:toy.density.compare} provides a visual illustration of different underlying marginal distributions $\QOzero$ and how they correspond to different SLOPEs for the median and the mean.

\clearpage
\section{SLOPE for Other Estimands}\label{supp.sec:slope.other.estimands}
In this section we provide examples of SLOPE with some other target functionals $\psi(\cdot)$ that were not included in the main text.

\subsection{SLOPE for Expectations}\label{app.sec:slope.expectation}
Suppose the target estimand is the expectation of a known function $\xi(O,X)$,
\begin{align}\label{eq:psi.xi}
    \psi^{\xi}(\QOX) &= \E_{\QOX}\{\xi(\O,X)\}.
\end{align}
Lemma \ref{lemma:slope.expectation} shows that SLOPE is the expectation (under $\QX$) of the conditional covariance (under $\POmidX)$ of $\O$ and $\xi(\O,X)$.
\begin{lemma}\label{lemma:slope.expectation}
    Suppose Condition \ref{condition:regular.z.slope} holds with $s(O,\psi^{\xi}) = \psi^{\xi}-\xi(O,X)$. Then SLOPE for the functional $\psi^{\xi}$ defined in \eqref{eq:psi.xi} is
    \begin{align*}
        \SI(\QOXzero,\psi^{\xi}) &= \E_{\QX}\left\{\cov_{\POmidX}\left[\O,\xi(\O,X)\mid X\right]\right\}.
    \end{align*}
\end{lemma}
For example, if $\xi(\O,X)=\O$, then $\psi^{\xi}=\psi^{\mean}$ and the SLOPE is the expectation of the conditional variance of $\O$. Additional examples include the centered moment  in Example \ref{ex:center.moment}, risks and excess risks in Section  \ref{supp.subsec:slope.risk}.

\begin{example}[Centered Moments]\label{ex:center.moment}
Consider $\psi(\QOX) = \E_{\QX}[\{\O-\mu(X)\}^k]$ for a positive integer $k$. Then the SLOPE is $\E_{\QOXzero} [\{\O-\mu(X)\}^{k+1}]$ provided that the expectation exists.
\end{example}

\subsection{SLOPE for Risks}\label{supp.subsec:slope.risk}
Suppose $\delta(X)$ is a (fixed) decision rule that maps from $X$ to the support of $O$. For a loss function $L(O,\cdot)$, define the risk of $\delta(X)$ on the target distribution as $R(\delta)=\E_{\QOX}[L(O,\delta(X))]$. In this section we consider  SLOPEs for risks and excess risks. Note that they can be viewed as special cases of the SLOPE for expectations in Lemma \ref{lemma:slope.expectation}  with different choices of $\xi$.
\subsubsection{SLOPE for Mean Squared Error}
Consider the squared loss with $L(O,\delta(X))=\{O-\delta(X)\}^2$. The target functional is the mean squared error of the decision $\delta(X)$ on the target population, i.e., $\psi = \E_{\QOX}[\{O-\delta(X)\}^2]$. Then the SLOPE is $\SI(\QOXzero,\psi) = \E_{\QOXzero}\left[\{\delta(X)-O\}^2\{\O-\mu(X)\}\right]$.

\subsubsection{SLOPE for Excess Risk under Squared Loss}
Consider again the squared loss $L(O,\delta(X))=\{O-\delta(X)\}^2$. Then $\mu(X)$ minimizes the corresponding risk. For any (fixed) $\delta(X)$, let the target functional be the excess risk, i.e., $\psi(\QOX) =R(\delta)-R(\mu) = \E_{\QOX}\left[\{\delta(X)-\O\}^2 - \{\mu(X)-\O\}^2\right]$. SLOPE for this excess risk under squared loss is
\begin{align*}
    \SI(\QOXzero,\psi) = 2\E_{\QX}\left[\{\mu(X)-\delta(X)\}\sigma^2(X)\right].
\end{align*}

\subsubsection{SLOPE for Excess Risk under 0-1 Loss}
Suppose $O\in\{1,-1\}$ and let  $\eta(X) = P(O=1\mid X)$. Consider the 0-1 loss, $L(O,\delta(X)) = \ind\{\delta(X)\neq O\}$, and the corresponding risk
\begin{align*}
R(\delta) = \E_{\QOX}\{L(O,\delta(X)\} = \E_{\QOX}\left[\ind\{\delta(X)\neq1\}\{2\eta(X)-1\}+1-\eta(X)\right].
\end{align*}
Then one of the Bayes classifiers is $\delta\star(X) = \ind\{\eta(X)\geq 1/2\} - \ind\{\eta(X)<1/2\}$. For any fixed $\delta(X)$, let the target functional be the excess risk, $\psi(\QOX) = R(\delta)-R(\delta\star)$. Then the SLOPE for the excess risk under 0-1 loss is
\begin{align*}
    \SI(\QOXzero,\psi) =\E_{\QX}\left[\sigma^2(X)\ind\{\delta(X)\neq\delta\star(X)\}\sign\{\eta(X)-1/2\}\right],
\end{align*}
where $\sign(\cdot)$ is the sign function such that $\sign(t)=1$ if $t>0$, $\sign(t)=0$ if $t=0$, and $\sign(t)=-1$ if $t<0$.

\subsection{SLOPE for Quantiles}\label{supp.sec:slope.quantile}
For $q\in(0,1)$, let $\psi(\QO) = F_{\QO}\inv(q)$ be the $q$ quantile of the marginal distribution $\QO$; for example when $q=1/2$, $\psi=\psi^{\med}$. Then SLOPE for the quantile, as a general case of the SLOPE for the median (Theorem \ref{thm:slope.median}), is presented in Theorem \ref{thm:slope.quantile}.
\begin{theorem}[SLOPE for Quantiles]\label{thm:slope.quantile}
    Suppose Condition \ref{condition:regular.quantile} holds. Then SLOPE for the $q$-th quantile is \begin{align*}
   \SI(\QOXzero,\psi) =& \dfrac{
        \E_{\QX}\left[F_{\POmidX}(m_q\mid X)\mu( X)\right] - \E_{\QOXzero}\left[\O\ind(\O\leq m_q)\right]
        }{f_{\QOzero}(m_q)},
    \end{align*}
    where $m_q$ satisfies $\int_{-\infty}^{m_q}d\QOzero=q$.
\end{theorem}

\subsection{SLOPE for $\alpha$-Trimmed Mean}
Let the target functional be the $\alpha$-trimmed mean, $\psi^{\trim}$, which is the mean of $\QO$ after trimming off the lower and upper $\alpha$ quantiles. Specifically, the $\psi^{\trim}$ is defined as
\begin{align}\label{eq:trimmed.mean}
    \psi^{\trim}(\QO) = \dfrac{1}{1-2\alpha}\int_{\alpha}^{1-\alpha}F_{\QO}\inv(p)dp,
\end{align}
for $\alpha\in(0,1/4)$.
\begin{theorem}[SLOPE for $\alpha$-Trimmed Mean]\label{thm:slope.trimmed.mean}
SLOPE for the $\alpha$-trimmed mean $\psi^{\trim}$ defined in \eqref{eq:trimmed.mean} is
\begin{align*}
         \SI(\QOXzero,\psi^{\trim}) =&
        \frac{F\inv_{\QOzero}(1-\alpha)}{1-2\alpha}\E_{\QOXzero}\left\{F_{\POmidX}( F_{\QOzero}\inv(1-\alpha))\mu(X) - \O\ind_{(-\infty,F_{\QOzero}\inv(1-\alpha)]}(\O)\right\}\\
        &-  \frac{F\inv_{\QOzero}(\alpha)}{1-2\alpha}\E_{\QOXzero}\left\{F_{\POmidX}(F_{\QOzero}\inv(\alpha))\mu(X) - \O\ind_{(-\infty,F_{\QOzero}\inv(\alpha)]}(\O)\right\}\\
        &+\dfrac{1}{1-2\alpha}\E_{\QOXzero}\left[
        \O\{\O-\mu(X)\}\ind_{\left(F_{\QOzero}\inv(\alpha),F_{\QOzero}\inv(1-\alpha)\right)}(\O)
        \right].
\end{align*}
Moreover, if $\POmidX\sim N\left(\mu(X),\sigma^2(X)\right)$, then the SLOPE is
\begin{align*}
    \SI(\QOXzero,\psi^{\trim}) =
        \dfrac{1}{1-2\alpha}
        \E_{\QX}\left[\sigma^2(X)\left\{
        \Phi\left(\frac{F_{\QOzero}\inv(1-\alpha)-\mu(X)}{\sigma(X)}\right)-
                \Phi\left(\frac{F_{\QOzero}\inv(\alpha)-\mu(X)}{\sigma(X)}\right)
        \right\}\right],
\end{align*}
where $\Phi(\cdot)$ is the cumulative distribution function of standard normal.
\end{theorem}

\subsection{SLOPE for OLS Coefficients}\label{supp.sec:ols}
Suppose $\O=Y$ is an outcome variable and $X$ is a vector of covariates that includes one. We are interested in the OLS coefficient of regressing $Y$ on $X$ in the target distribution, i.e., $\psi^{\ols}(Q_{Y,X})$ such that
 \begin{align}\label{eq:psi.ols}
\E_{Q_{Y,X}}\left[XX\trans\psi^{\ols} - XY\right]=0.
\end{align}

\subsubsection{SLOPE for OLS Coefficients with Omitted Variables}
We consider transferring OLS coefficients $\psi^{\ols}$ in  \eqref{eq:psi.ols} where $X$ is a vector of covariates and contains one as the intercept, $\O=Y$ is a scalar outcome variable.
The target estimand $\psi^{\ols}$ can be equivalently defined as the solution to the following least squares problem,
\begin{align}\label{eq:ols}
    \psi^{\ols}=\arg\min_{\psi^{\ols}}\E_{\QOX}\left[\left(Y-X\trans\psi^{\ols}\right)^2\right]
\end{align}
For generality, we also consider the OLS coefficient where only a subset $\Xsub\subset X$ that contains the intercept is being modeled:
\begin{align}\label{eq:ols.sub}
    \psi^{\olssub}=\arg\min_{\psi^{\olssub}}\E_{\QOX}\left[\left(Y-\Xsub\trans\psi^{\olssub}\right)^2\right].
\end{align}
Note that \eqref{eq:ols.sub} contains \eqref{eq:ols} as a special case by setting $\Xsub=X$.

\begin{theorem}[SLOPE for OLS Coefficient]\label{thm:slope.ols}
Suppose Condition \ref{condition:regular.z.slope} holds with $s(Y,\psi^{\olssub}) = \Xsub\Xsub\trans\psi^{\olssub}-\Xsub Y$.
    SLOPE for the OLS coefficient with a subset of covariates \eqref{eq:ols.sub} is
    \begin{align*}
        \SI(\QOXzero,\psi^{\olssub}) =
        \left\{\E_{\QX}\left(\Xsub\Xsub\trans\right)\right\}\inv\E_{\QX}\left\{\Xsub\sigma^2(X)\right\}.
    \end{align*}
\end{theorem}
In the special case where $\Xsub=1$, coefficient $\psi^{\olssub}$ becomes the outcome mean (Theorem \ref{thm:slope.mean}). As expected, SLOPE in Theorem \ref{thm:slope.ols} becomes $\E_{\QX}\{\sigma^2(X)\}$, which is identical to the SLOPE for the mean. On the contrary, in another special case where $\Xsub=X$, the SLOPE becomes $ \SI(\QOXzero,\psi^{\ols}) =
        \left\{\E_{\QX}\left(XX\trans\right)\right\}\inv\E_{\QX}\left\{X \sigma^2(X)\right\}$.

\subsubsection{SLOPE in ANCOVA}
Suppose $X=[1, A, L\trans]\trans$ where $A$ is a binary treatment and $L$ contains pre-treatment covariates. We are interested in the effect of $A$ on $Y$ in two models of analysis of covariance (ANCOVA)  in which \eqref{eq:ols.adj} adjusts for covariates $L$ and \eqref{eq:ols.unadj} does not:
\begin{align}
    (\alpha\adj,\tau\adj,\beta\adj)&=\arg\min_{(\widetilde\alpha\adj,\widetilde\tau\adj,\widetilde\beta\adj)}
    \E_{\QOXzero}\left[\left(Y-\widetilde\alpha\adj-\widetilde\tau\adj A-\widetilde\beta\adj\trans L\right)^2\right],\label{eq:ols.adj}\\
     (\alpha\unadj,\tau\unadj)&=\arg\min_{(\widetilde\alpha\unadj,\widetilde\tau\unadj)}
    \E_{\QOXzero}\left[\left(Y-\widetilde\alpha\unadj-\widetilde\tau\unadj A\right)^2\right]. \label{eq:ols.unadj}
\end{align}
Specifically, consider SLOPE for the regression slopes $\tau\adj$ and $\tau\unadj$.
We will see that the conditional variance $\sigma^2(X)$ still plays an important role. To reflect  its dependency on the treatment $A$ and covariates $L$, we slightly abuse the notation and let $\sigma^2(L,A)=\var_{P_{Y\mid X}}(Y\mid A,L)=\var_{P_{Y\mid X}}(Y\mid X)$.

\begin{remark}[Difference Between Transporting $Y\mid A$ and  Transporting $Y(A)$]
    Coefficients $\tau\adj$ and $\tau\unadj$ can be interpreted as the (causal) treatment effect of $A$ on $Y$ under modeling assumptions. Nevertheless, they are different from the causal effects in Section \ref{supp.sec:causal.functional} which are based on transporting potential outcomes, even under the same modeling assumptions.\\
    In Section \ref{supp.sec:causal.functional}, the transportation is for the potential outcome $\O=Y(A)$, based on the commonly observed pre-treatment covariates in the two populations (i.e., $X$). The SLOPE depends on target population only through  the distribution of pre-treatment covariates; notably, the treatment $A$ itself need not  be well defined on the target population.
    The SLOPE depends on the source population only through the distribution of the potential outcome $Y(A)$ given pre-treatment covariates. It does not directly depend on how $A$ has been randomized on the source population, as longs as the identification is guaranteed. This setting is widely adopted in generalizing a causal effect from one population where an experimental has taken place to another population where only baseline covariates are collected.\\
    In this Section, the transportation is for the observed outcome $\O=Y$, based on the commonly observed intervention and pre-treatment covariates (i.e., $X$ contains both $A$ and pre-treatment covariates). The intervention $A$ is observed on the target population and the SLOPE depends on the target population not only through covariates, but also through $A$.  This setting is less common than the previous one. One example of this setting is in surveys (or in general, missing data) where $P$ contains units whose outcome variables are observed and  $Q$ contains units whose outcome variables are not observed.
\end{remark}

The following  Theorem \ref{thm:slope.ancova} gives the SLOPE for regression slopes $\tau\adj$ and $\tau\unadj$.
\begin{theorem}[SLOPE in ANCOVA]\label{thm:slope.ancova}
    The SLOPEs for $\tau\unadj$ and $\tau\adj$ are
    \begin{align}
        \SI(\QOXzero, \tau\unadj) &= \dfrac{\cov_{Q_{L,A}}[A,\sigma^2(L,A)]}{\var_{Q_A}(A)},\label{eq:slope.tau.unadj}\\
         \SI(\QOXzero, \tau\adj) &=  \SI(\QOXzero, \tau\unadj) +
         \delta\trans V\inv\delta\cov_{Q_{L,A}}[A,\sigma^2(L,A)] - \delta\trans V\inv\cov_{Q_{L,A}}[L,\sigma^2(L,A)],\label{eq:slope.tau.adj}
    \end{align}
    respectively, where $\delta = \cov_{Q_{X}}[L,A]/\var_{Q_{A}}[A]$ and $V=\cov_{Q_L}(L)-\delta\delta\trans\var_{Q_A}(A)$.
\end{theorem}

From Theorem \ref{thm:slope.ancova}, if $\cov_{Q_{L,A}}(L,A)=0$ almost surely, which can happen  when the intervention $A$ has been randomized, then by definition $\delta=0$ and
    \begin{align*}
        \SI(\QOXzero, \tau\adj) =\SI(\QOXzero, \tau\unadj)=\cov_{Q_{L,A}}[A,\sigma^2(L,A)]/\var_{Q_{A}}(A).
    \end{align*}

\subsection{SLOPE for Scale Parameters}
\subsubsection{SLOPE for Variance}\label{supp.subsubsec:slope.var}
Let the target functional be the variance of $\O$, i.e., $\psi^{\text{var}} = \var_{\QO}(O)$. Then the SLOPE is
\begin{align*}
    \SI(\QOXzero,\psi^{\text{var}}) = \E_{\QX}\left\{\cov_{\POmidX}[\O^2,\O\mid X] \right\} - 2\E_{\QOzero}(\O)\E_{\QX}\{\sigma^2(X)\}.
\end{align*}

\subsubsection{SLOPE for MAD}
Let the target functional be the median absolute deviation from the median (MAD), $\psi^{\mad}(\cdot)$ such that
\begin{align}\label{eq:mad}
    \QO\left(\left|\O-\psi^{\med}(\QO)\right|\leq \psi^{\mad}\right) = 1/2,
\end{align}
where $\psi^{\med}$ is the functional that maps to the marginal median which has been defined in Example \ref{ex:transfer.median} of the main text. Recall that $m_{1/2}=\psi^{\med}(\QOzero)$ and to ease notation, let $\mad = \psi^{\mad}(\QOzero)$ be the MAD of $\QOzero$.

\begin{theorem}[SLOPE for MAD]\label{thm:slope.mad}
Suppose Condition \ref{condition:regular.hadamard} holds for $\psi^{\mad}$. Then SLOPE for the MAD defined through $\psi^{\mad}$ in \eqref{eq:mad} is
\begin{align*}
    \SI(\QOXzero,\psi^{\mad}) =&
    \dfrac{f_{\QOzero}\left(m_{1/2} - \mad\right) -
    f_{\QOzero}\left(m_{1/2} + \mad \right)
    }
    {f_{\QOzero}\left(m_{1/2} -\mad\right) +
    f_{\QOzero}\left(m_{1/2} + \mad\right)} \cdot\SI(\QOXzero,\psi^{\med}) \\
    &-
    \dfrac{
    \E_{\QOXzero} \left[
    \ind_{[m_{1/2} -\mad,m_{1/2} + \mad]}(\O)
    \{\O-\mu(X)\}
    \right]
    }{f_{\QOzero}\left(m_{1/2}-\mad\right) +
    f_{\QOzero}\left(m_{1/2} + \mad\right)},
\end{align*}
where $\SI(\QOXzero,\psi^{\med})$ is the SLOPE for median (Theorem \ref{thm:slope.median}).
\end{theorem}

\subsubsection{SLOPE for $\alpha$-Quantile Range}
The $\alpha$-quantile range of $\QO$ is $\psi^{\alpha\text{-range}}(\QO) = F_{\QO}\inv(1-\alpha)-F_{\QO}\inv(\alpha)$ for a given $\alpha\in(0,1/2)$. A common choice is to let $\alpha=1/4$ and the $\alpha$-quantile range is the interquartile range. By the SLOPE for quantiles (Theorem \ref{thm:slope.quantile}), the SLOPE for the $\alpha$-quantile range is
\begin{align*}
    \SI(\QOXzero,\psi^{\alpha\text{-range}}) = &
    \dfrac{
    \E_{\QOXzero}\left[(1-\alpha)\O - \ind\left(\O\leq F_{\QOzero}\inv(1-\alpha)\right)\O\right]
    }{f_{\QOzero}\left(F_{\QOzero}\inv(1-\alpha)\right)}\\
   & -
      \dfrac{
    \E_{\QOXzero}\left[\alpha\O - \ind\left(\O\leq F_{\QOzero}\inv(\alpha)\right)\O\right]
    }{f_{\QOzero}\left(F_{\QOzero}\inv(\alpha)\right)}.
\end{align*}

\subsection{SLOPE for  Pearson Correlation Coefficient}
Suppose $\O\in\mathbb{R}$ and let the target functional be the Pearson correlation between $\O$ and $X$:
\begin{align}
    \psi^{\cor}(\QOX) = \dfrac{
    \E_{\QOX}(X\O) - \E_{\QX}(X)\E_{\QO}(\O)
    }{
    \sqrt{\var_{\QX}(X)\var_{\QO}(\O)}
    }.
\end{align}
The SLOPE for the Pearson correlation coefficient is
\begin{align*}
    \SI(\QOXzero,\psi^{\cor})
    &=
    \dfrac{\cov_{\QX}[X,\sigma^2(X)]}{\sqrt{\var_{\QX}(X)\var_{\QOzero}(\O)}}
    -\dfrac{
    \cor_{\QOXzero}[X,\O]
    }{2\var_{\QOzero}(\O)}\cdot\SI(\QOXzero,\psi^{\var}),
\end{align*}
where $\SI(\QOXzero,\psi^{\var})$ is the SLOPE for the variance of $\O$ (Section \ref{supp.subsubsec:slope.var}).
We also note that the relationship between SLOPE and IF (Theorem \ref{thm:slope.if}) can be easily verified, where the IF is
\begin{align*}
    \IF\left(\O,X,\psi^{\cor}(\QOXzero)\right) =&
    \dfrac{X-\E_{\QX}(X)}{\sqrt{\var_{\QX}(X)}} \cdot
    \dfrac{\O-\E_{\QOzero}(\O)}{\sqrt{\var_{\QOzero}(\O)}}\\
    &-
    \dfrac{\cor_{\QOXzero}(X,\O)}{2}
    \left\{
    \dfrac{\{X-\E_{\QX}(X)\}^2}{\var_{\QX}(X)} +
    \dfrac{\{\O-\E_{\QOzero}(\O)\}^2}{\var_{\QOzero}(\O)}
    \right\}
\end{align*}
by \cite{devlin1975robust}.

\subsection{SLOPE for L-Estimands}\label{supp.sec:slope.L}

\subsubsection{L-Estimands and Their SLOPEs}
We discuss one important class of estimands whose sample counterparts correspond to L-estimates \citep[Section 3.3]{huber1981robust}. Specifically, consider a one-dimensional functional $\psi(\QOX)=\psi(\QO)$ defined on the marginal distribution of $\O$:
\begin{align}\label{eq:para.L}
     \psi(\QO) &= \int_{0}^1h(F_{\QO}\inv(p))l(p) d p.
\end{align}
Choosing a particular function $h(\cdot)$ and a density function $l$, both defined over the support $(0,1)$, determines a location parameter $\psi(\QO)$. For example, when $l(p)=1$, equation \eqref{eq:para.L} reduces to the mean of $\O$ in the target population, i.e., $\psi^{\mean}(\QO)$. When $l(p)=\ind_{\frac{1}{2}}(p)$ where
 $\ind_C(p)$ for a set $C$ is the indicator function such that $\ind_C(p)=1$ if $s\in C$ and $\ind_C(p)=0$ otherwise, equation \eqref{eq:para.L} becomes the marginal median, $\psi^{\med}(\QO)$.
When $l(p) = \ind_{[\alpha,1-\alpha]}(p)/(1-2\alpha)$, then the target functional becomes the marginal trimmed mean, $\psi^{\trim}$.

Theorem \ref{thm.slope:L.stat} derives the SLOPE
for estimands defined in \eqref{eq:para.L}.

\begin{theorem}[SLOPE for L-Estimands]\label{thm.slope:L.stat}
Suppose  Condition \ref{condition:regular.hadamard} holds with $\psi$ in \eqref{eq:para.L}.
Then SLOPE of $\psi$ defined in \eqref{eq:para.L} is
\begin{align*}
&\SI(\QOX^{0}, \psi)\\
=&\int_{0}^1
\dfrac{-\E_{\QOXzero}\left[O\ind(O\leq F_{\QOzero}\inv(p))\right] +\E_{\QOXzero}\left[
    \mu(X)\ind(\O\leq F_{\QOzero}\inv(p))
    \right]}{f_{\QOzero}\left(F_{\QOzero}\inv(p)\right)} h'\left(F_{\QOzero}\inv(p)\right)
    l(p) d p,
\end{align*}
where $h'(\cdot)$ is the derivative of $h(\cdot)$.
\end{theorem}

\subsubsection{Lemma \ref{lemma:slope.median.gaussian.two}}
\begin{lemma}[SLOPE of Median in Two-Component Gaussian Mixtures]\label{lemma:slope.median.gaussian.two}
    Suppose $X\in\{x_1,x_2\}$ and $q_1=Q_X(x_1)$ and $q_2=Q_X(x_2)$, $P_{O\mid X=x_1}\sim N(\mu_1,\sigma_1^2)$ and $P_{O\mid X=x_2}\sim N(\mu_2,\sigma_2^2)$ with $\sigma_2>\sigma_1$ fixed. We denote the SLOPE for the median as a function of $\Delta=\mu_2-\mu_1$: $\SI(\QOXzero,\psi\med)[\Delta]$. Then the followings hold.\\
    (i) If $0<q_1<q_2$, then $\SI(\QOXzero,\psi\med)[\Delta]$ is increasing with $\Delta$.\\
    (ii) If $q_1>q_2>0$, then $\SI(\QOXzero,\psi\med)[\Delta]$ is decreasing with $\Delta$.\\
    (iii) If $q_1=q_2=1/2$, then $\SI(\QOXzero,\psi\med)[\Delta]=\sigma_1\sigma_2$ does not depend on $\Delta$.
\end{lemma}
\pf{Lemma \ref{lemma:slope.median.gaussian.two}}
We start with parts (i) and (ii). Without loss of generality we suppose $\mu_1=0$ and $\sigma_1=1$ and therefore $\mu_2=\Delta$ and $\sigma_2>1$.

The SLOPE of the median can be expressed as
\begin{align*}
    \SI(\QOXzero,\psi\med)[\Delta] = w_1(\Delta)(\sigma_1^2-\sigma_2^2)+\sigma_2^2,
\end{align*}
The weight is
\begin{align*}
    w_1(\Delta) &= \dfrac{q_1f_{P_{Y\mid X=x_1}}(m_{1/2})}{q_1f_{Y\mid X=x_1}(m_{1/2}) + q_2f_{P_{Y\mid X=x_2}}(m_{1/2})}\\
    &=\dfrac{1}{1+
    \dfrac{q_2}{q_1\sigma_2}\dfrac{\varphi\left((m_{1/2}(\Delta)-\Delta)/\sigma_2\right)}{\varphi(m_{1/2}(\Delta))}
    }\\
    &=\dfrac{1}{1+
    \dfrac{q_2}{q_1\sigma_2}
\exp\left\{\frac{1}{2}\left[
\{m_{1/2}(\Delta)\}^2-\{m_{1/2}(\Delta)-\Delta\}^2/\sigma_2^2
\right]\right\}
    }\\
    &= \dfrac{1}{1+
    \dfrac{q_2}{q_1\sigma_2}\exp\{h(\Delta)\}
    }
\end{align*}
where $m_{1/2}=m_{1/2}(\Delta)$ is the marginal median that satisfies
\begin{align}\label{eq:lemma.pf.median.def}
    \int_{-\infty}^{m_{1/2}(\Delta)} \left[q_1\varphi(y) + q_2\varphi\left(\frac{y-\Delta}{\sigma_2}\right)/\sigma_2\right] d y=1/2.
\end{align}
and $h(\Delta) =\dfrac{1}{2}\left[\{m_{1/2}(\Delta)\}^2-\{m_{1/2}(\Delta)-\Delta\}^2/\sigma_2^2\right]$. Since $\dfrac{q_2}{q_1\sigma_2}>0$ and the function $\exp(\cdot)$ is increasing, to prove $w_1(\Delta) $ is monotonically decreasing (resp. increasing) with $\Delta$, it's sufficient to prove $h(\Delta)$ is monotonically increasing (resp. decreasing) with $\Delta$.

We start with case (i) when $0<q_1<q_2$. From \eqref{eq:lemma.pf.median.def} we know that the derivative of $m(\Delta)$ is
\begin{align*}
   m_{1/2}'(\Delta) = \dfrac{\dfrac{q_2}{\sigma_2}\varphi\left((m_{1/2}(\Delta)-\Delta)/\sigma_2\right)}
    {q_1\varphi(m_{1/2}(\Delta)) + \dfrac{q_2}{\sigma_2}\varphi\left((m_{1/2}(\Delta)-\Delta)/\sigma_2\right)}
\end{align*}
and it satisfies $ m_{1/2}'(\Delta)\in(0,1)$.
Then the derivative of $h(\Delta)$ is
\begin{align*}
    h'(\Delta) =&
    m_{1/2}(\Delta)\cdot m_{1/2}'(\Delta) - \dfrac{m_{1/2}(\Delta)-\Delta}{\sigma_2^2}\left[
    m_{1/2}'(\Delta)-1\right]\\
    =& \left[q_1\varphi(m_{1/2}(\Delta)) + \dfrac{q_2}{\sigma_2}\varphi\left((m_{1/2}(\Delta)-\Delta)/\sigma_2\right)\right]
    \cdot \\
    &
    \left[
  m_{1/2}(\Delta)  \dfrac{q_2}{\sigma_2}\varphi\left(\dfrac{m_{1/2}(\Delta)-\Delta}{\sigma_2}\right)
    + \dfrac{1}{\sigma_2^2}\left\{m_{1/2}(\Delta)-\Delta\right\} q_1\varphi\left(m_{1/2}(\Delta)\right)
    \right]\cdot\\
  =&  \underbrace{\dfrac{1}{\sigma_2}\left[q_1\varphi(m_{1/2}(\Delta)) + \dfrac{q_2}{\sigma_2}\varphi\left((m_{1/2}(\Delta)-\Delta)/\sigma_2\right)\right]}_{>0}
    \cdot \\
    &
    \left[
  q_2 m_{1/2}(\Delta)\varphi\left(\dfrac{m_{1/2}(\Delta)-\Delta}{\sigma_2}\right)
    - \dfrac{q_1}{\sigma_2}\left\{\Delta-m_{1/2}(\Delta)\right\}\varphi\left(m_{1/2}(\Delta)\right)
    \right].\\
\end{align*}
Since the first term $\dfrac{1}{\sigma_2}\left[q_1\varphi(m_{1/2}(\Delta)) + \dfrac{q_2}{\sigma_2}\varphi\left((m_{1/2}(\Delta)-\Delta)/\sigma_2\right)\right]>0$, the sign of $h'(\Delta)$ is determined by the sign of the second term. Hence, we have that
\begin{align*}
    h'(\Delta) >0 &\Longleftrightarrow
     q_2 m_{1/2}(\Delta)\varphi\left(\dfrac{m_{1/2}(\Delta)-\Delta}{\sigma_2}\right)
    - \dfrac{q_1}{\sigma_2}\left\{\Delta-m_{1/2}(\Delta)\right\}\varphi\left(m_{1/2}(\Delta)\right) >0 \\
    &\Longleftrightarrow
    \dfrac{q_2 m_{1/2}(\Delta)\varphi\left(\dfrac{m_{1/2}(\Delta)-\Delta}{\sigma_2}\right)}
    {
    \dfrac{q_1}{\sigma_2}\left\{\Delta-m_{1/2}(\Delta)\right\}\varphi\left(m_{1/2}(\Delta)\right)
    } >1\\
    &\Longleftrightarrow
   \dfrac{q_2}{q_1}\cdot
    \dfrac{
    \varphi\left(\dfrac{\Delta-m_{1/2}(\Delta)}{\sigma_2}\right)/\left\{\Delta-m_{1/2}(\Delta)\right\}
    }
    {
    \varphi\left(m_{1/2}(\Delta)\right)/m_{1/2}(\Delta)
    } >1,
\end{align*}
where the second line follows from the fact that both the numerator and the denominator are positive, and the third line follows by re-organizing terms.
In order to show $h'(\Delta) >0$, it's sufficient to show $ m_{1/2}(\Delta) > \dfrac{\Delta-m_{1/2}(\Delta)}{\sigma_2}$, because
\begin{align*}
    h'(\Delta) >0 &\Longleftarrow \dfrac{
    \varphi\left(\dfrac{\Delta-m_{1/2}(\Delta)}{\sigma_2}\right)/\left\{\Delta-m_{1/2}(\Delta)\right\}
    }
    {
    \varphi\left(m_{1/2}(\Delta)\right)/m_{1/2}(\Delta)
    } >1\\
    &\Longleftarrow m_{1/2}(\Delta) > \dfrac{\Delta-m_{1/2}(\Delta)}{\sigma_2},
\end{align*}
where the first line follows from the fact that $q_2/q_1>1$ and the second line follows since the function $\varphi(x)/x  = \exp(-x^2/2)/x$ is decreasing when $x>0$.

Therefore, we are left to show $m_{1/2}(\Delta) > \dfrac{\Delta-m_{1/2}(\Delta)}{\sigma_2}$.  Recall the definition of the median in \eqref{lemma:slope.median.gaussian.two}, we have
\begin{align*}
   & q_1\Phi\left(m_{1/2}(\Delta)\right) + q_2\Phi\left(\dfrac{m_{1/2}(\Delta)-\Delta}{\sigma_2}\right)=1/2\\
  \Longrightarrow  & q_1\left[\Phi\left(m_{1/2}(\Delta)\right)-1/2\right] + q_2\left[\Phi\left(\dfrac{m_{1/2}(\Delta)-\Delta}{\sigma_2}\right)-1/2\right]=0\\
   \Longrightarrow  &
    q_1\left[\Phi\left(m_{1/2}(\Delta)\right)-1/2\right]  = -q_2\left[\Phi\left(\dfrac{m_{1/2}(\Delta)-\Delta}{\sigma_2}\right)-1/2\right] = q_2\left[\Phi\left(\dfrac{\Delta-m_{1/2}(\Delta)}{\sigma_2}\right)-1/2\right]\\
    \Longrightarrow   &
    \dfrac{\Phi\left(m_{1/2}(\Delta)\right)-1/2}
    {\Phi\left(\dfrac{\Delta-m_{1/2}(\Delta)}{\sigma_2}\right)-1/2}=\dfrac{q_2}{q_1}>1\\
     \Longrightarrow &   m_{1/2}(\Delta) >\dfrac{\Delta-m_{1/2}(\Delta)}{\sigma_2}.
\end{align*}
From the above, we have $h'(\Delta)>0$ and in hence $w_1'(\Delta)<0$. The SLOPE $w_1(\Delta)(\sigma_1^2-\sigma_2^2) + \sigma_2^2$ is therefore increasing with $\Delta$. We have proven part (i).

The proof of part (ii) follows with a similar argument. When $q_2<q_1$, we have $m_{1/2}(\Delta)< \left[\delta-m_{1/2}(\Delta)\right]/\sigma_2$, and therefore $h'(\Delta)<0$. This in turn gives $w_1'>0$ and the SLOPE is decreasing with $\Delta$.

Part (iii) follows by noticing the median is $m_{1/2}=(\mu_1\sigma_2+\mu_2\sigma_1)/(\sigma_1+\sigma_2)$.

$\square$

\subsection{SLOPE for Z-Estimands}
For a Z-estimand defined through \eqref{eq:z.est} in the main text, the corresponding SLOPE is presented in the Corollary \ref{thm:slope.z} below, which is an immediate result of Theorem \ref{thm:slope.if}.
\begin{corollary}[SLOPE for Z-Estimands]\label{thm:slope.z}
    Under Condition \ref{condition:regular.if.hadamard}, the SLOPE for a Z-estimand is
    \begin{align}
        \SI(\QOXzero,\psi) = -\E_{\QOXzero}
       \{ \dot{s}(\O,X,\psi(\QOXzero))\}\inv \E_{\QOXzero}\left[s\left(\O,X,\psi(\QOXzero)\right)\{\O-\mu(X)\}\right].
    \end{align}
\end{corollary}

\section{SLOPE for Functionals of Potential Outcomes} \label{supp.sec:causal.functional}
In this section, we discuss how to use SLOPE to quantify the sensitivity of the conditional exchangeability assumption in generalizing parameters in causal inference. We provide a brief review on notation in causal inference in Section \ref{subsec:causal.notation} and  discuss the SLOPE in Section \ref{subsec:causal.trans}.

\subsection{Notation for Causal Inference}\label{subsec:causal.notation}
Let $Y$ be the observed outcome variable, $A \in \{0,1\}$ denote the observed treatment, and $X$ denote pre-treatment covariates. Let $Y(a)$ denote the potential outcome if the study unit, contrary to fact, was assigned treatment value $a \in \{0,1\}$. Throughout the paper, we assume that we observe only one of the potential outcomes based on the observed treatment assignment, i.e., $Y = Y(A)$ almost surely \citep{rubin1980randomization}. A central goal is in causal inference to identify some functional of $Y(a)$, for example, the mean of $Y(a)$, for $a\in\{0,1\}$.

\subsection{Transporting Functionals of a Potential Outcome}\label{subsec:causal.trans}
Suppose $P$ is the source population where researchers have completed a randomized experiment to study the effect of treatment $A\in\{0,1\}$ on an outcome $Y$. We adopt notation in Section \ref{subsec:causal.notation} and let  $Y(a)$ represent the potential outcome under treatment $a$. The goal is to estimate the ``average'' of the potential outcome $Y(a)$ on the target country, where only the marginal $\QX$ can be identified. In this problem, $\O=Y(a)$ is the potential outcome under the treated. To identify its average on the target distribution, in the literature it's common to impose the overlap condition between $\QX$ and $\PX$ and the conditional exchangeability condition concerns the counterfactual outcome between the source and target populations, i.e., Assumptions \ref{assump:overlap} and \ref{assump:exchange} with $O=Y(a)$. Specifically, Assumption \ref{assump:exchange} becomes:  $\QYaX(\cdot\mid X)$ is absolute continuous with respect to $\PYamidX(\cdot\mid X) $ and the Radon-Nikodym derivative satisfies
\begin{align}\label{eq:exchange.Y(a)}
   d\QYaX(\O,X)/d \PYamidX(\O,X)=1 \text{ almost everywhere } \PYamidX\times\QX
\end{align}
Accordingly, the sensitivity model \eqref{eq:sensitivity} becomes
\begin{align}\label{eq:sensitivity.Y(a)}
     \dfrac{f_{\QYamidXgamma}(\O,X)}{f_{\PYamidX}(\O,X)}\propto \exp(\gamma\cdot  \O) \text{, almost surely } \PYamidX\times\QX.
\end{align}
Following Theorems \ref{thm:slope.mean} and \ref{thm:slope.median}, the SLOPEs for the mean potential outcome and median potential outcome are

\begin{align}
     \SI(Q_{Y(a),X}^0,\psi^{\mean}) =& \E_{\QX}\left\{
     \E_{P_{Y(a)\mid X}}\left[\left\{Y(1)- \E_{P_{Y(a)\mid X}}[Y(a)\mid X]\right\}^2 \Bigg\vert X\right]
     \right\},\text{ and}\label{eq:slope.mean.Y(1)}\\
     \SI(Q_{Y(a),X}^0,\psi^{\med}) =& \dfrac{
        \E_{\QX}\{F_{P_{Y(a)\mid X}}( m_{1/2})\mu( X)\} - \E_{Q_{Y(a)}^0}\{Y(a)\ind(Y(a)\leq m_{1/2})\}
        }{f_{Q_{Y(a)}^0}(m_{1/2})},\label{eq:slope.median.Y(1)}
\end{align}
respectively, where we recall that $m_{1/2}=F_{Q_{Y(a)}^0}(1/2)$ is the marginal median under conditional exchangeability and $\mu(X) = \E_{P_{Y(a)\mid X}}\{Y(a)\mid X\}$ is the conditional mean function.

We make some remarks on the SLOPEs in \eqref{eq:slope.mean.Y(1)}  and \eqref{eq:slope.median.Y(1)} compared with their counterparts for an observed outcome in Theorems \ref{thm:slope.mean} and \ref{thm:slope.median}. First, the SLOPE for
functionals of a potential outcome is no different from in form from the SLOPE for
functionals of an observed outcome, except for a change in notation. This is because the identification strategy in transporting/generalizing a causal quantity is based on the counterfactual outcome $Y(a)$ (i.e., conditional exchangeability in \eqref{eq:exchange.Y(a)}, and so as its violation (i.e., the sensitivity model \eqref{eq:sensitivity.Y(a)}). In this perspective, the SLOPE for a potential outcome should exactly mimic the SLOPE for an observed outcome, as we have seen above.

Second, although the forms look identical, we remind the readers that working with potential outcomes requires more discretion because their SLOPEs (e.g., \eqref{eq:slope.mean.Y(1)}  and \eqref{eq:slope.median.Y(1)}) usually involve the counterfactual distribution $P_{Y(a)\mid X}$ which is  un-identified  without additional causal assumptions.
Fortunately, if the randomization procedure in the source population is known, then the identification is straightforward with additional justifiable assumptions. We discuss these assumption in the followings.

Suppose the randomization is known to depend on $X$ and another variable $V$ which are both measured in the source population, then by design, the strong ignorability assumption (Assumption \ref{assump:strong.ign}) holds on the source population, which in turn provides equivalence between  $P_{Y(a)\mid X,V}$ and  $P_{Y(a)\mid X,V,A=1}$.
Further, under SUTVA (Assumption \ref{assump:sutva}), $P_{Y(a)\mid X,V,A=1}$ equals to $P_{Y\mid X,V,A=1}$.
Consequently, one can replace $P_{Y(1)\mid X}$ in the SLOPE with $\int P_{Y\mid X,V,A=1} d P_{V\mid X, A=1}$, which no longer involves counterfactual quantities and thus can be identified. Under these identification assumptions, the SLOPEs can be expressed in observed quantities; see \eqref{eq:slope.mean.ya} and \eqref{eq:slope.median.ya} in Section \ref{supp.subsec:data.identify.slope}.
In the special case when the randomization solely depends on $X$ (i.e., all confounders can been observed in both the source and the target), we have $P_{Y(1)\mid X}=P_{Y(1)\mid X,A=1}$.

\begin{assumption}[Strong Ignorability on $P$; \citet{rosenbaum1983central}]\label{assump:strong.ign}
    $P_{X,V}(\cdot)$ is absolute continuous with respect to $P_{X,V\mid A=a}(\cdot)$ and $P_{Y(a)\mid X,V}(\cdot\mid x,v)=P_{Y(a)\mid X,V,A=a}(\cdot\mid x,v)$ almost everywhere $P_{X,V}$.
\end{assumption}

\begin{assumption}[Stable Unit Treatment Variable Assumption (SUTVA)]\label{assump:sutva}
    $Y=Y(a)$ if $A=a$ on the source population $P$.
\end{assumption}

\clearpage
\section{Supplementary Materials for Estimation}\label{supp.sec:est}
we provide supplementary results and some deferred discussions for the estimation of SLOPE presented in Section \ref{subsec:estimation} of the main text.
We start with defining some notations in Section \ref{supp.subsec:est.notation}. Then we discuss regularity conditions and asymptotic properties of the weighting estimator and the regression estimator in Sections \ref{supp.subsec:est.weighting} and \ref{supp.subsec:est.regression}, respectively.
Next, we detail   estimators of SLOPEs for the mean, OLS coefficients, and the median in Sections \ref{supp.subsec:slope.mean.est} to \ref{supp.subsec:slope.median.est}, including the weighting and regression estimators presented in the main text and  estimators based on the efficient influence function of the SLOPE (if exists). Then we discuss ways of estimating the nuisance function $\omega(X)$ in Section \ref{supp.subsec:est.nuisance}. Finally, we present a general statement of the efficient influence function for the SLOPE of scalar valued Z-estimands in Section \ref{supp.subsec:eif}.

\subsection{Notation and Setup}\label{supp.subsec:est.notation}
As stated in the main, suppose we have i.i.d. samples from the target population with size $\nq$ and  i.i.d. samples from the source population with size $\np$. We pool the samples together and denote by $T_i=1$ if the sample comes from the target distribution $Q$ and $T_i=0$ otherwise.
Therefore, we have a random sample $\{(X_i,\NA,T_i=1),i=1,\cdots,\nq\}\cup\{(X_i,\O_i,T_i=0),i=\nq+1,\cdots,n\}$ from the  pseudo population that combines $Q$ and $P$, where $\NA$ means unobserved. We will establish the asymptotic properties for the weighting and regression estimators by re-expressing these estimators as solutions to estimating equations, and then apply standard M-estimation theory \citep{van2000asymptotic}.

After introducing the pseudo population that combines $P$ and $Q$, in this section, we use $\pr(\cdot)$ and $\E(\cdot)$, $f(\cdot)$ without subscripts to denote probabilities, expectations, and densities on this pseudo population, and similarly for conditional quantities.
For quantities on a single population (e.g., expectations on $P$ or $Q$ only), there are two equivalent notations. One is to use subscripts, e.g., $\E_{\POmidX}(O\mid X)$ and $f_{\QX}(X)$, as adopted in the main text.
The other is to conditioning on $T$, e.g., $\E(O\mid X, T=0)$ and $f(X\mid T=1)$,  We prefer the first approach in order to be consistent with our convention of notation in the main text, while in this section we  use the second approach if parts of Section \ref{supp.subsec:est.nuisance} when the context is clearer.

\subsection{Asymptotic Properties for the Weighting Estimator}\label{supp.subsec:est.weighting}
To start with, we express the SLOPE in equation \eqref{eq:slope.z.scalar} as
\begin{align*}
   & \SI(\QOXzero,\psi) = -\eta_2\inv \eta_1, \text{ where}\\
   & \eta_2= \left\{\E_{\QX}\left(\E_{\POmidX}
       \left[ \dot{s}(\O,X,\psi)\mid X\right]\right)\right\},
       \text{ and }
       \eta_1 =    \E_{\QX}\left(\E_{\POmidX}\left[\{\O-\mu(X)\} s\left(\O,X,\psi\right)\mid X\right]\right).
\end{align*}
Similarly, we re-express the weighting estimator by the estimates of $\eta_1$ and $\eta_2$ as follows,
\begin{align*}
\wh\SI^{\wt}(\wh{Q}_{O,X}^0,\wh\psi) &= -\left\{\wh\eta_1^{\wt}\right\}\inv \wh\eta_2^{\wt}, \text{ where}\\
    \wh\eta_1^{\wt} &=\sum_{i=\nq+1}^{n}\wh{\omega}(X_i)\{O_i-\wh{\mu}(X_i)\}s(O_i,X_i,\wh{\psi}), \text{ and}
    \\
    \wh\eta_2^{\wt} &= \sum_{i=\nq+1}^{n}\wh{\omega}(X_i)\dot{s}(O_i,X_i,\wh{\psi}).
\end{align*}

In order to establish the  consistency and asymptotic normality of the weighting estimator, we re-express $\wh\eta_1^{\wt}$ and $\wh\eta_2^{\wt}$ as solutions to some estimating equations and then apply standard M-estimation theory.
Specifically,  consider a vector of  parameters $\eta^{\wt}=[\eta_1,\eta_2,\eta_3\trans,\eta_4\trans,\eta_5\trans]\trans$, and suppose these parameters are estimated by solving estimating equations $G^{\wt}=\left[\left(g_1^{\wt}\right)\trans,\left(g_2^{\wt}\right)\trans,\left(g_3^{\wt}\right)\trans,\left(g_4^{\wt}\right)\trans,\left(g_5^{\wt}\right)\trans\right]\trans$, i.e.,  by
\begin{align}\label{eq:est.eq.weighting}
    \sum_{i=1}^nG^{\wt}(T_i,O_i,X_i,\eta^{\wt})=0
\end{align}
 Specifically, elements in $\eta^{\wt}$  and estimating equations $G^{\wt}$ are defined as follows. The first two parameters, $\wh\eta_1^{\wt}$ and $\wh\eta_2^{\wt}$, have been defined above, and their corresponding estimating equations are
 \begin{align*}
  g_1^{\wt}(T_i,O_i,X_i,\eta^{\wt}) &= \dfrac{1-T_i}{\pr(T_i=0)} \omega(X_i,\eta_5) s(\O_i,X_i,\eta_3)\{\O_i-\mu(X_i,\eta_4)\}-\eta_1 = 0,\text{and}\\
 g_2^{\wt}(T_i,O_i,X_i,\eta^{\wt}) &= \dfrac{1-T_i}{\pr(T_i=0)} \omega(X_i,\eta_5) \dot{s}(\O_i,X_i,\eta_3)-\eta_2 = 0.
\end{align*}
Next, $\eta_3=\psi(\QOXzero)$ is defined as the target estimand under conditional exchangeability, and is estimated through
 \begin{align*}
     g_3^{\wt}(T_i,O_i,X_i,\eta^{\wt}) = \dfrac{1-T_i}{\pr(T_i=0)} \omega(X_i,\eta_5) s(\O_i,X_i,\eta_3).
 \end{align*}
 Finally, suppose parametric models of $\mu(x)$ and $\omega(x)$ are posited by the researcher with parameters $\eta_4$ and $\eta_5$, respectively, and let $g_4^{\wt}$ and $g_5^{\wt}$ be the corresponding estimating equations (e.g., score functions).
 Denote the two nuisance functions as $\mu(x,\eta_5)$ and $\omega(x,\eta_4)$ and hence their estimates are $\wh\mu(x)=\mu(x,\wh\eta_5)$ and  $\wh\omega(x)=\omega(x,\wh\eta_4)$. Suppose Let $g_4^{\wt}$ and $g_5^{\wt}$ are estimating equations (e.g., score functions) that correspond to $\mu(x)$ and  $\omega(x)$, respectively, via parametric models posited by the researcher.  To indicate the dependencies on nuisance parameters, we denote these two functions as $\mu(x,\eta_5)$ and $\omega(x,\eta_4)$, respectively, and hence their estimates are $\wh\mu(x)=\mu(x,\wh\eta_5)$ and  $\wh\omega(x)=\omega(x,\wh\eta_4)$.

With $G^W$ defined as above, the asymptotic properties of the weighting estimator for SLOPE, i.e.,
\begin{align}
\label{eq:main.est.weight}
\widehat{\SI}^{\wt} (\wh{Q}_{O,X}^0, \wh\psi)
&=- \left\{ \sum_{i=\nq+1}^{n}\wh{\omega}(X_i)\dot{s}(O_i,X_i,\wh{\psi})\right\}
\inv  \sum_{i=\nq+1}^{n}\wh{\omega}(X_i)\{O_i-\wh{\mu}(X_i)\}s(O_i,X_i,\wh{\psi}),
\end{align}
can be established under standard regularity conditions in M-estimation theory \citep{newey1994large,van2000asymptotic}.
Below, Condition \ref{assump:regular.weighted.consistent} parallels  the condition in Theorem 2.6 of \citet{newey1994large}  for consistency, and Condition \ref{assump:regular.weighted.an} parallels assumptions in Theorem 3.4 of \citet{newey1994large} and Theorem 5.31 of \citet{van2000asymptotic} for asymptotic normality.

\begin{condition}[Regularity Conditions for Consistency for Weighting Estimator]\label{assump:regular.weighted.consistent}\ \\
    (i) $\E\{G^{\wt}(T_i,\O_i,X_i,\eta)\}=0$ implies $\eta=\eta^{\wt}$.
    (ii) $\eta^{\wt}\in\Theta$ where $\Theta$ is compact. (iii)  $G^{\wt}$ is continuous at each $\eta\in\Theta$ with probability one and $\E\left\{\sup_{\eta\in\Theta}\lVert G^{\wt}(T_i,\O_i,X_i,\eta)\rVert \right\} < \infty$.
    (iv) $\eta_2=\E_{\QOXzero}\left[\dot{s}(O_i,X_i,\psi(\QOXzero)\right]\neq 0$ with probability one.
\end{condition}

\begin{condition}[Regularity Conditions for Asymptotic Normality for Weighting Estimator]\label{assump:regular.weighted.an}
    (i) $\eta^{\wt}$ lies in the interior of $\Theta$.
    (ii) $\E\lVert G^{\wt}(T_i,\O_i,X_i,\eta^{\wt})\rVert^2<\infty$.
    (iii) The function class $\left\{G^{\wt}: \lVert\eta-\eta^{\wt}\rVert<\delta\right\}$ is
    Donsker for some $\delta>0$ and $\E\lVert G^{\wt}(T_i,\O_i,X_i,\eta) - G^{\wt}(T_i,\O_i,X_i,\eta^{\wt})\rVert^2\to0$ as $\eta\to\eta^{\wt}$.
    (iv) The map $\eta\mapsto\E\{G^{\wt}(T_i,\O_i,X_i,\eta)\}$ is differentiable at $\eta^{\wt}$ with a non-singular derivative matrix $\Omega^{\wt}$ with inverse matrix $V^{\wt}$.
\end{condition}

With Conditions \ref{assump:regular.weighted.consistent} and \ref{assump:regular.weighted.an}, we establish the asymptotic properties of the weighting estimator.
\begin{theorem}[Weighting Estimator]\label{thm:est.weight}
   Let $\wh\SI^{\wt}$ be the weighting estimator in \eqref{eq:main.est.weight} where $\wh\eta_j$'s are estimated with \eqref{eq:est.eq.weighting} and we drop the notation in parentheses of the SLOPE for ease of communication.
   Suppose Condition \ref{assump:regular.weighted.consistent} holds, then $\wh\SI^{\wt}$ converges to $\SI$ in probability.
    Additionally suppose Condition \ref{assump:regular.weighted.an}  holds, then  $\sqrt n(\wh \SI^{\wt}-\SI)$ converges in distribution to a normal distribution with mean zero and variance
$(\eta_2)^2V_{11}^{\wt}/(\eta_1)^{4}+V_{22}^{\wt}/(\eta_1)^{2} - 2\eta_2V^W_{12}/(\eta_1)^{3}$, where $V^{\wt}$ is inverse of the derivative matrix of $\E\{G^{\wt}(T_i,O_i,X_i,\eta^{\wt})\}$ with respect to $\eta^{\wt}$, with $V^{\wt}_{ij}$ denoting its entry at the $i$-th row and $j$-th column.
\end{theorem}

 We note that depending on the target estimand, some nuisance parameters listed above may be trivial.
For example, for the SLOPE for the mean, $\eta_2=-1$ need not be estimated  since $\dot{s}=-1$ is constant.
In addition, we provide  example estimators for other estimands later in this section and  discuss ways of estimating $\omega(x)$ in Section \ref{supp.sec:est}.

Finally, we remark on the weighting estimator for SLOPE when the target functional $\psi(\cdot)$ is vector valued.
\begin{remark}[Weighting Estimator with a Vector Valued $\psi$]\label{remark:weight.est.vector}
With a vector valued target functional $\psi(\cdot)$, the target estimand and the SLOPE  become vectors of the same dimension, say $p$. The SLOPE formula derived from the IF (i.e., \eqref{eq:slope.z.scalar})  still holds and the weighting estimator \eqref{eq:main.est.weight} is still applicable. The difference is that $\eta_1$ becomes a vector of length $p$ and $\eta_2$ becomes a $p$ by $p$ matrix. To establish the asymptotic properties, we need to adjust the vector of nuisance parameter, $\eta^{\wt}$, mainly to include a vectorization of $\eta_2$ instead of $\eta_2$ itself. More concretely, let
    $\eta^{\wt}=[\eta_1\trans,\{\Vec(\eta_2)\}\trans,\eta_3\trans,\eta_4\trans,\eta_5\trans]\trans$, where $\Vec(\cdot)$ is vectorization of a matrix, and let $G^{\wt}$ be modified such that $g_2^{\wt}$ corresponds to $\Vec(\eta_2)$.  Then under the same regularity conditions on the updated $G^{\wt}$ and $\eta^{\wt}$, the weighting estimator is still consistent and asymptotically normal. The asymptotic variance is $A\Sigma^{\wt}A\trans$, where $\Sigma^{\wt}$ is the first $(p^2+p)$ diagonal matrix of $V^{\wt}$, and $A=\left[-(\SI)\trans\otimes(\eta_2)\inv, \eta_2\inv\right]$.
\end{remark}

\subsection{Asymptotic Properties for the Regression Estimator}\label{supp.subsec:est.regression}
Recall the regression estimator is
\begin{align}\label{eq:main.est.reg}
\widehat{\SI}^{\reg} (\wh{Q}_{O,X}^0, \wh\psi)
= &-\left[\sum_{i=1}^{\nq}\wh\E_{\POmidX}\{\dot{s}(O_i,X_i,\psi)\mid X_i\}\right]\inv
\sum_{i=1}^{\nq}\wh\E_{\POmidX}\left[\{O_i-\wh{\mu}(X_i)\}s(O_i,X_i,\psi)\mid X_i\right].
\end{align}
We re-express it with $\wh\eta_1^{\reg}$ and $\wh\eta^{\reg}_2$, the regression typed estimators for $\eta_1$ and $\eta_2$ respectively, as follows:
\begin{align*}
   \widehat{\SI}^{\reg} (\wh{Q}_{O,X}^0, \wh\psi)=
  -\left(\eta_2^{\reg}\right)\inv\eta_1^{\reg},&\text{ where}\\
  \eta_1=\sum_{i=1}^{\nq}\wh\E_{\POmidX}\left[\{O_i-\wh{\mu}(X_i)\}s(O_i,X_i,\psi)\mid X_i\right], &\text{ and }
\eta_2=\left[\sum_{i=1}^{\nq}\wh\E_{\POmidX}\{\dot{s}(O_i,X_i,\psi)\mid X_i\}\right].
\end{align*}
Next, we establish the asymptotic properties of the regression estimator in a similar way as done for the weighting estimator.
Specifically,  we consider nuisance parameters $\eta^{\reg}=[\eta_1,\eta_2,\eta_3,\eta_4\trans,\eta_6\trans,\eta_7\trans,\eta_8\trans]\trans$ estimated through
\begin{align}\label{eq:est.eq.regress}
    \sum_{i=1}^nG^{\reg}(T_i,O_i,X_i,\eta^{\reg})=0
\end{align}
with estimating equations $G^{\reg}=\left[g_1^{\reg},g_2^{\reg},g_3^{\reg},\left(g_4^{\reg}\right)\trans,\left(g_6^{\reg}\right)\trans,\left(g_7^{\reg}\right)\trans,\left(g_8^{\reg}\right)\trans\right]\trans$. Specifically, $\eta_1$ to $\eta_4$ have been defined previously with these new estimating equations based on regressions on the target samples instead of weighting on the source samples,
\begin{align*}
    g_1^{\reg}(T_i,O_i,X_i,\eta^{\reg}) &= \dfrac{T_i}{\pr(T_i=1)}\E_{\POmidX}\left[s(O_i,X_i,\eta_3)\{O_i-\mu(X_i)\}\mid X_i,\eta_6\right] - \eta_1,\\
     g_2^{\reg}(T_i,O_i,X_i,\eta^{\reg}) &= \dfrac{T_i}{\pr(T_i=1)} \E_{\POmidX}\left\{\dot{s}(O_i,X_i,\eta_3)\mid X_i,\eta_7\right\} -\eta_2,\\
     g_3^{\reg}(T_i,O_i,X_i,\eta^{\reg}) &= \dfrac{T_i}{\pr(T_i=1)} \E_{\POmidX}\{s(O_i,X_i,\eta_3)\mid X_i,\eta_8\},
\end{align*}
and $g_4^{\reg}=g_4^{\wt}$ is the estimating equation for the regression function $\mu(x)$ which is now denoted as $\mu(x,\eta_4)$ as in Section \ref{supp.subsec:est.weighting}.
As denoted in the proceeding formulas, the additional nuisance parameters, $\eta_6$, $\eta_7$, and $\eta_8$ represent the nuisance parameters in parametric models for $\E_{\POmidX}\{s(O_i,X_i,\eta_3)\{O_i-\mu(X_i)\mid X\}$, $\E_{\POmidX}\{\dot{s}(O_i,X_i,\eta_3)\mid X\}$ and $\E_{\POmidX}\{s(O_i,X_i,\eta_3)\mid X\}$, respectively, with estimating equations $g_6^{\reg}$, $g_7^{\reg}$ and $g_8^{\reg}$.
Depending on the specific estimand, some nuisance parameters may be trivial and the estimation is simpler than what has been shown. That being said, using regression based estimator typically involves more nuisance functions/parameters than the weighting estimator.

The regularity conditions for the regression estimator are standard and are similar to the conditions for the weighting estimator. Condition \ref{assump:regular.regress.consistent}  parallels  the condition in Theorem 2.6 of \citet{newey1994large}  for consistency and Condition \ref{assump:regular.regress.an} parallels assumptions in Theorem 3.4 of \citet{newey1994large} and Theorem 5.31 of \citet{van2000asymptotic} for asymptotic normality.

 \begin{condition}[Regularity Conditions for Consistency for Regression Estimator]\label{assump:regular.regress.consistent}\ \\
    (i) $\E\{G^{\reg}(T_i,\O_i,X_i,\eta)\}=0$ implies $\eta=\eta^{\reg}$.
    (ii) $\eta^{\reg}\in\Theta$ where $\Theta$ is compact.
    (iii)  $G^{\reg}$ is continuous at each $\eta\in\Theta$ with probability one and $\E\left\{\sup_{\eta\in\Theta}\lVert G^{\reg}(T_i,\O_i,X_i,\eta)\rVert \right\} < \infty$.
     (iv) $\eta_2=\E_{\QOXzero}\left[\dot{s}(O_i,X_i,\psi(\QOXzero)\right]\neq 0$ with probability one.
\end{condition}

\begin{condition}[Regularity Conditions for Asymptotic Normality for Regression Estimator]\label{assump:regular.regress.an}
    (i) $\eta^{\reg}$ lies in the interior of $\Theta$.
    (ii) $\E\lVert G^{\reg}(T_i,\O_i,X_i,\eta^{\reg})\rVert^2<\infty$.
    (iii) The function class $\left\{G^{\reg}: \lVert\eta-\eta^{\reg}\rVert<\delta\right\}$ is
    Donsker for some $\delta>0$ and $\E\lVert G^{\reg}(T_i,\O_i,X_i,\eta) - G^{\reg}(T_i,\O_i,X_i,\eta^{\reg})\rVert^2\to0$ as $\eta\to\eta^{\reg}$.
    (iv) The map $\eta\mapsto\E\{G^{\reg}(T_i,\O_i,X_i,\eta)\}$ is differentiable at $\eta^{\reg}$ with a non-singular derivative matrix $\Omega^{\reg}$ with inverse matrix $V^{\reg}$.
\end{condition}

Under  regularity conditions listed above, the consistency and asymptotic normality of the regression estimator is presented as follows.
\begin{theorem}[Regression Estimator]\label{thm:est.regress}
 Let $\wh\SI^{\reg}$ be the regression estimator in \eqref{eq:main.est.reg} where $\eta_j$'s are estimated with \eqref{eq:est.eq.regress} and we drop the notation in parentheses of the SLOPE for ease of communication.
    Suppose Condition \ref{assump:regular.regress.consistent} in the Appendix holds, then $\wh\SI^{\reg}$ converges to $\SI$ in probability.
    Additionally suppose Condition \ref{assump:regular.regress.an} in the Appendix holds, then  $\sqrt n(\wh \SI^{\reg}-\SI)$ converges in distribution to a Gaussian distribution with mean zero and variance
    $(\eta_2)^2V_{11}^{\reg}/(\eta_1)^{4}+V_{22}^{\reg}/(\eta_1)^{2} - 2\eta_2V^{\reg}_{12}/(\eta_1)^{3}$, where $V^{\reg}$ is the derivative matrix of $\E\{G^{\reg}(T_i,O_i,X_i,\eta^{\reg})\}$ with respect to $\eta^{\reg}$, with $V^{\reg}_{ij}$ denoting its entry at the $i$-th row and $j$-th column.
\end{theorem}

\begin{remark}[Regression Estimator with a Vector Valued $\psi$]\label{remark:regression.est.vector}
When the target functional is vector valued, the regression estimator is still applicable with a similar argument as in Remark \ref{remark:weight.est.vector} for the weighting estimator.
\end{remark}

\subsection{Estimating SLOPE for the Mean}\label{supp.subsec:slope.mean.est}
We elaborate on the estimators of the SLOPE for the mean, $\SI(\QOXzero,\psi^{\mean})$.
In Section \ref{subsec:estimation}, we have presented two estimators, a weighting estimator and a regression estimator,
\begin{align}
     \wh{\SI}^{\wt}(\wh Q_{O,X}^{0},\wh\psi^{\mean})
    =&
    \dfrac{1}{\np}\sum_{i=\nq+1}^{n}\wh{\omega}(X_i)\{O_i-\wh{\mu}(X_i)\}(O_i-\wh\psi^{\mean}),
    \label{eq:est.weighting.slope.mean}
    \\
     \wh{\SI}^{\reg}(\wh Q_{O,X}^{0},\wh\psi^{\mean})
    =&\dfrac{1}{\nq}\sum_{i=1}^{\nq}\wh\sigma^2(X_i). \label{eq:est.regression.slope.mean}
\end{align}
To implement the two estimators, one can resort to Section \ref{supp.subsec:est.nuisance} for estimating $\omega(x)$ and any regression method for estimating $\mu(x)$. In addition, $\wh\psi^{\mean}$ can be obtained by a weighted average of outcomes over source samples, i.e., $\sum_{i=\nq+1}^n\wh\omega(X_i)O_i/{\np}$, and $\wh\sigma^2(x)$ can be obtained by regressing the squared residual, $\{O_i-\wh\mu(X_i)\}^2$ over $X_i$.

Next we motivate the estimator based on the efficient influence function. By expressing the SLOPE explicitly,
\begin{align*}
    \SI(\QOXzero,\psi^{\mean}) = \E_{\QX}\{\sigma^2(X)\} = \E_{\QX}\left(\E_{\POmidX}\left[\{O-\mu(X)\}^2\mid X\right]\right),
\end{align*}
an alternative  weighting estimator is naturally motivated:
\begin{align} \label{eq:est.weighting.alt.slope.mean}
     \wh{\SI}^{\text{W,Alt}}(\wh Q_{O,X}^{0},\wh\psi^{\mean})
    =&
    \dfrac{1}{\np}\sum_{i=\nq+1}^{n}\wh{\omega}(X_i)\{O_i-\wh{\mu}(X_i)\}^2.
\end{align}
Moreover, a fourth estimator is motivated by combining properties of the alternative weighting estimator \eqref{eq:est.weighting.alt.slope.mean} with the regression estimator \eqref{eq:est.regression.slope.mean}, i.e.,
\begin{align}
    \wh{\SI}^{\text{EIF}}(\wh Q_{O,X}^{0},\wh\psi^{\mean})
    =&      \wh{\SI}^{\text{W,Alt}}(\wh Q_{O,X}^{0},\wh\psi^{\mean}) -  \dfrac{1}{\np}\sum_{i=\nq+1}^{n}\wh\omega(X_i)\wh\sigma^2(X_i) +
   \dfrac{1}{\nq}\sum_{i=1}^{\nq}\wh\sigma^2(X_i) \nonumber\\
   =& \dfrac{1}{\np}\sum_{i=\nq+1}^{n}\wh{\omega}(X_i) \left[\{O_i-\wh{\mu}(X_i)\}^2 - \wh\sigma^2(X_i)\right] +
    \dfrac{1}{\nq}\sum_{i=1}^{\nq}\wh\sigma^2(X_i).
    \label{eq:est.eif.slope.mean}
\end{align}
Since this estimator  \eqref{eq:est.eif.slope.mean} can be naturally motivated from the efficient influence function (EIF) of the SLOPE (see Section \ref{supp.subsec:eif} below), we refer to it as the EIF-based estimator and denote it using the superscript ``EIF''. Under standard regularity conditions, the EIF-based estimator is consistent and asymptotically normal,
\begin{align*}
    \sqrt{n}\left\{ \wh{\SI}^{\text{EIF}}(\wh Q_{O,X}^{0},\wh\psi^{\mean}) -  {\SI}(Q_{O,X}^{0},\psi^{\mean})\right\}\to_d N\left(0, \E\{\EIF^2(T,O,X,\SI)\}\right),
\end{align*}
where $\EIF(T,O,X,\SI)$ is the efficient influence function of $\SI(\QOXzero,\psi^{\mean})$ and it takes the form of
\begin{align*}
    \EIF(T,O,X,\SI) &= \dfrac{1-T}{\pr(T=0)}\omega(X)\left[\{O-\mu(X)\}^2 - \sigma^2(X)\right]+ \dfrac{T}{\pr(T=1)}\sigma^2(X)-\SI,
\end{align*}
where $\pr(T=1)$ is the probability limit of $\nq/n$ and $\pr(T=0)=1-\pr(T=1)$; see Section \ref{supp.subsec:eif} (in particular, Proposition \ref{prop:eif.slope}) for a general formula of the EIF for the SLOPE. Therefore, the asymptotic variance can be estimated with
\begin{align}\label{eq:se.est.eif.slope.mean}
\dfrac{1}{\nq}\sum_{i=1}^{\nq}\left\{\wh\sigma^2(X_i)-\wh{\SI}^{\text{EIF}}\right\}^2+
 \dfrac{1}{\np}\sum_{i=\nq+1}^{n}\wh\omega^2(X_i)
 \left[\{O_i-\wh\mu(X_i)\}^2-\wh\sigma^2(X_i)\right]^2,
\end{align}
which is consistent under regularity conditions.
Consequently, the standard error (SE) of the EIF-based estimator can be  estimated with
\begin{align*}
 \dfrac{1}{\sqrt{n}}
 \sqrt{
  \dfrac{1}{\nq}\sum_{i=1}^{\nq}\left\{\wh\sigma^2(X_i)-\wh{\SI}^{\text{EIF}}\right\}^2+
 \dfrac{1}{\np}\sum_{i=\nq+1}^{n}\wh\omega^2(X_i)
 \left[\{O_i-\wh\mu(X_i)\}^2-\wh\sigma^2(X_i)\right]^2
 }.
\end{align*}

\subsection{Estimating SLOPE for OLS Coefficients}\label{supp.subsec:slope.ols.est}
 Suppose $\O$ is an outcome variable and $X$ is a vector of covariates that includes $1$ as the first component. We are interested in the OLS coefficient of regressing $O$ on $X$ in the target distribution, i.e., $\psi^{\ols}(Q_{O,X})$ such that $\E_{Q_{O,X}}\left[XX\trans\psi^{\ols} - XO\right]=0$. Then according to Theorem \ref{thm:slope.ols}, the SLOPE for the OLS coefficient $\psi^{\ols}$ is
 \begin{align*}
     \SI(\QOXzero,\psi^{\ols}) &= \left\{\E_{\QX}(XX\trans)\right\}\inv\E_{\QX}\{X\sigma^2(X)\}.
 \end{align*}

To estimate the SLOPE, we consider three estimators based on weighting, regression, and the efficient influence function:
\begin{align}
&\wh{\SI}^{\wt}\left(\wh{Q}_{O,X}^0 ,\wh\psi^{\ols}\right)=\left(\meannq X_iX_i\trans\right)\inv \meannp\wh\omega(X_i)\wh r_i^2X_i,\label{eq:est.weighting.slope.ols}\\
  & \wh{\SI}^{\reg}\left(\wh{Q}_{O,X}^0 ,\wh\psi^{\ols}\right)=\left(\meannq X_iX_i\trans\right)\inv
    \meannq\wh\sigma^2(X_i)X_i\label{eq:est.regression.slope.ols}\\
& \wh{\SI}^{\EIF}\left(\wh{Q}_{O,X}^0 ,\wh\psi^{\ols}\right)=\left(\meannq X_iX_i\trans\right)\inv\left[
    \meannp\wh\omega(X_i)\{\wh r_i^2-\wh\sigma^2(X_i)\}X_i +
    \meannq X_i\wh\sigma^2(X_i)
    \right],\label{eq:est.eif.slope.ols}
\end{align}
where $\wh r_i=O_i-\wh\mu(X_i)$ is the regression residuals on the source data.

In addition, for the EIF-based estimator, the variance can be consistently estimated by
\begin{align}\label{eq:se.est.eif.slope.ols}
    \meann\wh\EIF\left(T_i,O_i,X_i,\wh\SI^{\EIF}\right)\left\{\wh\EIF(T_i,O_i,X_i,\wh\SI^{\EIF})\right\}\trans,
\end{align} where
\begin{align*}
  \wh\EIF\left(T_i,O_i,X_i,\wh\SI^{\EIF}\right)=& \left(\meannq X_iX_i\trans\right)\inv
    \left[\meannp \wh\omega(X_i)X_i\{\wh r_i^2-\wh\sigma^2(X_i)\} \right]\\\
    &+ \left(\meannq X_iX_i\trans\right)\inv
    \left[\meannq \left\{-X_i\trans \wh{\SI}^{\EIF}+\wh\sigma^2(X_i)\right\}X_i\right].
\end{align*}

\subsection{Estimating SLOPE for the Median}\label{supp.subsec:slope.median.est}

In this section, we consider estimating the SLOPE for the median.
We start with a simpler case when $\POmidX$ is  Gaussian, i.e., $\POmidX\sim N\left(\mu(X),\sigma^2(X)\right)$. Then by part (ii) of Theorem \ref{thm:slope.median}, the SLOPE for the median is
\begin{align*}
\SI(\QOXzero,\psi^{\med}) = \dfrac{\E_{\QX}\left[\sigma(X)\varphi(\{m_{1/2}-\mu(X)\}/\sigma(X))\right]}{\E_{\QX}\left[\varphi(\{m_{1/2}-\mu(X)\}/\sigma(X))/\sigma(X)\right]}.
\end{align*}
This motivates the following weighting and regression estimators:
\begin{align}
    \wh{\SI}^{\wt}\left(\wh{Q}_{O,X}^0 ,\wh\psi^{\med}\right)=&\dfrac{\meannp \wh\omega(X_i)\wh\sigma(X_i)\varphi\left(\{\wh m_{1/2}-\wh\mu(X_i)\}/\wh\sigma(X_i)\right)}
    {\meannp\wh\omega(X_i)\varphi\left(\{\wh m_{1/2}-\wh\mu(X_i)\}/\wh\sigma(X_i)\right)/\wh\sigma(X_i)},
    \label{eq:est.weighting.slope.median.normal}\\
   \wh{\SI}^{\reg}\left(\wh{Q}_{O,X}^0 ,\wh\psi^{\med}\right)=&
    \dfrac{
    \meannq \wh\sigma(X_i)\varphi\left(\{\wh m_{1/2}-\wh\mu(X_i)\}/\wh\sigma(X_i)\right)
    }
    {\meannq \varphi\left(\{\wh m_{1/2}-\wh\mu(X_i)\}/\wh\sigma(X_i)\right)
    /\wh\sigma(X_i)},\label{eq:est.regression.slope.median.normal}
\end{align}
where $\varphi(\cdot)$ is the density of the standard normal distribution.

Next, we consider the general case with SLOPE given by \eqref{eq:slope.median}. Since it involves conditional densities, the efficient influence function does not exist in general. We will consider the weighting estimator and the regression estimator only. Let $\wh F_{\POmidX}$, $\wh f_{\POmidX}$ and $\wh\E_{\POmidX}\left\{O\ind(O\leq \wh m_{1/2})\right\}$ be estimates of the c.d.f. $F_{\POmidX}$, the p.d.f. $f_{\POmidX}$, and the truncated mean $\E_{\POmidX}\left\{O\ind(O\leq \wh m_{1/2})\right\}$, respectively, and $\wh m_{1/2}$ be an estimate of $m_{1/2}$ where all estimates are based on parametric models. Then the weighting and regression estimators are
\begin{align*}
     \wh{\SI}^{\wt}\left(\wh{Q}_{O,X}^0 ,\wh\psi^{\med}\right)=&
          \dfrac{
       \meannp \wh\omega(X_i) \left\{\wh  F_{\POmidX}(\wh m_{1/2}\mid X_i)\wh\mu(X_i)- O_i\ind(O_i\leq \wh{m}_{1/2})\right\}
      }{
      \meannp \wh\omega(X_i)\wh f_{\POmidX}(\wh m_{1/2}\mid X_i)
      }, \text{ and}
     \\
      \wh{\SI}^{\reg}\left(\wh{Q}_{O,X}^0 ,\wh\psi^{\med}\right)=&
           \dfrac{
       \meannq \wh F_{\POmidX}(\wh m_{1/2}\mid X_i)\wh\mu(X_i)-  \wh\E_{\POmidX}\left\{O_i\ind(O_i\leq \wh{m}_{1/2})\mid X_i\right\}
      }{
      \meannq \wh f_{\POmidX}(\wh m_{1/2}\mid X_i)
      },
\end{align*}
respectively. In practice, one may impose parametric assumptions on $\POmidX$ and estimate $F_{\POmidX}$, $f_{\POmidX}$ and $\E_{\POmidX}\left\{O\ind(O\leq \wh m_{1/2})\right\}$ accordingly. Then the marginal $m_{1/2}$ can be estimated by  numerically solving either of the following equations using bisection,
\begin{align*}
  \sumnq  \wh F_{\POmidX}(m_{1/2}\mid X_i) &= 1/2, \text{ or}\\
  \sumnp \wh\omega(X_i)\wh F_{\POmidX}(m_{1/2}\mid X_i) &= 1/2.
\end{align*}

\subsection{Example Estimating Equations for $\omega(x)$}\label{supp.subsec:est.nuisance}
In this section, we discuss some common approaches to estimating the density ratio $\omega(x)$, including the logistic regression, entropy balancing, and a method based on empirical distributions when the support of $\QX$ (i.e., $\SX$) is finite and discrete.
Specifically, following the notation in Section \ref{supp.subsec:est.weighting}, we build estimating equations $g_5^W$ for the nuisance parameter $\eta_5$ and after obtaining the estimate $\wh\eta_5$, we present $\wh\omega(x)$ in terms of $\wh\eta_5$. For the sake of exposition, we take the convention that $X$ does not include one (i.e., a constant that can serve as an intercept in regression models).

\subsubsection{Logistic Regression}
 Let $\pr(T=1\mid x) = \pr(T=1\mid X=x)$ be the probability of being included in the target sample.
Since the density ratio $\omega(x)$ can be re-formulated in terms of $\pr(T=1\mid x)$ via Bayes rule,
\begin{align*}
    \omega(x) &= \dfrac{f(x\mid T=1)}{f(x\mid T=0)} & \text{ (By definition)}\\
    &=
    \dfrac{\pr(T=1\mid x)}{\pr(T=0\mid x)}\dfrac{\pr(T=0)}{\pr(T=1)}
     & \text{ (By Bayes rule)},
\end{align*}
one common strategy to estimate $\omega(x)$ is to first estimate the conditional probability function $\pr(T=1\mid x)$ using some binary classification/regression model, and then plug in the estimates $\wh\pr(T=1\mid x)$ to obtain $\wh\omega(x)$,
\begin{align}\label{supp.eq:omega.est.lr}
    \wh\omega(x)  =   \dfrac{\wh\pr(T=1\mid x)}{\wh\pr(T=0\mid x)}\dfrac{\wh\pr(T=0)}{\wh\pr(T=1)}
     =\dfrac{\wh\pr(T=1\mid x)}{\wh\pr(T=0\mid x)}\dfrac{\np}{\nq},
\end{align}
where $\wh\pr(T=0)=\np/n$ and $\wh\pr(T=1) = \nq/n$.

Next,  we list the estimation equations based on maximum likelihood estimation  when $\pr(T=1\mid x)$ is modeled by logistic regression, one of the most popular binary regression models. This can be implemented in R using the built-in  function $\texttt{glm}$.
To demonstrate the estimating equation for the nuisance parameter $\eta_5$, we let $\eta_5=[\alpha_{\LR},\beta_{\LR}\trans]\trans$ where $\alpha_{\LR}$ and $\beta_{\LR}$ are the intercept and slope coefficients in the logistic regression model. Then the likelihood can be expressed as
\begin{align*}
    l_{\LR}(\alpha_{\LR},\beta_{\LR}) &= \sum_{T_i=0}(\alpha_{\LR}+\beta_{\LR}\trans X_i) -
    \sum_{i=1}^n \log\left(1 + \exp\{\alpha_{\LR} + \beta_{\LR}\trans X_i\}\right),
\end{align*}
By setting $\partial l_{\LR}/\partial\alpha_{\LR}=0$ and $\partial l_{\LR}/\partial\beta_{\LR}=0$ and checking the second-order conditions, parameters $\alpha_{\LR}$ and $\beta_{\LR}$ can be estimated by $\wh\alpha_{\LR}$ and $\wh\beta_{\LR}$ which are solutions to the following equations,
\begin{align*}
   \np-\sum_{i=1}^n \dfrac{\exp(\wh\alpha_{\LR} + \wh\beta_{\LR}\trans X_i)}{1 +\exp(\wh\alpha_{\LR} + \wh\beta_{\LR}\trans X_i)} = 0,
  \text{ and}
   \sum_{T_i=0}X_i - \sum_{i=1}^n
   \dfrac{X_i\exp(\wh\alpha_{\LR} + \wh\beta_{\LR}\trans X_i)}{1 + \exp(\wh\alpha_{\LR} + \wh\beta_{\LR}\trans X_i)}.
\end{align*}
Therefore, the estimating equation $g_5^{\wt}$ is
\begin{align*}
    g_5^{\wt}(T,O,X,\eta^{\wt}) &= \begin{bmatrix}
        \ind(T=0) - \dfrac{\exp(\alpha_{\LR} + \beta_{\LR}\trans X)}{1+\exp(\alpha_{\LR} + \beta_{\LR}\trans X)}\\
        {}\\
        \ind(T=0)X - \dfrac{X\exp(\alpha_{\LR}+\beta_{\LR}\trans X)}{1+\exp(\alpha_{\LR}+ \beta_{\LR}\trans X)}
    \end{bmatrix},
\end{align*}
and the estimate $\wh\eta_5 =[\wh\alpha_{\LR},\wh\beta_{\LR}\trans]\trans$ is obtained by setting $\sum_{i=1}^ng_5^{\wt}(T_i,O_i,X_i,\eta^{\wt})=0$. After estimating these parameters, by \eqref{supp.eq:omega.est.lr}, the resulting estimate for the density ratio is
\begin{align*}
    \wh\omega(x) &= \exp\left\{
    -\left[\wh\alpha_{\LR}-\log(\np/\nq)\right]-\wh\beta_{\LR}\trans x
    \right\}.
\end{align*}

\subsubsection{Entropy Balancing}\label{supp.subsec:w.est.EB}
Note that for any measurable function $h(x)$, the density ratio $\omega(x)$ satisfies that
\begin{align*}
    \E\{\omega(X)h(X)\mid T=0\} = \E\{h(X)\mid T=1\}.
\end{align*}
In other words, $\omega(x)$ reweighs the source population in order to match the target population. This property motivates estimating $\omega(x)$ by balancing functions (usually moments) of the source samples so that they match the target samples.

In this section, we introduce the method of entropy balancing and aim at balancing the first moments of $X$.
Specifically,
for  source samples with $i=1+\nq,\cdots,n$, suppose $\omega_i = \omega(X_i)$ are the weights. Entropy balancing seeks for $\omega_i$'s that maximizes  their entropy as well as  balances the first moment of $X$:
\begin{align}\label{eq:EB}
 \begin{split}
     &\underset{w_i}{\rm argmin} \sum_{T_i=0}w_i\log(w_i), \quad{}
    \textnormal{s.t. }\dfrac{1}{\np}\sum_{T_i=0}w_i X_i = \dfrac{1}{\nq}\sum_{T_i=1}X_i.
     \end{split}
 \end{align}
 According to \citet{lee2023improving,chen2023entropy}, solutions to \eqref{eq:EB}, denoted as $\wh\omega_i$, can be expressed by
 \begin{align}\label{supp.eq:omega.EB}
     \wh\omega_i=\wh\omega(X_i) = \exp\left(-\wh\alpha_{\EB}- \wh\beta_{\EB}\trans X_i\right),
 \end{align}
 where $\wh\alpha_{\EB}$ and $\wh\beta_{\EB}$ satisfy
 \begin{align*}
     \sum_{T_i=0}\exp(\wh\alpha_{\EB} -\wh\beta_{\EB}\trans X_i) = \np,
     \quad
     \dfrac{1}{\np}\sum_{T_i=0}\exp\left(\wh\alpha_{\EB}-\wh\beta_{\EB}\trans X_i\right) =
     \dfrac{1}{\nq}\sum_{T_i=1}X_i.
 \end{align*}
Following the notation in the main text, let the nuisance parameter be $\eta_5 = [\alpha_{\EB},\beta_{\EB}\trans]\trans$, then the estimating equation $g_5^{\wt}$ is
\begin{align*}
    g_5^{\wt}(T,O,X,\eta^{\wt}) =
    \begin{bmatrix}
       \ind(T=0)\exp(-\alpha_{\EB} - \beta_{\EB}\trans X)- \ind(T=0)\\
        \ind(T=0)X_i\exp(-\alpha_{\EB}-\beta_{\EB}\trans X) - \dfrac{\np}{\nq}\ind(T=1)X
    \end{bmatrix},
\end{align*}
and the estimate $\wh\eta_5 =[\wh\alpha_{\EB},\wh\beta_{\EB}\trans]\trans$ is obtained by setting $\sum_{i=1}^ng_5^{\wt}(T_i,O_i,X_i,\eta^{\wt})=0$. After estimating these parameters, by \eqref{supp.eq:omega.EB}, the resulting estimate for the density ratio is
\begin{align*}
    \wh\omega(x) &= \exp\left(
\wh\alpha_{\EB}-\wh\beta_{\EB}\trans x
    \right).
\end{align*}
We note that the two methods, entropy balancing and logistic regression, are related in that when models are correctly specified, $\beta_{\EB}=\beta_{\LR}$ and $\alpha_{\EB}=\alpha_{\LR}-\log(\pr(T=0)/\pr(T=1))$, while their estimates are numerically different due to different first-order conditions \citep{zhao2017entropy}. In practice, we recommend using entropy balancing over logistic regression since by enabling covariate balancing, entropy balancing is more robust when models are slightly mis-specified \citep{imai2014covariate,zhao2019covariate}.

\subsubsection{Estimation for Discrete Covariates}\label{supp.subsec:w.est.discrete}
Suppose $X$ is discrete with a finite support, $\SX=\{x_1,x_2,\cdots, x_K\}$ where $K$ is fixed. Then $\omega(x)$ can be estimated by the ratio of the empirical distributions of $f_{\QX}(x)$ and $f_{\PX}(x)$ for $x\in\SX$, i.e.,
\begin{align}\label{supp.eq:est.omega.discrete}
    \wh\omega(x) =
    \dfrac{\sum_{T_i=1}\ind(X_i=x)/\nq}{\sum_{T_i=0}\ind(X_i=x)/\np}.
\end{align}
Following the notation in Section \ref{supp.subsec:est.weighting}, let $\eta_5=[w_1,w_2,\cdots,w_K]\trans$ where $w_k=f_{\QX}(x_k)/f_{\PX}(x_k)$ for $k=1,2,\cdots,K$. Then by defining the estimating equations as
\begin{align*}
    g_5^{\wt}(T,O,X,\eta^{\wt}) &=
    \begin{bmatrix}
        \dfrac{T\ind(X=x_1)}{\nq} - w_1\cdot\dfrac{(1-T)\ind(X=x_1)}{\np}\\
        \ldots\\
        \dfrac{T\ind(X=x_K)}{\nq} - w_K\cdot\dfrac{(1-T)\ind(X=x_K)}{\np}
    \end{bmatrix},
\end{align*}
the nuisance parameter can be estimated by $\wh\eta_5=[\wh w_1,\wh w_2,\cdots,\wh w_5]$ where $\wh w_k$'s are estimated by setting $\sum_{i=1}^ng_5^{\wt}(T_i,O_i,X_i,\eta^{\wt})=0$. Then $\wh\omega(x_k)$ can be estimated with $\wh w_k$ for $x_k\in\SX$.

\subsection{General Statements for the Efficient Influence Function}\label{supp.subsec:eif}
In this section, we derive the efficient influence function (EIF) of $\SI$ when the target functional (and therefore the SLOPE) is scalar valued. Examples with $\psi=\psi^{\mean}$ and $\psi=\psi^{\ols}$ can be found in Sections \ref{supp.subsec:slope.mean.est} and \ref{supp.subsec:slope.ols.est}, respectively.

Let $\EIF(T,O,X,\SI)$ be EIF for $\SI(\QOXzero,\psi)$. It is provided by Proposition  \ref{prop:eif.slope}. Additionally, the next Proposition \ref{prop:eif.psi} gives the EIF of the target functional, denoted as $\EIF(T,O,X,\psi)$.
\begin{proposition}[EIF of SLOPE]\label{prop:eif.slope}
    Suppose $\psi$ and $s$ are one-dimensional. The efficient influence function of  $\SI(\QOXzero,\psi)$ is
   {\footnotesize{
   \begin{align*}
   & \E_{\QOXzero}\{\dot{s}(O,X,\psi)\}\cdot \EIF(T,O,X,\SI) \\
    =&     \dfrac{(1-T)\omega(X)}{\pr(T=0)}\left[ -s(O,X,\psi)\{O-\mu(X)\} + \cov_{\POmidX}[s(O,X,\psi),O\mid X] +
    \E_{\POmidX}\{s(O,X,\psi)\mid X\}\{O-\mu(X)\}\right]\\
    & -\dfrac{T}{\pr(T=1)}\E_{\POmidX}[s(O,X,\psi)\{O-\mu(X)\}\mid X]\\
    & -\SI(\QOXzero,\psi)\left(
    \dfrac{(1-T)\omega(X)}{\pr(T=0)}\left[\dot{s}(O,X,\psi) - \E_{\POmidX}\{\dot{s}(O,X,\psi)\mid X\}\right]  -
    \dfrac{T}{\pr(T=1)}\E_{\POmidX}\{\dot{s}(O,X,\psi)\mid X\}\right)\\
    &-\EIF(T,O,X,\psi)\left(
    \E_{\QOXzero}\left[\dot{s}(O,X,\psi)\{O-\mu(X)\}\right] + \SI(\QOXzero,\psi)\cdot\E_{\QOXzero}\{\ddot{s}(O,X,\psi)\}
    \right),
    \end{align*}
    }}
     where  $\EIF(T,O,X,\psi)$ be the EIF of the target functional $\psi(\QOXzero)$ (see Proposition \ref{prop:eif.psi}).
\end{proposition}
\begin{proposition}[EIF of The Target Functional]\label{prop:eif.psi}
    The efficient influence function of  $\psi(\QOXzero)$ is
   {\footnotesize{
   \begin{align*}
\EIF(T,O,X,\psi) =&
-\left[\E_{\QOXzero}\{\dot{s}(O,X,\psi)\}\right]\inv \dfrac{(1-T)\omega(X)}{\pr(T=0)}\left[\{s(O,X,\psi) - \E_{\POmidX}\{s(O,X,\psi)\mid X\}\right]\\
&- \left[\E_{\QOXzero}\{\dot{s}(O,X,\psi)\}\right]\inv
\dfrac{T}{\pr(T=1)}\E_{\POmidX}\{s(O,X,\psi)\mid X\}.
    \end{align*}
    }}
\end{proposition}

We elaborate these two propositions on a few examples.
First, suppose the target functional is $\psi^{\mean}$ with $s(O,X,\psi^{\mean})=O-\psi^{\mean}$. Then $\dot{s}(O,X,\psi)=-1$ and $\ddot{s}(O,X,\psi)=0$. With Proposition \ref{prop:eif.psi}, the EIF for the target functional is
\begin{align}\label{supp.eq:mean.eif.psi}
    \EIF(T,O,X,\psi^{\mean})&= \dfrac{1-T}{\pr(T=0)}\omega(X)\{O-\mu(X)\}
    +\dfrac{T}{\pr(T=1)}\left\{\mu(X)-\psi^{\mean}(\QOXzero)\right\}.
\end{align}
This is identical to the EIF derived by \citet{zeng2025efficient} in their special case when the source and target samples share the same set of covariates. Estimators for the mean based on this EIF were  proposed by \cite{dahabreh2020extending} and then by \citet{zeng2025efficient}  in a more general setting.

Further, by Proposition \ref{prop:eif.slope}, the EIF of $\SI(\QOXzero,\psi^{\mean})$ is
\begin{align}\label{supp.eq:mean.eif.slope}
\begin{split}
    \EIF(T,O,X,\SI) =& \dfrac{1-T}{\pr(T=0)}\omega(X)\left[\{O-\mu(X)\}^2 - \sigma^2(X)\right] \\
    &+
    \dfrac{T}{\pr(T=1)} \left\{\sigma^2(X) - \SI(\QOXzero,\psi^{\mean})\right\}.
    \end{split}
\end{align}
\eqref{supp.eq:mean.eif.slope} consists of two parts. The first part is indexed by $1-T$ and it is a weighted, mean-zero term involving the source data. The weight is the density ratio $\omega(X)$ that reweighs the source covariates to match the target population, and the mean zero part can be viewed as the residual of estimating $\sigma^2(X)=\E_{\POmidX}\left[\{O-\mu(X)\}^2\mid X\right]$. The second part is indexed by $T$ and it can be viewed as the conditional variance $\sigma^2(X)$ re-centered over the target population.
By estimating nuisance functions $\mu(X)$, $\omega(X)$, and $\sigma^2(X)$ and setting the summation of the empirical EIFs to zero, we obtain an EIF-based estimator that was presented in \eqref{eq:est.eif.slope.mean}.

For the second example, suppose the target functional is the OLS coefficient $\psi^{\ols}$ where $s(O,X) =\E_{Q_{O,X}}\left[XX\trans\psi^{\ols} - XO\right]=0 $ with $X$ includes an intercept term as its first component. We notice that although Proposition \ref{prop:eif.slope} has been stated for one-dimensional parameters, it is also valid for the $\SI(\QOXzero,\psi^{\ols})$ because the corresponding $\ddot{s}(O,X,\psi^{\ols}) = 0$. Consequently, the EIF for $\SI(\QOXzero,\psi^{\ols})$  is
\begin{align*}
    \EIF(T,O,X,\SI) =& \left\{\E_{\QX}(XX\trans)\right\}\inv\left\{
    \dfrac{1-T}{\pr(T=0)}\omega(X) \left[\{O-\mu(X)\}^2-\sigma^2(X)\right] +
    \dfrac{T}{\pr(T=1)}\sigma^2(X)
    \right\}.
\end{align*}

After presenting these examples of the EIF, finally, we comment on some limitations for estimators based on the EIF. First, the EIF for the SLOPE does not always exist. For example, when the target estimand is the median, $\psi^{\med}$, the SLOPE involves the conditional density $f_{\POmidX}$, and hence the EIF does not exist.
Second, unlike many well-known examples in the literature, the  SLOPE estimators based on the EIF often do not enjoy the double robustness property. In the simplest case when the target functional is the mean, $\psi^{\mean}$, the EIF-based estimator \eqref{eq:est.eif.slope.mean} involves three nuisance functions, $\omega(X)$, $\mu(X)$, and $\sigma^2(X)$. Roughly speaking, this estimator is \emph{conditionally} doubly robust in that as long as $\wh\mu(x)$ is consistent, the estimator will be consistent when either $\wh\omega(x)$ or $\wh\sigma^2(x)$ is consistent. A similar result for the conditional rate double robustness holds under standard regularity conditions or cross fitting procedures. Due to these limitations and the complexity of the EIF, in practice we recommend using weighting or regression estimators  presented in Section \ref{subsec:estimation} of the main text.
Meanwhile, we implement all three estimators in the numeric simulations (Section \ref{supp.sec:simulation}) and we recognize the development of robust estimators of the SLOPE as an import future direction.

\clearpage

\section{Supplementary Materials for Data Application}\label{supp.sec:data}
In this section, we provide supplementary results for the data application in Section \ref{sec:data} of the main text.
\subsection{Causal Identification}\label{supp.subsec:data.identify.slope}
On the source population, we assume SUTVA (Assumption \ref{assump:sutva})  holds  for potential outcome $Y(a)$ under treatment ($a=1$) and control ($a=0$). Let $X$ be the baseline measurement and $V$ be the village that a household belongs to. We assume the strong ignorability assumption  holds with $X$ and $V$, i.e., Assumption \ref{assump:strong.ign}; note that this holds by the design of \citet{banerjee2015multifaceted}.
Under these identification assumptions, the SLOPE for the mean becomes
\begin{align}
    &\SI(\QYaXzero,\psi^{\mean})\nonumber\\
    =&
    \E_{\QX}\left\{\var_{\PYamidX}[Y(a)\mid X]\right\}\nonumber\\
    =& \E_{\QX}
    \E_{P_{V\mid X}}\left\{\var_{P_{Y\mid X,V,A=a}}[Y\mid X,V,A=a]\mid X
    \right\} +
    \var_{P_{V\mid X}}\left[\mu_a(X,V)\mid X\right],\label{eq:slope.mean.ya}
\end{align}
where we let $\mu_a(X,V) = \E_{P_{Y\mid X,V,A=a}}(Y\mid X,V,A=a)$.

The SLOPE for the median becomes
\begin{align}
    &\SI(\QYaXzero,\psi^{\med}) \nonumber\\
    =&
 \dfrac{
        \E_{\QX}\left[F_{\PYamidX}(m_{1/2}\mid X)\mu( X)\right] - \E_{\QYaXzero}\left[Y(a)\ind\{Y(a)\leq m_{1/2}\}\right]
        }{f_{\QYazero}(m_{1/2})}.
    \nonumber\\
    =& \dfrac{
    \E_{\QX}\left(\E_{P_{V\mid X}}
\left[
    F_{P_{Y\mid X,V,A=a}}(m_{1/2}\mid X,V) \mu_a(X,V) - \E_{Y\mid X,V,A=a}\{Y\ind(Y\leq m_{1/2}\}
    \Bigg\vert X\right]\right)
    }
    {
    \E_{\QX}\left(\E_{P_{V\mid X}}
\left[
  f_{Y\mid X,V,A=a}(m_{1/2}\mid X,V)
     \Bigg\vert X\right]\right)\label{eq:slope.median.ya}
    },
\end{align}
where $m_{1/2}$ is the marginal median such that
\begin{align}
    \E_{\QX}\left[\E_{P_{V\mid X}}\left\{F_{P_{Y\mid X,V,A=a}}(m_{1/2}\mid X,V)\mid X\right\}\right]=1/2.
\end{align}

\subsection{Estimation of the SLOPE}\label{supp.subsec:data.est}

With a slight abuse of notation, we let $\mu_a(X,V)=\E_{P_{Y\mid X,V,A=a}}(Y\mid X,V,A=a)$ and $\mu_a(X)=\E_{P_{V\mid X}}\{\mu_a(X,V)\mid X\}$. In notation we keep $a\in\{0,1\}$ for generality while in the main text we have focused on $a=1$. For estimation, we assume a linear model whereas for individual $i$ in treatment group $a$, i.e. $A_i=a$, we  have
\begin{align}\label{eq:data.linear.model}
 \mu_a(x,v) = \alpha_v+ \beta_x + \delta_a + (\beta\delta)_{xa},
\end{align}
where $v$ is a discrete variable ranges in all villages in the source country and $x\in\SX$ ranges in the categories of the baseline variable, and regression coefficients are constrained to guarantee identification: $\beta_1=0$, $\delta_0=0$, $(\beta\delta)_{x0}=0$ for $x\in\SX$ and $(\beta\delta)_{0a}=0$ for $a\in\{0,1\}$. In addition, we assume that the outcome variance depends on the baseline measurement $x$ as below.
\begin{align*}
    \sigma^2_a(x) := \var[Y\mid X=x,V=v,A=a].
\end{align*}

In Sections \ref{subsec:data.choice.source} and \ref{subsec:data.weight}, the SLOPE for the mean is estimated by a simple plug-in estimator of \eqref{eq:slope.mean.ya}:
\begin{align*}
    \sum_{x\in\SX}
    \left[\wh{Q}_X(x)
    \left\{
    \wh\sigma^2_a(x) +
    \sum_{v\in\S_{V}}\wh{P}_{V\mid X=x}(v\mid x)\wh\mu_a(x,v)\right\}
    \right],
\end{align*}
where $\wh\mu_a(x,v)$ is estimated by the least squares estimator that regresses $Y$ on $X$ and $V$ as in \eqref{eq:data.linear.model}, $\wh\sigma^2_a(x)$ is estimated by the sample variance of regression residuals, $\wh{Q}_X(x)$ and $\wh{P}_{V\mid X=x}(v\mid x)$ are estimated by the empirical distribution of the corresponding distributions.
Since $X$ is discrete, this plug-in estimator can alternatively be viewed as a weighting estimator with weights estimated by empirical distributions of $X$ in the source and target populations.

To estimate the SLOPE for the median in Section \ref{subsec:data.choice.source}, we assume $Y_i-\mu_a(X_i,V_i)$ follows a normal distribution.  Under this conditional normality assumption, the SLOPE in \eqref{eq:slope.median.ya} can be estimated by a plug-in (or equivalently, weighting) estimator as follows:
\begin{align*}
   &\sum_{x\in\SX}\wh Q_{X}(x)
   \left[
   \sum_{v\in\S_{V}} \wh P_{V\mid X}(v\mid x)\Phi\left(
   \dfrac{\wh m_{1/2}-\wh\mu_a(x,v)}{\wh\sigma_a^2(x)}
   \right)
   \right]
   \left[
   \sum_{v\in\S_V}\wh P_{V\mid X}(v\mid x)\wh\mu_a(x,v)
   \right]\\
   &- \sum_{v\in\S_V}\wh P_{V\mid X}(v\mid x)
   \left[
   \wh\mu_a(x,v)
   \Phi \left( \dfrac{\wh m_{1/2}-\wh\mu_a(x,v)}{\wh\sigma_a^2(x)}\right)
   -
   \wh\sigma_a(x)
   \varphi\left( \dfrac{\wh m_{1/2}-\wh\mu_a(x,v)}{\wh\sigma_a^2(x)} \right)
   \right],
\end{align*}
where $\varphi(\cdot)$ and $\Phi(\cdot)$ are the probability density functional and cumulative distribution function of the standard normal distribution, and $\wh m_{1/2}$ is the solution of the following  equation using bisection search,
\begin{align*}
    \sum_{x\in\SX}\wh Q_{X}(x)\sum_{v\in\S_V}\wh P_{V\mid X}(v\mid x)
    \Phi\left(\dfrac{\wh m_{1/2}-\wh\mu_a(x,v)}{\wh\sigma_a^2(x)}\right)=1/2.
\end{align*}

\subsection{Auxiliary Data Information and Results for Section \ref{subsec:data.choice.source}}
\label{supp.subsec:data.choice.source}
In this section, we provide additional data information and analysis results for Section \ref{subsec:data.choice.source} where the outcome variable is the log-transformed per capita consumption.
\subsubsection{Auxiliary Data Information}
Table \ref{tab:X.consumption} gives the distribution of the categorized baseline measurement of the log-transformed per-capita consumption across countries on the overlapped sample.
\begin{table}[!h]
    \centering
     \caption{Baseline Measurement $X$ of the log-transformed per-capita consumption.}
    \label{tab:X.consumption}
    \begin{tabular}{llllll}
    \hline
     &  Peru      &   Pakistan   & India  &     Honduras  &   Ghana \\
  Sample size &              1768    &     829    &     771       &  2152     &    2379     \\
  \hline
  Category of $X$  &        &&&&\\
     (-0.41,3.65]  &  97 ( 5.5\%)  & 50 ( 6.0\%) & 450 (58.4\%) & 1026 (47.7\%) &  1001 (42.1\%)  \\
     (3.65,4.24] &  449 (25.4\%) & 170 (20.5\%)  &260 (33.7\%)   &817 (38.0\%)  & 832 (35.0\%)    \\
     (4.24,8]    & 1222 (69.1\%) & 609 (73.5\%) &  61 (7.9\%)  & 309 (14.4\%)&   546 (23.0\%)\\
     \hline
    \end{tabular}
\end{table}

\begin{figure}
    \centering
    \includegraphics[width=1\linewidth]{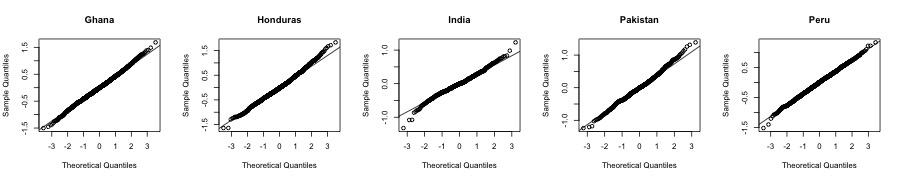}
    \caption{QQ-plots for the residuals of the linear model \eqref{eq:data.linear.model} across source countries. Each plot is generated by \texttt{qqplot} in R where the straight line generated by \texttt{qqline} passes through the first and third quartiles.}
    \label{fig:consumption.normal}
\end{figure}

Figure \ref{fig:consumption.normal} provides the normal diagnostics for the conditional normal assumption imposed in Section \ref{subsec:data.choice.source} when estimating the SLOPE for the median.
From these QQ-plots, the normality assumption is reasonable well for most countries.

\subsubsection{Additional Results}\label{supp.subsubsec:data.choice.source.result}
First, we present one hypothesis on why SLOPE for the average per capita consumption is lower in India and Peru. For India, the experiment  was located at West Bengal, an area  abutting Bangladesh and shares a language and a culture. Therefore, the unique cultural and locational features may have caused the uniformity of the underlying population. For Peru, according to \citet{banerjee2015multifaceted}, there has already been a consumption support program implemented on part of the households. This may has led to a higher homogeneity among the treated households.

Second, as mentioned in the main text, Table \ref{tab:data.choice.source.mean.median} presents the mean and median of the transported per-capita consumption.
\begin{table}[!h]
 \caption{
 The estimated mean and median (not their SLOPEs) for transporting the counterfactual log-transformed per capita consumption under intervention from a source country (i.e., the rows of the table) to a target country (i.e., the columns of the table). Bootstrap standard errors are in the parentheses.
 }
    \label{tab:data.choice.source.mean.median}
    \centering
   {\footnotesize \begin{tabular}{c | c | ccccc }
    \hline
    \multirow{2}{*}{Estimand ($\psi$)} & \multirow{2}{*}{Source ($\POmidX$)} & \multicolumn{5}{c}{Target ($\QX$)} \\
    &  & Ghana & Honduras & India & Pakistan & Peru \\
    \hline
    \multirow{5}{*}{Mean}
&Ghana     &             & 3.47 (0.02) & 3.44 (0.02) & 3.67 (0.03) & 3.67 (0.03) \\
&Honduras  & 4.23 (0.02)   &           & 4.15 (0.02) & 4.42 (0.04) & 4.42 (0.04) \\
&India     & 4.07 (0.03)   & 4.04 (0.02) &           & 4.24 (0.07) & 4.24 (0.07) \\
&Pakistan  & 4.24 (0.05)   & 4.21 (0.05) & 4.17 (0.06) &           & 4.42 (0.02) \\
&Peru      & 4.77 (0.03)   & 4.74 (0.03) & 4.71 (0.04) & 4.93 (0.02) &  \\
\hline
     \multirow{5}{*}{Median}
&Ghana     &             & 3.48 (0.02) & 3.44 (0.02) & 3.68 (0.03) & 3.67 (0.03) \\
&Honduras  & 4.23 (0.02)   &           & 4.16 (0.02) & 4.42 (0.04) & 4.41 (0.04) \\
&India     & 4.05 (0.02)   & 4.03 (0.02) &           & 4.23 (0.07) & 4.22 (0.06) \\
&Pakistan  & 4.24 (0.04)   & 4.21 (0.05) & 4.18 (0.06) &           & 4.42 (0.02) \\
&Peru      & 4.76 (0.03)   & 4.73 (0.03) & 4.70 (0.04) & 4.94 (0.02) &  \\
    \hline
    \end{tabular}}
\end{table}

\subsubsection{First-Order Approximation of Bias}\label{supp.subsec:data.approx.bias}
\begin{figure}
    \centering
    \includegraphics[width=0.7\linewidth]{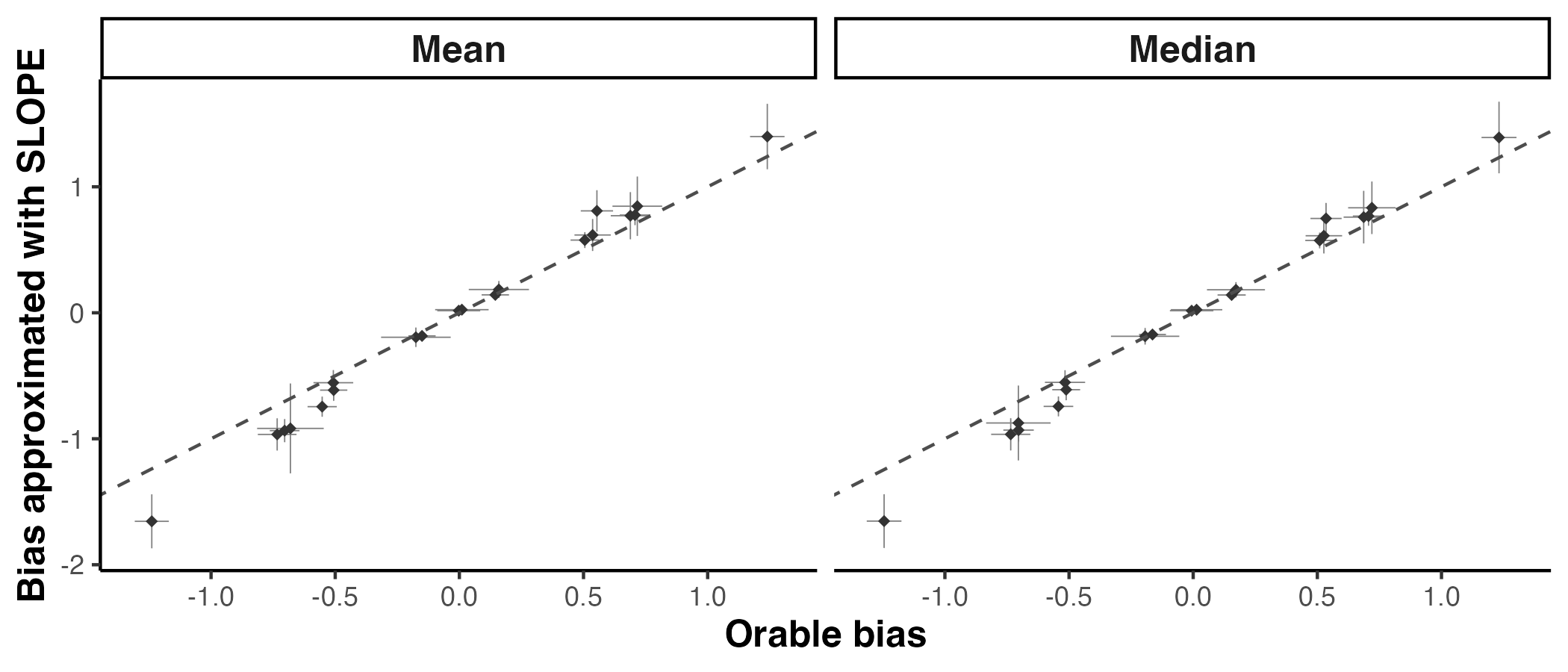}
    \caption{Bias approximation with SLOPE for mean (left) and median (right). Each panel plots the bias approximated with SLOPE in the y-axis and the oracle bias in the x-axis in dots and the corresponding bootstrapped $95\%$ confidence interval in bars. The dashed straight line is $y=x$.}
    \label{fig:data.bias.slope.consumption}
\end{figure}

For each pair of source and target countries, we estimate the oracle bias for the target country, i.e., the left hand side of \eqref{eq:first.order.approx}. In specific, we estimate the mean/median of the potential outcome in the target country by either transporting from the source country or using the target data (with the outcome information) itself. The difference between the two estimates is treated as the ``oracle bias'' from conditional exchangeability  since it represent the bias one my occur by directly assuming conditional exchangeability when outcome information in the target is unavailable. The confidence intervals of the oracle bias (i.e., horizontal bars) in Figure \ref{fig:data.bias.slope.consumption} are obtained via bootstrap.

Next, we estimate the bias approximated with SLOPE, i.e., the right hand side of \eqref{eq:first.order.approx}. In addition to estimating SLOPE as described in Section \ref{supp.subsec:data.est}, we also estimate the sensitivity parameter $\gamma$ as follows. First, suppose the normal assumption holds, i.e., $Y_i-\mu_a(X_i,V_i)$ is conditionally normally distributed with  mean zero and variance $\sigma^2_a(X_i)$. Then the sensitivity model \eqref{eq:sensitivity} enlists a location shift in the errors between the source and the target, i.e., $Q_{Y\mid X,V,A=a}\sim N\left(\mu_a(X,V)+\gamma\sigma_a^2(X),\sigma_a^2(X)\right)$. Therefore, by method of moment, we estimate $\gamma$ through the following formula,
\begin{align*}
 \sum_{i=1}^{\np}\left[ \wh\mu_a(X_i,V_i) +    \wh\gamma\cdot \wh\sigma_a^2(X_i) \right]= \sum_{i=\np+1}^{n}\wh\sigma_a^2(X_i),
\end{align*}
where $\wh\mu_a$ and $\wh\sigma_a^2$ are estimates of $\mu_a$ and $\sigma_a^2$, respectively.
Therefore, the approximated bias is the product of $\wh\gamma$ and the estimate of SLOPE.
For confidence intervals, we fix $\wh\gamma$ as obtained as above from the original source and target samples, and construct $95\%$ confidence intervals for the SLOPE  by bootstrapping  the source and the target samples. Therefore, the vertical bars shown in Figure \ref{fig:data.bias.slope.consumption} do not include the randomness of estimating $\gamma$. Such construction is to align with the common understanding in sensitivity analysis that the sensitivity parameter  is  taken as a pre-specified fixed value instead of a random quantity.

The results are shown in Figure \ref{fig:data.bias.slope.consumption}, where the two panels represent the mean and median, and each panel plots the approximated bias with SLOPE against the oracle bias. These dots roughly lie on the $y=x$ line, thereby validating the bias approximation with SLOPE.

\subsection{Auxiliary Data Information for Section \ref{subsec:data.weight}}\label{supp.subsec:data.weight}
For the data analysis in Section \ref{subsec:data.weight}, all health variables and the corresponding physical health index were measured at individual level. To keep the sample units as households, we average the individual level measurements over households.
Although we changed the outcome variable in terms of the weights, we kept the pre-treatment covariate the same to guarantee fair comparison across weighting schemes.
The pre-treatment covariate $X$ is the categorized baseline measurement of the physical health index (i.e., the average of  three health variables mentioned in Section \ref{subsec:data.weight}).
Table \ref{tab:X.health} gives the distribution of $X$ across countries on the overlapped sample.
\begin{table}[!h]
    \centering
     \caption{Baseline Measurement $X$ of physical health index.}
    \label{tab:X.health}
    \begin{tabular}{l lll}
    \hline
     &  Peru      &   India  &      Ethiopia \\
  Sample size &              1307    &     740    &     785    \\
  \hline
  Category of $X$ &&& \\
  (-1.36, 0.235] & 979 (74.9\%) & 523 (70.7\%) & 350 (44.6\%)\\
  (0.235,0.818] &328 (25.1\%) &  217 (29.3\%)  & 435 (55.4\%)\\
     \hline
    \end{tabular}
\end{table}

\clearpage
\section{Simulations}\label{supp.sec:simulation}
In this section, we validate the asymptotic properties of the proposed estimators in synthetic datasets that were generated to mimic the real data.

\subsection{Simulation Setting}
 Suppose $O$ is a continuous variable and $X$ is a random variable that is either binary and continuous.
In the case when $X$ is binary, we suppose the support $\SX=\{1,2\}$ and $P_{O\mid X=j}\sim N(\mu_j,\sigma_j^2)$ with $j=1,2$. In the case when $X$ is continuous, we suppose both $\PX$ and $\QX$ are Gaussian and $\POmidX\sim N(\alpha_m + \beta_mX, \alpha_v+\beta_mX^2)$.
Data generation parameters were estimated from real data in Section \ref{subsec:data.choice.source}. Specifically, the conditional distribution $\POmidX$ is estimated from the per-capita consumption in the treated group of Pakistan.   $\QX$ is estimated from Pakistan (i.e., no covariate shift) and Honduras (i.e., covariate shift). To construct a binary $X$, we dichotomize the log-transformed  baseline measurement by whether it is below the median; to construct a continuous $X$ we use the original log-transformed measurement.  The exact numbers of the (semi-)synthetic distributions are given in Table \ref{tab:sim.dgp}.

For both data generation procedures, we are interested in the SLOPE for mean and median. When $X$ is continuous, we also consider the regression coefficients in simple linear regression which regresses $O$ on $X$. For SLOPE of the mean and OLS coefficients, we consider three estimators: weighting estimator (\eqref{eq:est.weighting.slope.mean} and \eqref{eq:est.weighting.slope.ols}), the regression estimator (\eqref{eq:est.regression.slope.mean} and \eqref{eq:est.regression.slope.ols}), and the efficient influence function  based estimator \eqref{eq:est.eif.slope.mean}.
For the SLOPE of the median, we consider  the weighting estimator \eqref{eq:est.weighting.slope.median.normal} and the regression estimator \eqref{eq:est.regression.slope.median.normal}.
During estimation we suppose  $\POmidX$ is Gaussian.
For nuisance functions/parameters,  $\omega(X)$ is estimated with empirical distributions when $X$ is binary (see Section \ref{supp.subsec:w.est.discrete}) and entropy balancing when $X$ is continuous (see Section \ref{supp.subsec:w.est.EB}), $\mu(X)$ is estimated with linear regression, $\sigma^2(X)$ is estimated with subgroup sample variance when $X$ is binary and is estimated by  regressing the squared residuals $\{\O-\wh\mu(X)\}^2$ on $X^2$ when $X$ is continuous. The target mean is estimated by a weighted average $\sumnp\O_i\wh\omega(X_i)/\np$ for weighting estimators and by regression $\sumnq\wh\mu(X_i)/\nq$ for regression estimators. For the target median, when $X$ is binary it is estimated with the \texttt{qmixnorm} function in the R package \texttt{KScorrect} where the group mean and variance of $P_{\O\mid X=j}$ are estimated by  maximum likelihood estimation; when $X$ is continuous it is estimated by the numerical solution to
\begin{align*}
    \meannq \dfrac{1}{\wh\sigma(X_i)}\varphi\left(\dfrac{m_{1/2}-\wh\mu(X_i)}{\wh\sigma(X_i)}\right)=1/2
\end{align*}
via bisection, where $\varphi(\cdot)$ is the p.d.f. of the standard Gaussian.
Inference for the weighting and regression estimators is based on bootstrap with 1000 times of resampling and inference for the EIF-based estimator is based on the second moment of the EIF  with nuisances plugged in (i.e., \eqref{eq:se.est.eif.slope.mean} and \eqref{eq:se.est.eif.slope.ols}).

\begin{table}[!h]
    \centering
     \caption{Data generation in simulated data.}
    \label{tab:sim.dgp}
     {\footnotesize  \begin{tabular}{cc | cc | cc| cccc }
         \hline
$X$ &Cov. shift &\multicolumn{2}{c|}{ $\QX$}  &\multicolumn{2}{c|}{ $\PX$} & \multicolumn{4}{c}{ $\POmidX$}   \\
  \hline
  \multirow{3}{*}{Binary}& & $Q(X=1)$  & $Q(X=2)$ & $P(X=1)$ & $P(X=2)$ & $\mu_1$ & $\mu_2$ & $\sigma_1$ & $\sigma_2$\\
&  No & $0.1258 $ & $0.8742$ &  \multirow{2}{*}{$0.1258$} &  \multirow{2}{*}{$0.8742$} &  \multirow{2}{*}{$4.1816$} &  \multirow{2}{*}{$4.4773$} &  \multirow{2}{*}{$0.4761$} &  \multirow{2}{*}{$0.4524$}\\
  &Yes & $0.6597$ & $0.3403$ &&&&&\\
  \hline
  \multirow{3}{*}{Continuous}  &     & $\E_{\QX}(X)$ & $\sqrt{\var_{\QX}(X)}$  & $\E_{\PX}(X)$ & $\sqrt{\var_{\PX}(X)}$  & $\alpha_m$ & $\beta_m$ & $\alpha_v$ & $\beta_v$ \\
       & No & $4.5803$ & $0.5970$&
        \multirow{2}{*}{$4.5803$} & \multirow{2}{*}{$0.5970$} &
        \multirow{2}{*}{$3.1304$} &
        \multirow{2}{*}{$0.2766 $} &
        \multirow{2}{*}{$0.1924$} &
        \multirow{2}{*}{$-0.0003$} \\
       & Yes& $3.7054$ & $0.5340$ &&&&\\
        \hline
    \end{tabular}}
\end{table}

\subsection{Simulation Result}
Simulations are based on 1000 replicates. In each data setting, we consider $\np=\nq\in\{1000,2000\}$ and report the bias (bias), root mean squared error (rMSE), empirical standard deviation (empSD), the average estimated standard error (avgSE) and coverage rate (rate). Tables \ref{tab:sim.X.binary} and \ref{tab:sim.X.cont} list the simulation results for the SLOPE of mean and median.
For the OLS coefficients, Table \ref{tab:sim.ols.slope} and Table \ref{tab:sim.ols.intercept} provides results for the slope coefficient and  the intercept coefficient ,respectively.
As these results show, in all cases, the bias becomes closer to zero and sample sizes increases and the rMSE decays with root $\np$. The average of estimated standard error is close to the empirical standard deviation. The coverage rate is closer to $95\%$. Overall, the simulation results suggest that the estimators are $\sqrt{n}$-consistent and the standard error estimates are consistent.

\begin{table}[!h]
    \centering
      \caption{Simulation results for SLOPE of mean and median when $X$ is binary. All numbers have been multiplied with $100$.}
    \label{tab:sim.X.binary}
    {\footnotesize   \begin{tabular}{c c | ccc | ccc | cc | cc}
    \hline
     \multicolumn{2}{c|}{Estimand}  & \multicolumn{6}{c|}{Mean} & \multicolumn{4}{c}{Median}\\
     \hline
    \multicolumn{2}{c|}{Covariate shift} & \multicolumn{3}{c|}{Yes} & \multicolumn{3}{c|}{No}
       & \multicolumn{2}{c|}{Yes} & \multicolumn{2}{c}{No}\\
       \hline
        \multicolumn{2}{c|}{Estimator} & Regress & Weight & EIF & Regress & Weight & EIF  & Regress & Weight& Regress & Weight \\
       \hline
 \multirow{5}{*}{$\np=1000$}
 &bias    &0.01 & -0.10 & -0.10 &  0.03 & -0.01 & -0.02 &  0.03 &  0.01 & -0.01 & -0.02 \\
 &rMSE & 1.66 &  1.65 &  1.65 &  1.01 &  1.00 &  1.04 &  1.86 &  1.90 &  1.01 &  1.04 \\
 &empSD  & 1.66 &  1.65 &  1.65 &  1.01 &  1.00 &  1.04 &  1.86 &  1.90 &  1.01 &  1.04 \\
  &avgSE  &1.67 &  1.66 &  1.66 &  1.01 &  1.01 &  1.01 &  1.78 &  1.78 &  1.01 &  1.01 \\
 &rate & 94.7\% & 94.6\% & 94.4\% & 94.8\% & 94.8\% & 94.7\% & 93.7\% & 92.0\% & 94.5\% & 93.9\% \\
 \hline

\multirow{5}{*}{$\np=2000$}
 &bias    &-0.01 & -0.07 & -0.07 &  0.00 & -0.02 & -0.02 & -0.11 & -0.13 & -0.02 & -0.03 \\
 &rMSE & 1.18 &  1.18 &  1.18 &  0.68 &  0.68 &  0.68 &  1.29 &  1.33 &  0.72 &  0.73 \\
 &empSD  & 1.18 &  1.18 &  1.18 &  0.68 &  0.68 &  0.68 &  1.29 &  1.32 &  0.72 &  0.73 \\
   &avgSE  &1.19 &  1.19 &  1.19 &  0.72 &  0.72 &  0.72 &  1.26 &  1.26 &  0.72 &  0.72 \\
 &rate & 95.3\% & 94.9\% & 94.8\% & 95.7\% & 95.8\% & 95.8\% & 94.5\% & 93.6\% & 94.8\% & 94.4\% \\
\hline
    \end{tabular}}
\end{table}

\begin{table}[!h]
    \centering
      \caption{Simulation results for SLOPE of mean and median when $X$ is continuous. All numbers have been multiplied with $100$.}
    \label{tab:sim.X.cont}
     {\footnotesize  \begin{tabular}{c c | ccc | ccc | cc | cc}
    \hline
     \multicolumn{2}{c|}{Estimand}  & \multicolumn{6}{c|}{Mean} & \multicolumn{4}{c}{Median}\\
     \hline
    \multicolumn{2}{c|}{Covariate shift} & \multicolumn{3}{c|}{Yes} & \multicolumn{3}{c|}{No}
       & \multicolumn{2}{c|}{Yes} & \multicolumn{2}{c}{No}\\
       \hline
        \multicolumn{2}{c|}{Estimator} & Regress & Weight & EIF & Regress & Weight & EIF  & Regress & Weight& Regress & Weight \\
       \hline
 \multirow{5}{*}{$\np=1000$}
& bias     & -0.09 & -0.15 & -0.09 & -0.09 & -0.09 & -0.09 & -0.06 & -0.07 & -0.06 & -0.06 \\

& rmse     & 0.86 &  2.16 &  2.06 &  0.86 &  0.86 &  0.86 &  0.87 &  1.48 &  0.87 &  0.89 \\

& empSD & 0.85 &  2.16 &  2.05 &  0.85 &  0.85 &  0.86 &  0.86 &  1.48 &  0.86 &  0.89 \\

& avgSE  &0.84 &  2.24 &  1.92 &  0.84 &  0.85 &  0.84 &  0.84 &  1.46 &  0.84 &  0.88 \\

& rate& 93.9\%  & 91.6\%  & 91.9\%  & 93.9\%  & 94.1\%  & 94.0\%  & 94.3\%  & 93.6\%  & 94.3\%  & 94.5\%  \\

\hline

\multirow{5}{*}{$\np=2000$}
& bias     & -0.01 & -0.14 & -0.11 & -0.01 & -0.01 & -0.01 &  0.01 & -0.01 &  0.01 &  0.01 \\

& rmse     & 0.61 &  1.53 &  1.44 &  0.61 &  0.61 &  0.61 &  0.59 &  1.07 &  0.59 &  0.62 \\

& empSD  & 0.61 &  1.53 &  1.44 &  0.61 &  0.61 &  0.61 &  0.59 &  1.07 &  0.59 &  0.62 \\

& avgSE  & 0.60 &  1.44 &  1.38 &  0.60 &  0.60 &  0.60 &  0.60 &  1.04 &  0.60 &  0.62 \\

& rate& 94.2\%  & 92.4\%  & 93.2\%  & 94.2\%  & 94.4\%  & 94.7\%  & 95.3\%  & 93.7\%  & 95.3\%  & 95.5\%  \\

\hline
    \end{tabular}}
\end{table}

\begin{table}[!h]
    \centering
      \caption{Simulation results for SLOPE of the slope  coefficient in simple linear regression. All numbers have been multiplied with $100$.}
    \label{tab:sim.ols.slope}
    {\footnotesize   \begin{tabular}{c c | ccc | ccc }
    \hline
     \multicolumn{2}{c|}{Estimand}  & \multicolumn{6}{c}{Slope} \\
     \hline
    \multicolumn{2}{c|}{Covariate shift} & \multicolumn{3}{c|}{Yes} & \multicolumn{3}{c}{No}\\
       \hline
        \multicolumn{2}{c|}{Estimator} & Regress & Weight & EIF & Regress & Weight & EIF  \\
       \hline
 \multirow{5}{*}{$\np=1000$}
& bias      &-0.02 & -0.03 & -0.02 & -0.02 & -0.02 & -0.02\\
& rmse      & 0.23 &  0.47 &  0.47 &  0.19 &  0.19 &  0.19 \\
& empSD  & 0.23 &  0.47 &  0.47 &  0.18 &  0.19 &  0.19 \\
& avgSE   & 0.22 &  0.45 &  0.45 &  0.18 &  0.18 &  0.19 \\
& rate  & 93.9\%  & 91.0\%   & 92.4\%  & 93.9\%  & 94.1\%  & 95.6\%  \\

\hline

\multirow{6}{*}{$\np=2000$}
& bias      & 0    & -0.03 & -0.03 &  0    &  0    &  0    \\
& rmse      & 0.16 &  0.33 &  0.33 &  0.13 &  0.13 &  0.13 \\
& empSD   &  0.16 &  0.33 &  0.33 &  0.13 &  0.13 &  0.13 \\
& avgSE   &0.16 &  0.32 &  0.32 &  0.13 &  0.13 &  0.14 \\
& coverage  & 94.5\%  & 92.9\%  & 94.4\%  & 94.4\%  & 94.3\%  & 95.9\%  \\
\hline
    \end{tabular}}
\end{table}

\begin{table}[!h]
    \centering
      \caption{Simulation results for SLOPE of the intercept  coefficient in simple linear regression. All numbers have been multiplied with $100$.}
    \label{tab:sim.ols.intercept}
     {\footnotesize  \begin{tabular}{c c | ccc | ccc}
    \hline
     \multicolumn{2}{c|}{Estimand}  & \multicolumn{6}{c}{Intercept coefficient} \\
     \hline
    \multicolumn{2}{c|}{Covariate shift} & \multicolumn{3}{c|}{Yes} & \multicolumn{3}{c}{No}\\
       \hline
        \multicolumn{2}{c|}{Estimator} & Regress & Weight & EIF & Regress & Weight & EIF  \\
       \hline
 \multirow{5}{*}{$\np=1000$}
& bias      &-0.09 & -0.12 & -0.09 & -0.09 & -0.09 & -0.09 \\
& rmse      &  0.86 &  2.09 &  2.06 &  0.86 &  0.86 &  0.86 \\
& empSD   &  0.85 &  2.08 &  2.05 &  0.85 &  0.86 &  0.86 \\
& avgSE  & 0.84 &  1.93 &  1.94 &  0.84 &  0.84 &  0.90 \\
& rate  &93.9\%  & 90.6\%  & 92.3\%  & 93.9\%  & 94.0\%  & 95.2\%  \\

\hline

\multirow{6}{*}{$\np=2000$}
& bias      & -0.01 & -0.13 & -0.11 & -0.01 & -0.01 & -0.01 \\
& rmse      &  0.61 &  1.46 &  1.44 &  0.61 &  0.61 &  0.61 \\
&empSD  &0.61 &  1.46 &  1.44 &  0.61 &  0.61 &  0.61 \\
& avgSE   &  0.60 &  1.38 &  1.39 &  0.60 &  0.60 &  0.64 \\
& rate  &94.2\%  & 92.4\%  & 93.4\%  & 94.2\%  & 94.3\%  & 96.4\%  \\
\hline
    \end{tabular}}
\end{table}

\clearpage

\section{Proof for The Derivation of the SLOPE}\label{supp.sec:proof.slope}

\subsection{Proof of Theorem \ref{thm:slope.median}}
Since Theorem \ref{thm:slope.median} is a special case of Theorem \ref{thm:slope.quantile} with $q=1/2$, please find the proof of Theorem \ref{thm:slope.quantile} in Section \ref{supp.subsec:proof.slope.quantile}.

\subsection{Proof of Theorem \ref{thm:slope.if}}
We recall some notation defined in Section \ref{supp.subsec:exist.notation} for the support under $\QOXzero$: denote the support of $\QOXzero$ as $\SOX$ and the supports of the marginals $\QX$ and $\QOzero$ as $\SX$ and $\SO$, respectively.

\pf{Theorem \ref{thm:slope.if}}
The proof proceeds in two steps. First, we show that fixing $\QOX=\QOXzero $, the (partial) derivative of $\phi$ with respect to $\gamma$  at $\gamma=0$ is
\begin{align*}
   \phi'_{\gamma}(0)= \widetilde{Q}_{O,X}\in l^{\infty}(\SOX), \text{ such that }
    \int_Bd\widetilde{Q}_{O,X}=\int_{B} \{O-\E_{\POmidX}(O\mid X)\}d\QOXzero ,
\end{align*}
for any measurable set $B\in\SOX$.

Second, we show that for an  $H\in l^{\infty}(\SOX)$ such that $\int dH=0$, the Hadamard derivative of $\psi$ with respect to $\QOX $ at $\QOXzero$ in the direction of $H$ is
\begin{align*}
\psi'_{\QOXzero }(H) = \int \IF\left(O,X,\psi(\QOXzero )\right)dH.
\end{align*}
Then, using chain rule of Hadamard derivative (Theorem 20.4 of \cite{van2000asymptotic}), we have the derivative of the composite function $\psi\circ\phi$ with respect to $\gamma$ at $\gamma=0$ being:
\begin{align*}
   \psi'_{\QOX }\left(\phi'_{\gamma}(0)\right)= \int \IF \left(O,X,\psi(\QOXzero )\right) \{O-\E_{\POmidX}(O\mid X)\}d\QOXzero .
\end{align*}

We prove the two steps in order.

For the first step, we note that for any $h\in\mathcal {R}$ and $t\downarrow0$,
\begin{align*}
\int_Bd\QOX ^{th} - \int_Bd\QOXzero  &= \int_B\left(
    \dfrac{d\QOX ^{th}}{d\QOXzero } - 1
    \right) d\QOXzero \\
    &=\int_B\left[\dfrac{\exp(th)}{\E_{\POmidX}\left\{\exp(thO)\mid X\right\}}\right] d\QOXzero .
\end{align*}
Let $\Delta_{t,h}= \dfrac{d\QOX ^{th}}{d\QOXzero }-1$, then
\begin{align*}
    \Delta_{t,h} &= \dfrac{1+thO + O(t^2)}{1 + th\E_{\POmidX}(O\mid X)+O(t^2)} - 1\\
    &= \left\{1 + thO + O(t^2)\right\} \left\{1-th\E_{\POmidX}(O\mid X) + O(t^2)\right\} -1 + O(t^2)\\
    &= th\{O-\E_{\POmidX}(O\mid X)\} + O(t^2).
\end{align*}
Therefore,
\begin{align*}
    \dfrac{
    \int_B d\QOX ^{th} - \int_Bd\QOXzero
    }{t}
    =& \dfrac{\int_B \Delta_{t,h}d\QOXzero }{t}\\
    =& \dfrac{
    \int_B th\{O-\E_{\POmidX}(O\mid X)\} d\QOXzero  + O(t^2)
    }{t}\\
    \to&h\int_B\{O-\E_{\POmidX}(O\mid X)\} d\QOXzero .
\end{align*}

For the second step, we note that under Condition \ref{condition:regular.if.hadamard}, $\IF\left(o,x,\psi(\QOXzero )\right)=\psi'_{\QOXzero}[\delta_{o,x}-\QOXzero ]$. Then
\begin{align*}
    \int \IF\left(o,x,\psi(\QOXzero )\right)dH(o,x) &= \int \psi'_{\QOXzero }[\delta_{o,x}-\QOXzero ] dH(o,x)\\
    &=\psi'_{\QOX }\left[\int \delta_{o,x}-\QOXzero dH(o,x)\right]\\
    &= \psi'_{\QOX }[H],
\end{align*}
where the second equality follows from the linearity and continuity of Hadamard derivative, and the last equation follows from the observation that for $B\in l^{\infty}(\SOX)$,
\begin{align*}
    \int \delta_{o,x}(B)-\QOXzero (B)dH(o,x) = H(B) - \QOXzero (B)\cdot 0 = H(B).
\end{align*}
Hence, both two steps have been proved.
$\square$

\subsection{Proof of Theorem \ref{thm:slope.if.rho}}

\pf{Theorem \ref{thm:slope.if.rho}}
The proof follows the same procedure as the proof of Theorem \ref{thm:slope.if} except for replacing the exponential function $\exp(\gamma O)$ with $\rho(O,X,\gamma)$. Specifically, the conclusion of the first step becomes
\begin{align*}
   \phi'_{\gamma}(0)= \widetilde{Q}_{O,X}\in l^{\infty}(\SOX), \text{ such that }
    \int_Bd\widetilde{Q}_{O,X}=\int_{B} \left[\dot{\rho}(O,X,0)-\E_{\POmidX}\{\dot{\rho}(O,X,0)\mid X)\right]d\QOXzero.
\end{align*}
In the derivation of the first step, we use the Taylor expansion on $\rho(O,X,\gamma)$ around $\gamma=0$,
\begin{align*}
    dQ_{O,X}^{th}=1 + th\dot{\rho}(O,X,0) + O(t^2),
\end{align*}
which holds under Condition \ref{condiiton:regular.rho.slope}. The rest of the proof follows the same procedure as the proof of Theorem \ref{thm:slope.if}.

\subsection{Proof of Lemma \ref{lemma:slope.expectation}}
\pf{Lemma \ref{lemma:slope.expectation}}

Under sensitivity model \eqref{eq:sensitivity}, the target estimand is
\begin{align*}
    \psi^{\xi}(\QOXgamma) = \E_{\QX}\left[
    \dfrac{\E_{\POmidX}\left\{\exp(\gamma\O)\xi(\O,X)\mid X\right\}}{\E_{\POmidX}\left\{\exp(\gamma\O)\mid X\right\}}
    \right].
\end{align*}
Taking derivative with respect to $\gamma$ at $\gamma=0$, we have
\begin{align*}
    \SI(\QOXzero,\psi^{\xi}) &=
    \E_{\QX}\left[
    \E_{\POmidX}\left\{\O\xi(\O,X)\mid X\right\} -
    \E_{\POmidX}\left\{\xi(\O,X)\mid X\right\}-\E_{\POmidX}(\O\mid X)
    \right]\\
    &= \E_{\QX}\left\{\cov_{\POmidX}\left[\O,\xi(\O,X)\mid X\right]\right\}.
\end{align*}

$\square$

\subsection{Proof of Theorem \ref{thm:slope.quantile}}\label{supp.subsec:proof.slope.quantile}
\pf{Theorem \ref{thm:slope.quantile}}
Let $\kappa(\gamma)=\E_{\POmidX}\{\exp(\gamma O)\mid X\}$.
Under sensitivity model \eqref{eq:sensitivity}, the $q$-th quantile of $O$ under $\QOzero$ can be defined as
\begin{align*}
    q &= F_{\QOgamma}\left(\psi(F_{\QOgamma})\right)\\
    &= \int_{-\infty}^{F_{\QOgamma}\inv(q)}
    \int
    \dfrac{\exp(\gamma\O)}{\E_{\POmidX}\{\exp(\gamma\O)\mid X\}}
    d \QX d \POmidX\\
   &=  \int\int_{-\infty}^{F_{\QOgamma}\inv(q)}
      \dfrac{\exp(\gamma\O )}{\E_{\POmidX}\{\exp(\gamma\O)\mid X)\}}
     d\POmidX d \QX.
\end{align*}
On both sides, we take derivative with respect to $\gamma$. Applying the Leibniz rule, we have
\begin{align*}
    0 &= \int
     \left(\dfrac{\partial}{\partial\gamma}
    \int_{-\infty}^{F_{\QOzero}\inv(q)}
     \left[
     \dfrac{\exp(\gamma\O)}{\E_{\POmidX}\{\exp(\gamma\O)\mid X\}}
     \right]
   d\POmidX
    \right)\Bigg\vert_{\gamma=0}
       d\QX\\
       &=\int \left[
f_{\POmidX}(m_q\mid X)\cdot \SI(\QOXzero,\psi) +
\int_{-\infty}^{m_q}\{\O-\mu(X)\} d\POmidX \right]
       d\QX\\
    &=\SI(\QOXzero,\psi) \cdot \int f_{\POmidX}(m_q\mid X) d\QX +
    \E_{\QOzero}\{\O\ind(\O\leq m_q\} - \E_{\QX}\left\{
    F_{\POmidX}(m_q)
    \mu(X)\right\}\\
    &=\SI(\QOXzero,\psi)  f_{\QOzero}(m_q)+ \E_{\QOzero}\{\O\ind(\O\leq m_q)\} - \E_{\QX}\left\{
    F_{\POmidX}(m_q)
    \mu(X)\right\}.
\end{align*}
Reorganize terms, we have
\begin{align*}
    \SI(\QOXzero,\psi)
    =&\dfrac{
    \E_{\QX}\left\{
    F_{\POmidX}(m_q)
    \mu(X)\right\}-
     \E_{\QOzero}\{\O\ind(\O\leq m_q)\}
    }
    {f_{\QOzero}(m_q)}.
\end{align*}

\subsection{Proof of Theorem \ref{thm:slope.trimmed.mean}}
\pf{Theorem \ref{thm:slope.trimmed.mean}}
Under the sensitivity model \eqref{eq:sensitivity}, the target estimand is
\begin{align*}
    \psi^{\trim}(\QOgamma) &= \dfrac{1}{1-2\alpha}
    \int_{F_{\QOgamma}\inv(\alpha)}^{F_{\QOgamma}\inv(1-\alpha)} \O d\QOgamma\\
    &= \dfrac{1}{1-2\alpha}
    \int_{F_{\QOgamma}\inv(\alpha)}^{F_{\QOgamma}\inv(1-\alpha)}
    \dfrac{ \O\exp(\gamma \O)}{\E_{\POmidX}\{\exp(\gamma \O)\mid X\}}
    d\QOXzero.
\end{align*}
Taking derivative with respect to $\gamma$ at $\gamma=0$, we obtain the SLOPE as follows:
\begin{align*}
\SI(\QOXzero,\psi^{\trim}) =&
    \dfrac{\partial  \psi^{\trim}(\QOgamma)}{\partial\gamma}\Bigg\vert_{\gamma=0}\\
    =& \dfrac{1}{1-2\alpha}
    \int
    \left\{
    F_{\QOzero}\inv(1-\alpha)f_{\QOzero}(F_{\QOzero}\inv(1-\alpha)) \left[\dfrac{\partial F_{\QOgamma}\inv(1-\alpha)}{\partial\gamma}\Bigg\vert_{\gamma=0}\right]
    \right\}
   d\QX\\
    &-\dfrac{1}{1-2\alpha}
 \int
    \left\{
    F_{\QOzero}\inv(\alpha)f_{\QOzero}(F_{\QOzero}\inv(\alpha)) \left[\dfrac{\partial F_{\QOgamma}\inv(\alpha)}{\partial\gamma}\Bigg\vert_{\gamma=0}\right]
    \right\}
    d\QX\\
    &+\dfrac{1}{1-2\alpha}
    \int
    \left[
   \int_{F_{\QOzero}\inv(\alpha)}^{F_{\QOzero}\inv(1-\alpha)}
  \O \{\O-\mu(X)\}
  d\POmidX
    \right]
    d\QX\\
    =& \dfrac{1}{1-2\alpha}F_{\QOzero}\inv(1-\alpha)
    \E_{\QOXzero}\left[
    \P(\O\leq F_{\QOzero}\inv(1-\alpha)\mid X)\mu(X) - \O\ind(\O\leq F_{\QOzero}\inv(1-\alpha))
    \right]\\
    &- \dfrac{1}{1-2\alpha}F_{\QOzero}\inv(\alpha)
    \E_{\QOXzero}\left[\P(\O\leq F_{\QOzero}\inv(\alpha)\mid X)\mu(X) - \O\ind(\O\leq F_{\QOzero}\inv(\alpha))\right]\\
    &+\dfrac{1}{1-2\alpha}\E_{\QX}\left(
    \E_{\POmidX}\left[
   \O \{\O-\mu(X)\}\ind_{[F_{\QOzero}\inv(\alpha),F_{\QOzero}\inv(1-\alpha)]}(\O)
    \mid X\right]
    \right),
\end{align*}
where the last equality follows from the SLOPE for quantiles (Theorem \ref{thm:slope.quantile}).

$\square$

\subsection{Proof of Theorem \ref{thm:slope.ols}}
\pf{Theorem \ref{thm:slope.ols}}
This theorem is a special case of Corollary \ref{thm:slope.z} with $s(Y,\psi^{\olssub}) = \Xsub\Xsub\trans\psi^{\olssub}-\Xsub Y$.
$\square$

\subsection{Proof of Theorem \ref{thm:slope.ancova}}

\pf{Theorem \ref{thm:slope.ancova}}

We start with $\tau_a$.  Suppose $\psi^{\ols}$ satisfies $\E_{\QOXzero}\left(XX\trans\psi^{\ols}-XY\right)$. From Theorem \ref{thm:slope.ols}, the SLOPE for $\tau_a$ is the second entry of $\SI(\QOXzero,\psi^{\ols})$, which is
\begin{align*}
  &\SI(\QOXzero,\psi^{\ols})\\
  =&   \left\{\E_{\QX}(XX\trans)\right\}\inv\E_{\QX}\left[X\sigma^2(X)\right]\\
  =& \left\{
   \begin{bmatrix}
  1 & \E_{Q_A}(A) & \E_{Q_L}(L\trans)\\
  \E_{Q_A}(A) & \E_{Q_A}(A^2) & \E_{\QX}(AL\trans)\\
  \E_{Q_L}(L) & \E_{\QX}(LA) & \E_{Q_L}(LL\trans)
   \end{bmatrix}
  \right\}\inv
  \begin{bmatrix}
      &\E_{\QX}\{\sigma^2(X)\}\\
      &\E_{\QX}\{A\sigma^2(X)\}\\
     & \E_{\QX}\{L\sigma^2(X)\}
  \end{bmatrix}\\
  =&
  \begin{bmatrix}
        1-b\trans S\inv b & \star & \star \\
        -\frac{\E_{Q_A}(A))}{\var_{Q_A}(A)} + \delta\trans V\inv\E_{Q_A}(A)\delta - \delta\trans V\inv\E_{Q_L}(L) & 1/\var_{Q_A}(A) + \delta\trans V\inv\delta & \star\\
        V\inv\delta\E_{Q_A}(A) - V\inv\E_{Q_L}(L) & -V\inv\delta & -V\inv
    \end{bmatrix}
    \begin{bmatrix}
      &\E_{\QX}\{\sigma^2(X)\}\\
      &\E_{\QX}\{A\sigma^2(X)\}\\
     & \E_{\QX}\{L\sigma^2(X)\}
  \end{bmatrix},
\end{align*}
where  $b = [\E_{Q_A}(A), \E_{Q_L}(L\trans)]$, $\delta = \cov_{Q_{L,A}}[L,A]/\var_{Q_{A}}[A]$, $V=\cov_{Q_X}(X)-\delta\delta\trans\var_{Q_A}(A)$,  $S = \begin{bmatrix}
    \var_{Q_A}(A) & \cov_{Q_X}[A,L\trans]\\
    \cov_{Q_X}[L,A]& \cov_{Q_L}(L)
\end{bmatrix}$, and entries as ``$\star$'' are omitted because that matrix is symmetric.
Then $\SI(\QOXzero,\tau\adj)$ in \eqref{eq:slope.tau.adj} follows by expanding terms in the formula above and taking the second entry.

The SLOPE for $\tau\unadj$ in the unadjusted model can be obtained similarly and is hence omitted.
$\square$

\subsection{Proof of Theorem \ref{thm:slope.mad}}
\pf{Theorem \ref{thm:slope.mad}}
Under sensitivity model \eqref{eq:sensitivity}, the MAD satisfies
\begin{align*}
    1/2 =& \int_{\psi^{\med}(\QOgamma) - \psi^{\mad}(\QOgamma)}
    ^{\psi^{\med}(\QOgamma) + \psi^{\mad}(\QOgamma)}
    d\QOXgamma\\
    =&\E_{\QX}\left[
    \int_{\psi^{\med}(\QOgamma)-\psi^{\mad}(\QOgamma)}^{\psi^{\med}(\QOgamma)+\psi^{\mad}(\QOgamma)}
    \dfrac{\exp(\gamma\O)}{\E_{\POmidX}\{\exp(\gamma \O)\mid X\}}
    d\POmidX
    \right].
\end{align*}
Taking derivative with respect to $\gamma$ at $\gamma=0$, we have
\begin{align*}
    0=& \E_{\QX}\left[ f_{\POmidX}\left(m_{1/2}+\mad\mid X\right)
    \left\{
    \dfrac{\partial\left\{\psi^{\med}(\QOgamma) + \psi^{\mad}(\QOgamma)\right\}}{\partial\gamma}\Bigg\vert_{\gamma=0}
    \right\}    \right]\\
    &-
    \E_{\QX}\left[f_{\POmidX}\left(m_{1/2}-\mad\mid X\right)
     \left\{
    \dfrac{\partial\left\{ \psi^{\med}(\QOgamma) - \psi^{\mad}(\QOgamma)\right\}}{\partial\gamma}\Bigg\vert_{\gamma=0}
    \right\}
    \right]\\
    &+ \E_{\QX}
    \left[\int_{m_{1/2}-\mad}^{m_{1/2}+\mad} \{\O-\mu(X)\}d\POmidX\right]\\
    =& \left\{
    f_{\QOzero}(m_{1/2}+\mad) + f_{\QOzero}(m_{1/2}-\mad)
    \right\}\cdot\SI(\QOXzero,\psi^{\mad}) \\
    &+
     \left\{
    f_{\QOzero}(m_{1/2}+\mad) - f_{\QOzero}(m_{1/2}-\mad)
    \right\}\cdot\SI(\QOXzero,\psi^{\med})\\
    &+
    \E_{\QOXzero}
    \left[\ind_{[m_{1/2}-\mad,m_{1/2}+\mad]}(\O)\{\O-\mu(X)\}\right].
\end{align*}
Then the result follows from re-organizing this equation.
$\square$

\subsection{Proof of Theorem \ref{thm.slope:L.stat}}
\pf{Theorem \ref{thm.slope:L.stat}}
For the L-estimand
\begin{align*}
     \psi(\QO) &= \int_{0}^1h(F_{\QO}\inv(p))l(p) d p,
\end{align*}
the corresponding SLOPE is (assuming differentiation and integration can exchange)
\begin{align*}
&\SI(\QOXzero,\psi) \\
=&     \dfrac{\partial \psi(\QOgamma)}{\partial\gamma}\Bigg\vert_{\gamma=0}\\
    =&  \int_{0}^1h'(F_{\QOzero}\inv(p))\cdot \frac{\partial F_{\QOgamma}\inv(p)}{\partial\gamma}\Bigg\vert_{\gamma=0}l(p) d p\\
    =& \int_{0}^1 h'(F_{\QOzero}\inv(p))\cdot
\left(\dfrac{-\E\{O\ind(O\leq F_{\QO}\inv(p))\}}{f_{\QOzero}(F_{\QOzero}\inv(p))} +
    \dfrac{\E_{\QX}\left[
    \mu(X)\E_{\POmidX}\left\{\ind(O\leq F_{\QOzero}\inv(p))\mid X\right\}
    \right]}{f_{\QOzero}(F_{\QOzero}\inv(p))}
    \right)
    l(p) d p,
\end{align*}
where the second line follows from the SLOPE for quantiles (Theorem \ref{thm:slope.quantile}).

$\square$

We note that SLOPE of L-estimands can be alternatively derived from the IF below  \citep[(3.11)]{huber1981robust} using Theorem \ref{thm:slope.if},
\begin{align*}
    \IF(o,x,\psi(\QOXzero)) &= \int \frac{ph'(F_{\QOzero}\inv(p))}{f_{\QOzero}(F_{\QOzero}\inv(p))}l(p) d p -
    \int_{F_{\QOzero}(\o)}^1\frac{h'(F_{\QOzero}\inv(p))}{f_{\QOzero}(F_{\QOzero}\inv(p))}l(p) d p.
\end{align*}

\section{Proof for the Estimation Theory} \label{supp.sec:proof.estimation}

\subsection{Proof of Theorem \ref{thm:est.weight} and Theorem \ref{thm:est.regress}}
The proof for Theorems \ref{thm:est.weight} and \ref{thm:est.regress} follows from standard M-estimation theory in \citet{newey1994large} and \citet{van2000asymptotic}. We next prove Theorem \ref{thm:est.weight} and omit the proof of Theorem \ref{thm:est.regress} since the proof for both theorems follows the same procedure.

\pf{Theorem \ref{thm:est.weight}}
We begin by proving consistency under Condition \ref{assump:regular.weighted.consistent} using the standard techniques in \citet[Theorem 2.1]{newey1994large}.
Let
\begin{align*}
    M(\eta)&=-\E\left[G^W(T_i,O_i,X_i,\eta)\right]\trans \E\left[G^W(T_i,O_i,X_i,\eta)\right],\text{ and}\\
    \wh{M}_n(\eta) &= -\left\{\dfrac{1}{n}G^W(T_i,O_i,X_i,\eta)\right\}\trans \left\{\dfrac{1}{n}G^W(T_i,O_i,X_i,\eta)\right\}.
\end{align*}
First, under  Condition \ref{assump:regular.weighted.consistent}(i), $\eta^W$ uniquely maximizes $M(\eta)$.
Next, using Lemma 2.4 of \citet{newey1994large}, under Condition \ref{assump:regular.weighted.consistent}(ii) and (iii), we have the uniform convergence of $G^W$:
\begin{align*}
    \sup_{\eta\in\Theta}\left\lVert\dfrac{1}{n}\sum_{i=1}^nG^W(T_i,O_i,X_i,\eta)-\E\{G^W(T_i,O_i,X_i,\eta)\}\right\rVert \to_p 0,
\end{align*}
and that the function $\E[G^W(T_i,O_i,X_i,\eta)]$ is continuous with respect to $\eta$, where $\to_p$ denotes convergence in probability. Then $M(\eta)$ is also continuous with respect to $\eta$. Next we show the uniform convergence of $\wh M_n$. Note that  the compactness of $\Theta$ implies the boundedness of $\E[G^W(T_i,O_i,X_i,\eta)]$. Then for any $\eta\in\Theta$,
\begin{align*}
    &\left|\wh{M}_n(\eta)-M(\eta)\right| \\
    \leq &
    \left\lVert \E\left[G^W(T_i,O_i,X_i,\eta)\right]- \dfrac{1}{n}\sum_{i=1}^nG^W(T_i,O_i,X_i,\eta)\right\rVert
    \cdot \
    \left\lVert  \E\left[G^W(T_i,O_i,X_i,\eta)\right]+ \dfrac{1}{n}\sum_{i=1}^nG^W(T_i,O_i,X_i,\eta) \right\rVert
\end{align*}
implies the uniform convergence,
\begin{align}\label{pf.eq:est.unif.converge}
     \sup_{\eta\in\Theta}| M_n(\eta)-M(\eta)|\to_p0.
\end{align}
With all regularity conditions checked above, we prove the consistency mimicking the proof of Theorem 2.1 in \citet{newey1994large}.
For any $\varepsilon>0$. with probability approaching one, we have
\begin{align}\label{pf.eq:est.maximum}
    M(\wh\eta^W) > \wh M_n(\wh\eta^W) - \varepsilon/3
    > \wh M_n(\eta^W)-\varepsilon/3-\varepsilon/3
    > M(\eta^W) - \varepsilon/3-\varepsilon/3  - \varepsilon/3,
    = M(\eta^W) - \varepsilon.
\end{align}
where the first inequality and third equality hold by the uniform convergence \eqref{pf.eq:est.unif.converge}, and   the second inequality holds because $\wh\eta^W$ uniquely maximizes $\wh M_n(\cdot)$.
Let $\mathcal{N}$ be an open subset of $\Theta$ that contains $\eta^W$. By the compactness of $\Theta$ and continuity of $M$, we have $\sup_{\eta\in\Theta\cap\mathcal{N}^C}M(\eta)<M(\eta^W)$ with probability approaching one. Let $\varepsilon=M(\eta^W)-\sup_{\eta\in\Theta\cap\mathcal{N}^C}M(\eta)$, then \eqref{pf.eq:est.maximum} implies that $\wh\eta^W\in\mathcal{N}$ with probability approaching one. Hence, with Condition \ref{assump:regular.weighted.an}(iv), the consistency of the SLOPE follows from the continuous mapping theorem.

Next, we prove the  asymptotic normality. It directly follows from Theorem 5.31 of \citet{van2000asymptotic} that under Conditions \ref{assump:regular.weighted.consistent}-\ref{assump:regular.weighted.an}, $\wh\eta^W$ is asymptotically normal in that $\sqrt{n}\left(\wh\eta^W-\eta^W\right)\to_dN\left(0, V^W (V^W)\trans\right)$. Then the asymptotic normality of the SLOPE estimate follows from delta method (Theorem 3.1 of \citet{van2000asymptotic}).

$\square$

\subsection{Proof of the Derivation of Efficient Influence Functions}
\subsubsection{Preliminaries}
Let $f_{P_{X}}$, $f_{\QX}$, and $f_{\POmidX}$ be the density functions of the corresponding random variables on the subscripts. For a generic observation, the log-likelihood of the observed data on the joint population can be written as
\begin{align*}
    l(T,O,X) =& (1-T) \log f_{\POmidX}(O\mid\X) + (1-T)\log f_{\PX}(X) + T\log f_{\QX}(X)\\
    &+
(1-T)\log (\pr(T=0)) + T\log(\pr(T=1)).
\end{align*}

Consider the Hilbert space $\calH$ that contains all one-dimensional zero-mean measurable functions of the observed data with finite variance.
Consider $f_{\PX}$, $f_{\QX}$, and $f_{\POmidX}$ as nuisance functions and denote their nuisance tangent spaces as $\calT_{\PX}$, $\calT_{\QX}$, and $\calT_{\POmidX}$, respectively.
Then  $\calH$ can be decomposed as
\begin{align*}
    \calH&=\calT_{\PX} \oplus \calT_{\QX} \oplus\calT_{\POmidX},\text{ where}\\
    \calT_{\PX} &= \left\{ (1-T) a_1(O,X): \E_{\POmidX}\{a_1(O,X)\mid X\}=0, \var_{\POmidX}\left[a_1(O,X)\mid X\right]<\infty\right\},\\
    \calT_{\QX} &=  \left\{ T a_2(X): \E_{\QX}\{a_2(X)\}=0, \var_{\QX}\left[a_2(X)\right]<\infty\right\},\\
     \calT_{\POmidX}&=  \left\{ (1-T)a_3(X): \E_{\PX}\{a_3(X)\}=0, \var_{\PX}\left[a_3(X)\right]<\infty\right\}.
\end{align*}
We consider parametric submodels $f_{\POmidX}(O\mid X,\xi_1)$ and $f_{\QX}(X,\xi_2)$ where $\xi_1=0$ and $\xi_2=0$ correspond to the underlying truth. We also let
\begin{align*}
    \SC_{\xi_1}(O,X) = \dfrac{\partial\log f_{\POmidX}(O\mid X,\xi_1)}{\partial\xi_1},
    \quad
    \SC_{\xi_2}(X) = \dfrac{\partial\log f_{\QX}(X,\xi_2)}{\partial\xi_2}
\end{align*}
be the score functions.

We revise the notation of the target functional  to indicate the dependency on the nuisance parameters $\xi_1$ and $\xi_2$. Specifically, let $\psi(\xi_1,\xi_2)$ be the target functional. Then the efficient influence function for the target functional, $\EIF(T,O,X,\psi)$, satisfies
\begin{align}
    \dfrac{\partial\psi(\xi_1,\xi_2)}{\partial\xi_1}\Bigg\vert_{\xi_1=0} &= \E\left[(1-T)\cdot \EIF(T,O,X,\psi)\cdot \SC_{\xi_1}(O,X)\right],\label{supp.eq:pf.eif.psi.xi1}\\
    \dfrac{\partial\psi(\xi_1,\xi_2)}{\partial\xi_2}\Bigg\vert_{\xi_2=0} &= \E\left\{T\cdot \EIF(T,O,X,\psi)\cdot \SC_{\xi_2}(X)\right\},\label{supp.eq:pf.eif.psi.xi2}
\end{align}

Similarly, we revise the notation of the SLOPE and its component: let  $\SI(\xi_1,\xi_2)$ be the  SLOPE which takes the form of
\begin{align*}
    \SI(\xi_1,\xi_2) & = -\dfrac{\eta_1(\xi_1,\xi_2)}{\eta_2(\xi_1,\xi_2)}, \text{ where}\\
    \eta_1(\xi_1,\xi_2) &= \E_{\QOXzero}\{s(O,X,\psi)\{O-\mu(X)\},\\
    \eta_2(\xi_1,\xi_2) &= \E_{\QOXzero}\{\dot{s}(O,X,\psi)\}.
\end{align*}

Likewise, the efficient influence function for the SLOPE, $\EIF(T,O,X,\SI)$, satisfies
\begin{align}
    \E\left[(1-T)\EIF(T,O,X,\SI)\cdot \SC_{\xi_1}(O,X)\right]
    &=\dfrac{\partial\SI(\xi_1,\xi_2)}{\partial\xi_1}\Bigg\vert_{\xi_1=0}\nonumber\\
    &=
   - \dfrac{
    \dfrac{\partial\eta_1(\xi_1,\xi_2)}{\partial\xi_1}\cdot\eta_2(\xi_1,\xi_2) -
    \dfrac{\partial\eta_2(\xi_1,\xi_2)}{\partial\xi_1}\cdot\eta_1(\xi_1,\xi_2)
    }
    {\left\{\eta_2(\xi_1,\xi_2)\right\}^2}
    \Bigg\vert_{\xi_1=0}
    ,\label{supp.eq:pf.eif.slope.xi1}\\
      \E\left\{T\cdot \EIF(T,O,X,\SI)\cdot \SC_{\xi_2}(X)\right\} &=\dfrac{\partial\SI(\xi_1,\xi_2)}{\partial\xi_2}\Bigg\vert_{\xi_2=0} \nonumber\\
      &=
   - \dfrac{
    \dfrac{\partial\eta_1(\xi_1,\xi_2)}{\partial\xi_2}\cdot\eta_2(\xi_1,\xi_2) -
    \dfrac{\partial\eta_2(\xi_1,\xi_2)}{\partial\xi_2}\cdot\eta_1(\xi_1,\xi_2)
    }
    {\left\{\eta_2(\xi_1,\xi_2)\right\}^2}
    \Bigg\vert_{\xi_2=0},\label{supp.eq:pf.eif.slope.xi2}
\end{align}

\subsubsection{Proof of Proposition \ref{prop:eif.slope}}
\pf{Proposition \ref{prop:eif.slope}}
Suppose the efficient influence functions for the target functional and the SLOPE take the following form,
\begin{align*}
\EIF(T,O,X,\psi) = (1-T)e_1(O,X) + Te_2(X),\quad
    \EIF(T,O,X,\SI) = (1-T)a_1(O,X) + Ta_2(X),
\end{align*}
where $(1-T)e_1(O,X),(1-T)a_1(O,X)\in\calT_{\POmidX}$ and $Te_2(X),Ta_2(X)\in\calT_{\QX}$. The specific forms of $e_1(O,X)$ and $e_2(X)$ are given in Proposition \ref{prop:eif.psi}.

In light of \eqref{supp.eq:pf.eif.slope.xi1} and \eqref{supp.eq:pf.eif.slope.xi2}, our goal is to solve the following equations for $a_1$ and $a_2$,
\begin{align}
    \E\left\{(1-T)a_1(O,X)\cdot\SC_{\xi_1}(O,X)\right\} &=    \dfrac{\partial\SI(\xi_1,\xi_2)}{\partial\xi_1}\Bigg\vert_{\xi_1=0}\nonumber\\
    &=   - \dfrac{
    \dfrac{\partial\eta_1(\xi_1,\xi_2)}{\partial\xi_1}\cdot\eta_2(\xi_1,\xi_2) -
    \dfrac{\partial\eta_2(\xi_1,\xi_2)}{\partial\xi_1}\cdot\eta_1(\xi_1,\xi_2)
    }
    {\left\{\eta_2(\xi_1,\xi_2)\right\}^2}
    \Bigg\vert_{\xi_1=0}, \label{supp.eq:pf.eif.slope.a1}\\
    \E\left\{Ta_2(X)\cdot\SC_{\xi_2}(X)\right\} &= \dfrac{\partial\SI(\xi_1,\xi_2)}{\partial\xi_2}\Bigg\vert_{\xi_2=0}\nonumber\\
    &=   - \dfrac{
    \dfrac{\partial\eta_1(\xi_1,\xi_2)}{\partial\xi_2}\cdot\eta_2(\xi_1,\xi_2) -
    \dfrac{\partial\eta_2(\xi_1,\xi_2)}{\partial\xi_2}\cdot\eta_1(\xi_1,\xi_2)
    }
    {\left\{\eta_2(\xi_1,\xi_2)\right\}^2}
    \Bigg\vert_{\xi_2=0}.\label{supp.eq:pf.eif.slope.a2}
\end{align}

We start with \eqref{supp.eq:pf.eif.slope.a1}. In order to calculate the right hand side (RHS), we first calculate $\partial\eta_1(\xi_1,\xi_2)/\partial\xi_1$ and $\partial\eta_2(\xi_1,\xi_2)/\partial\xi_1$; we have
\begin{align*}
    &\dfrac{\partial\eta_1(\xi_1,\xi_2)}{\partial\xi_1}\Bigg\vert_{\xi_1=0}\\
 =& \dfrac{\partial}{\partial\xi_1} \E_{\QOXzero} \left[s(O,X,\psi)\{O-\mu(X)\}\right]  \\
 =& \E_{\QOXzero} \left[s(O,X,\psi)\{O-\mu(X)\}\cdot\SC_{\xi_1}(O,X)\right] +
    \E_{\QOXzero}\left[\dot{s}(O,X,\psi)\{O-\mu(X)\}\right]\cdot\dfrac{\partial\psi(\xi_1,\xi_2)}{\partial\xi_1}\\
&-\E_{\QOXzero}\left[s(O,X,\psi)\dfrac{\partial}{\partial\xi_1}\int Of_{\POmidX}(O\mid X,\xi_1)dO\right]\\
=&\E_{\QOXzero} \left[s(O,X,\psi)\{O-\mu(X)\}\cdot\SC_{\xi_1}(O,X)\right] \\
& +  \E_{\QOXzero}\left[\dot{s}(O,X,\psi)\{O-\mu(X)\}\right]\cdot
\pr(T=0)\cdot \E_{\QOXzero}\cdot\left\{\dfrac{1}{\omega(X)}e_1(O,X,\psi)\cdot\SC_{\xi}(O,X)\right\}\\
&-\E_{\QOXzero}\left[\E_{\POmidX}\{s(O,X,\psi)\mid X\}O\cdot\SC_{\xi_1}(O,X)\right]\\
=& \E_{\QOXzero}\left\{c_1(O,X)\cdot\SC_{\xi_1}(O,X)\right\},
\end{align*}
where
\begin{align*}
    c_1(O,X) =& s(O,X,\psi)\{O-\mu(X)\} +  \E_{\QOXzero}\left[\dot{s}(O,X,\psi)\{O-\mu(X)\}\right]\cdot \dfrac{\pr(T=0)}{\omega(X)}e_1(O,X,\psi)\\\
    &-\E_{\POmidX}\{s(O,X,\psi)\mid X\}O,
\end{align*}
and
\begin{align*}
    &\dfrac{\partial\eta_2(\xi_1,\xi_2)}{\partial\xi_1}\\
    =&  \E_{\QOXzero}\{\ddot{s}(O,X,\psi)\} \cdot\dfrac{\partial\psi(\xi_1,\xi_2)}{\psi_1} +
    \E_{\QOXzero}\left\{\dot{s}(O,X,\psi)\cdot\SC_{\xi_1}(O,X)\right\} +
    \E_{\QOXzero}\left\{\dot{s}(O,X,\psi)\cdot\SC_{\xi_1}(O,X)\right\}\\
    =& \E_{\QOXzero}\{\ddot{s}(O,X,\psi)\} \cdot\pr(T=0)\E_{\QOXzero}\left\{\dfrac{1}{\omega(X)}e_1(O,X,\psi)\cdot\SC(O,X,\psi)\right\} \\
    &+ \E_{\QOXzero}\left\{\dot{s}(O,X,\psi)\cdot\SC_{\xi_1}(O,X)\right\}\\
    =& \E_{\QOXzero}\{c_2(O,X)\cdot\SC_{\xi_1}(O,X)\},
\end{align*}
where
\begin{align*}
    c_2(O,X) &= \dfrac{\pr(T=0)}{\omega(X)}\E_{\QOXzero}\{\ddot{s}(O,X,\psi)\}\cdot e_1(O,X,\psi)+ \dot{s}(O,X,\psi)
\end{align*}
Therefore, the RHS of \eqref{supp.eq:pf.eif.slope.a1} is
\begin{align*}
  \dfrac{\partial\SI(\xi_1,\xi_2)}{\partial\xi_1}&=   - \dfrac{
    \dfrac{\partial\eta_1(\xi_1,\xi_2)}{\partial\xi_1}\cdot\eta_2(\xi_1,\xi_2) -
    \dfrac{\partial\eta_2(\xi_1,\xi_2)}{\partial\xi_1}\cdot\eta_1(\xi_1,\xi_2)
    }
    {\{\eta_2(\xi_1,\xi_2)\}^2}
\\
    &= \dfrac{
    \E_{\QOXzero}
    \left[
    \left\{-c_1(O,X)- \SI(\xi_1,\xi_2)\cdot c_2(O,X)\right\}\cdot \SC_{\xi_1}(O,X)
    \right]
    }
    {\E_{\QOXzero}\{\dot{s}(O,X,\psi)\}}.
\end{align*}
Equalizing both sides of  \eqref{supp.eq:pf.eif.slope.a1}, we have
\begin{align*}
    0 =& \E\left\{(1-T)a_1(O,X)\cdot\SC_{\xi_1}(O,X)\right\} -  \dfrac{\partial\SI(\xi_1,\xi_2)}{\partial\xi_1}\\
   =& \E_{\QOXzero}\left[\dfrac{\pr(T=0)}{\omega(X)}a_1(O,X)\cdot\SC_{\xi_1}(O,X)\right]\\
   &-
   \dfrac{
    \E_{\QOXzero}
    \left[
    \left\{-c_1(O,X)- \SI(\xi_1,\xi_2)\cdot c_2(O,X)\right\}\cdot \SC_{\xi_1}(O,X)
    \right]
    }
    {\E_{\QOXzero}\{\dot{s}(O,X,\psi)\}}\\
    =& \E_{\QOXzero}
\left[\left\{
    \dfrac{\pr(T=0)}{\omega(X)}a_1(O,X) -
    \dfrac{-c_1(O,X)-\SI(\xi_1,\xi_2)\cdot c_2(O,X)}{\E_{\QOXzero}\{\dot{s}(O,X,\psi)\}}
    \right\}\cdot\SC_{\xi_1}(O,X)
    \right].
\end{align*}
Noticing that $\SC_{\xi_1}(O,X)\in\calT_{\POmidX}$ (and hence $\E_{\POmidX}\{\SC_{\xi}(O,X)\mid X\}=0$), we have
\begin{align*}
   &\dfrac{\pr(T=0)}{\omega(X)} a_1(O,X) \\
   =&
   \dfrac{-c_1(O,X)-\SI(\xi_1,\xi_2)\cdot c_2(O,X)}{\E_{\QOXzero}\{\dot{s}(O,X,\psi)\}} -
   \E_{\POmidX}\left[
   \dfrac{-c_1(O,X)-\SI(\xi_1,\xi_2)\cdot c_2(O,X)}{\E_{\QOXzero}\{\dot{s}(O,X,\psi)\}}
   \Bigg\vert X\right].
\end{align*}
Therefore, by plugging in $c_1(O,X)$ and $c_2(O,X)$ at $\xi_j=0$ ($j=1,2$) and reorganizing terms, $a_1(O,X)$ can be written as
\begin{align*}
    &a_1(O,X)\\
    =&  \dfrac{\omega(X)}{\pr(T=0)}\dfrac{
    -s(O,X,\psi)\{O-\mu(X)\}+ \cov_{\POmidX}[s(O,X,\psi),O\mid X]+
    \E_{\POmidX}\{s(O,X,\psi)\mid X\}\{O-\mu(X)\}
    }{\E_{\QOXzero}\{\dot{s}(O,X,\psi)\}} \\
    &-\dfrac{\omega(X)\cdot\SI}{\pr(T=0)}
    \dfrac{
    \dot{s}(O,X,\psi) - \E_{\POmidX}\{\dot{s}(O,X,\psi)\mid X\}
    }{\E_{\QOXzero}\{\dot{s}(O,X,\psi)\}}\\
    & - \SI\cdot\E_{\QOXzero}\{\ddot{s}(O,X,\psi)\}e_1(O,X,\psi).
\end{align*}

Next, we consider \eqref{supp.eq:pf.eif.slope.a2} for $a_2(O,X)$. For the RHS of \eqref{supp.eq:pf.eif.slope.a2}, we calculate $\partial\eta_1(\xi_1,\xi_2)/\partial\xi_2$ and $\partial\eta_2(\xi_1,\xi_2)/\partial\xi_2$; we have
\begin{align*}
     \dfrac{\partial\eta_1(\xi_1,\xi_2)}{\partial\xi_2}
    =& \E_{\QOXzero}\left[s(O,X,\psi)\{O-\mu(X)\}\cdot\SC_{\xi_2}(X)\right] +
    \E_{\QOXzero}\left[\dot{s}(O,X,\psi)\{O-\mu(X)\}\right]\cdot
    \dfrac{\partial\psi(\xi_1,\xi_2)}{\partial\xi_2}\\
    =& \E_{\QOXzero}\left[s(O,X,\psi)\{O-\mu(X)\}\cdot\SC_{\xi_2}(X)\right]\\
    &+
    \E_{\QOXzero}\left[\dot{s}(O,X,\psi)\{O-\mu(X)\}\right]\cdot \pr(T=1)\E_{\QX}\{e_2(X)\cdot\SC_{\xi_2}(X)\}\\
    =& \E_{\QX}\{b_2(X)\cdot\SC_{\xi_2}(X)\}, \text{ where}\\
    b_2(X) =& \E_{\POmidX}\left[s(O,X,\psi)\{O-\mu(X)\}\mid X\right]\\
    &+ \E_{\QOXzero}\left[\dot{s}(O,X,\psi)\{O-\mu(X)\}\right]\cdot\pr(T=1)\cdot e_2(X),
\end{align*}
and
\begin{align*}
    \dfrac{\partial\eta_2(\xi_1,\xi_2)}{\partial\xi_2}
    =& \E_{\QX}\left[\E_{\POmidX}\{\dot{s}(O,X,\psi)\mid X\}\cdot\SC_{\xi_2}(X)\right] +
    \E_{\QOXzero}\left[\ddot{s}(O,X,\psi)\right]\dfrac{\partial\psi(\xi_1,\xi_2)}{\partial\xi_2}\\
    =& \E_{\QX}\left[\E_{\POmidX}\{\dot{s}(O,X,\psi)\mid X\}\cdot\SC_{\xi_2}(X)\right]\\
    & +   \E_{\QOXzero}\left[\ddot{s}(O,X,\psi)\right]\pr(T=1)\E_{\QX}\{e_2(X)\cdot\SC_{\xi_2}(X)\} \\
    =& \E_{\QX}\{b_2(X)\cdot\SC_{\xi_2}(X)\},\text{ where }\\
    b_2(X)=& \E_{\POmidX}\{\dot{s}(O,X,\psi)\mid X\} +
     \E_{\QOXzero}\left[\ddot{s}(O,X,\psi)\right]\pr(T=1)e_2(X).
\end{align*}
Therefore, the RHS of \eqref{supp.eq:pf.eif.slope.a2} can be written as
\begin{align*}
    \dfrac{\partial\SI(\xi_1,\xi_2)}{\partial\xi_2}
    &=   - \dfrac{
    \dfrac{\partial\eta_1(\xi_1,\xi_2)}{\partial\xi_2}\cdot\eta_2(\xi_1,\xi_2) -
    \dfrac{\partial\eta_2(\xi_1,\xi_2)}{\partial\xi_2}\cdot\eta_1(\xi_1,\xi_2)
    }
    {\left\{\eta_2(\xi_1,\xi_2)\right\}^2}\\
    &= \E_{\QX}\left\{
    \left[
    \dfrac{-b_1(X)-b_2(X)\cdot\SI}{\E_{\QOXzero}\{\dot{s}(O,X,\psi)\}}
    \right]\cdot\SC_{\xi_2}(X)\right\}.
\end{align*}
Equalizing both sides of \eqref{supp.eq:pf.eif.slope.a2}, we have
\begin{align*}
    0 =& \E_{\QX}\left[  \left(\pr(T=1)a_2(X) -  \left[
    \dfrac{-b_1(X)-b_2(X)\cdot\SI}{\E_{\QOXzero}\{\dot{s}(O,X,\psi)\}}
    \right]\right)\cdot\SC_{\xi_2}(X) \right].
\end{align*}
Since $\SC_{\xi_2}(X)\in\calT_{\QX}$ and hence $\E_{\QX}\{\SC_{\xi_2}(X)\}=0$, we have
\begin{align*}
    a_2(X) =& \dfrac{
     \left[
    \dfrac{-b_1(X)-b_2(X)\cdot\SI}{\E_{\QOXzero}\{\dot{s}(O,X,\psi)\}}
    \right] -
    \E_{\QX}\left\{ \left[
    \dfrac{-b_1(X)-b_2(X)\cdot\SI}{\E_{\QOXzero}\{\dot{s}(O,X,\psi)\}}
    \right]\right\}
    }{\pr(T=1)}\\
    =& \dfrac{-\SI\cdot\E_{\POmidX}\{\dot{s}(O,X,\psi)\mid X\} - \E_{\POmidX}\{O-\mu(X)\mid X\}}
    {\pr(T=1)\E_{\QOXzero}\{\dot{s}(O,X,\psi)\}}\\
    & - e_2(X) \cdot
    \dfrac{
    \E_{\QOXzero}\left[\dot{s}(O,X,\psi)\{O-\mu(X)\}\right] + \SI\cdot\E_{\QOXzero}\{\ddot{s}(O,X,\psi)\}
    }
    {\E_{\QOXzero}\{\dot{s}(O,X,\psi)\}}.
\end{align*}
Finally, by plugging in $a_1(O,X)$ and $a_2(X)$ into $\EIF(T,O,X,\SI) = (1-T)a_1(O,X)+Ta_2(X)$ and noticing $(1-T)e_1(O,X) + Te_2(X)=\EIF(T,O,X,\psi)$, we obtain the EIF stated in Proposition \ref{prop:eif.slope}.
$\square$

\subsubsection{Proof of Proposition \ref{prop:eif.psi}}
\pf{Proposition \ref{prop:eif.psi}}
Suppose the efficient influence function of the target functional is
\begin{align*}
    \EIF(T,O,X,\psi) = (1-T)e_1(O,X) + Te_2(X),
\end{align*}
where $(1-T)e_1(O,X)\in\calT_{\POmidX}$ and $Te_2(X)\in\calT_{\QX}$. Then \eqref{supp.eq:pf.eif.psi.xi1}-\eqref{supp.eq:pf.eif.psi.xi2} can be re-expressed as
\begin{align}
    \E\left\{(1-T)e_1(O,X)\cdot\SC_{\xi_1}(O,X)\right\} &= \dfrac{\partial\psi(\xi_1,\xi_2)}{\partial\xi_1},
    \label{supp.eq:pf.eif.psi.e1}\\
    \E\left\{Te_2(X)\cdot\SC_{\xi_2}(X)\right\} &=\dfrac{\partial\psi(\xi_1,\xi_2)}{\partial\xi_2}.
    \label{supp.eq:pf.eif.psi.e2}
\end{align}

First, consider \eqref{supp.eq:pf.eif.psi.e1}. The LHS is
\begin{align*}
    \E\left\{(1-T)e_1(O,X)\cdot\SC_{\xi_1}(O,X)\right\}
    &= \E_{\QOXzero}\left[\dfrac{\pr(T=0)}{\omega(X)}e_1(O,X)\right],
\end{align*}
and the RHS is
\begin{align*}
    \dfrac{\partial\psi(\xi_1,\xi_2)}{\partial\xi_1}
    &= \left[\E_{\QOXzero}\{\dot{s}(O,X,\psi)\}\right]\inv\E_{\QOXzero}\left\{s(O,X,\psi)\cdot\SC_{\xi_1}(O,X)\right\}.
\end{align*}
We equalize them and then have
\begin{align*}
    e_1(O,X) = -\left[\E_{\QOXzero}\{\dot{s}(O,X,\psi)\}\right]\inv\dfrac{\omega(X)}{\pr(T=0)}
    \left[s(O,X,\psi) - \E_{\POmidX}\{s(O,X,\psi)\mid X\}\right].
\end{align*}

Next, consider \eqref{supp.eq:pf.eif.psi.e2}. The LHS is
\begin{align*}
    \E\left\{Te_2(X)\cdot\SC_{\xi_2}(X)\right\} &= \pr(T=1)\E_{\QX}\left\{ e_2(X)\cdot\SC_{\xi_2}(X)\right\},
\end{align*}
and the RHS is
\begin{align*}
    \dfrac{\partial\psi(\xi_1,\xi_2)}{\partial\xi_2} &=
    \left[\E_{\QOXzero}\{\dot{s}(O,X,\psi)\}\right]\inv \E_{\QOXzero}\left\{s(O,X,\psi)\cdot\SC_{\xi_2}(X)\right\}.
\end{align*}
We equalize them and have
\begin{align*}
    e_2(X)= -\dfrac{
    \E_{\POmidX}\{s(O,X,\psi)\mid X\} - \E_{\QOXzero}\{s(O,X,\psi)\}
    }
    {\pr(T=1)\E_{\QOXzero}\{\dot{s}(O,X,\psi)\}}.
\end{align*}
Therefore, by plugging in $e_1(O,X)$ and $e_2(X)$ to the EIF of the target functional, we have
\begin{align*}
    \EIF(T,O,X,\psi) =& -\dfrac{(1-T)\omega(X)}{\pr(T=0)} \left[\E_{\QOXzero}\{\dot{s}(O,X,\psi)\}\right]\inv
    \left[s(O,X,\psi)-\E_{\POmidX}\{s(O,X,\psi)\mid X\}\right]\\
    &-\dfrac{T}{\pr(T=1)} \left[\E_{\QOXzero}\{\dot{s}(O,X,\psi)\}\right]\inv  \E_{\POmidX}\{s(O,X,\psi)\mid X\}.
\end{align*}
$\square$

\section{Extended Remarks and Results}\label{supp.sec:extended.remarks}
This section details some extended remarks deferred from the main text. Section \ref{supp.subsec:slope.vector} discusses vector valued SLOPE.
Section \ref{supp:subsec:benchmark} discusses a way to benchmark the value of SLOPE in practice.
Section \ref{supp.sec:extend.app} discusses additional application scenarios of SLOPE, including cases when multiple source populations are available and using SLOPE to compare the robustness between target populations given a source population.
Section \ref{supp.subsec:disc.slope.practice} includes some extended remarks from Section \ref{sec:discussion}.
Section \ref{supp.sec:slope.other.contexts} discusses defining SLOPE for other types of conditional exchangeability assumptions, including the no unmeasured confounding assumption and the missing at random assumption.
Section \ref{supp.subsec:MIE} states the mathematical connection between SLOPE and the marginal interventional effect \citep{zhou2022marginal} with the incremental propensity score intervention \citep{kennedy2019nonparametric}.

\subsection{SLOPE for a Vector Valued $\psi(\cdot)$}\label{supp.subsec:slope.vector}
When the functional $\psi(\cdot)$ is vector valued with dimension $p$, the corresponding SLOPE, as defined in Definition \ref{def:proposal}, is also a vector of $p$ elements. Each dimension of SLOPE represents the robustness of the corresponding element of the target estimand when conditional exchangeability is ``near'' violated. For example, when $X$ is $p$-dimensional and the target estimand is a $p$-dimensional vector of  ordinary least squares (OLS) coefficients of regressing $O$ on $X$, then each element of SLOPE describes the robustness of  the corresponding coefficient; see Section \ref{supp.sec:ols} for a formal presentation of the SLOPE and Section \ref{supp.subsec:slope.ols.est} for the estimation. More generally, for SLOPE as vectors,  the connection between SLOPE and IF still holds (i.e., Theorem \ref{thm:slope.if}). In addition, the two proposed estimators are still applicable, as mentioned in Remarks \ref{remark:weight.est.vector} and \ref{remark:regression.est.vector}.

The main difference between a vector of SLOPE and a scalar SLOPE is on the interpretation, and in particular, the informed guidance in designs. Specifically, as mentioned in Section \ref{subsec:slope.def}, when SLOPE becomes a vector and all elements are of scientific interest, then practitioners need to find an appropriate summary of the vector in order to compare SLOPEs across study designs. This could be a visual summary, such as plotting the SLOPE of each dimension, or a quantitative measure such as a norm of the SLOPE.

\begin{example}[SLOPE for OLS Coefficient]
    Suppose $X$ is a $p$-dimensional vector and $O$ is an outcome variable of interest. The target estimand is defined as the regression coefficients of regressing $O$ on $X$. In other words, consider $\psi^{\ols}$ such that
    \begin{align*}
        \E_{\ols}\left(XX\trans \psi^{\ols}-XO\right)=0,
    \end{align*}
    which we suppose the uniqueness of $\psi^{\ols}(\QOX)$. Following Section \ref{supp.sec:ols}, we know the SLOPE for $\psi^{\ols}$ is
    \begin{align*}
        \SI(\QOXzero,\psi^{\ols}) &= \left\{\E(XX\trans)\right\}\inv\E_{\QX}\left\{\sigma^2(X)X\right\},
    \end{align*}
    which is a $p$-dimensional vector. \\
    To compare SLOPE across designs, we define the magnitude of SLOPE in two scenarios. First, suppose $X$ has been standardized within the source and the target population. Then we can define the norm of SLOPE as the $L_2$ norm of $\SI(\QOXzero,\psi^{\ols})$. With this notion of the magnitude, interpretations of SLOPE based on its magnitude that were discussed in Section \ref{subsec:slope.if} are applicable. Second, suppose $X$ has not been standardized, then we consider the Mahalanobis distance with covariance matrix $\{\E_{\QX}(XX\trans)\}\inv$. More specifically, we define the magnitude of SLOPE as
    \begin{align*}
      & \left\{ \SI(\QOXzero,\psi^{\ols}) \right\}\trans \E_{\QX}(XX\trans)\left\{\SI(\QOXzero,\psi^{\ols})\right\}\\
       =&
       \left[  \E_{\QX}\left\{\sigma^2(X)X\right\}\right]\trans
       \left\{\E_{\QX}(XX\trans)\right\}\inv
       \E_{\QX}\left\{\sigma^2(X)X\right\}.
    \end{align*}
    In practice, the SLOPE needs to be estimated. Example estimators of the SLOPE for OLS coefficients are provided  Section \ref{supp.subsec:slope.ols.est}.
\end{example}

\subsection{Benchmarking SLOPE}\label{supp:subsec:benchmark}

We describe one way to benchmark SLOPE as follows.
Motivated by the first-order bias approximation \eqref{eq:first.order.approx}, i.e., $ \psi(\QOXgamma)- \psi(\QOXzero) \approx  \gamma\cdot\SI(\QOXzero, \psi)$, a reasonable approximation of the SLOPE is
\begin{align*}
    \SI\approx \{\psi(\QOXgamma)- \psi(\QOXzero)\}/\gamma.
\end{align*}
Hence, one can benchmark SLOPE by benchmarking $\psi(\QOXgamma)- \psi(\QOXzero)$ and $\gamma$ separately, and then taking their ratio. The first term, $\psi(\QOXgamma)- \psi(\QOXzero)$, is the conjectured generalization bias induced by the violation of conditional exchangeability. Domain knowledge can be used to benchmark it against the largest bias one is willing to tolerate. For example, if $O$ is a clinical measurement and $\psi$  is its mean or median, then a natural choice is the precision of the instrument producing the measurement, or a threshold below which a difference in $O$ carries no clinical meaning.
Alternatively, this term can also be set  to a multiple of the estimated standard error (SE) of the estimator of the target functional under conditional exchangeability, i.e., the estimated SE of $\wh\psi(\QOXgamma)$.
The second term, $\gamma$, is a sensitivity parameter and benchmarking it follows the common practice of benchmarking sensitivity parameters in a sensitivity analysis.

\subsection{Extended Remarks on the Applications of SLOPE}\label{supp.sec:extend.app}
We elaborate the discussion on widening the application of SLOPE.

First, for a fixed target population and a fixed functional $\psi$, one may be interested in finding multiple source populations that minimize SLOPE. One way to achieve this goal is through an optimization over all possible combinations of more than one source populations. Specifically, let $[K]=\{1,\ldots,K\}$ index the $K$ candidate source populations, and define
    \begin{align*}
       \mathcal{A}
        =\left\{A\subseteq[K]:2\leq |A|\leq K\right\}.
    \end{align*}
   Then $\mathcal{A}$ is the collection of all subsets of $[K]$ with at least two elements. For each $A\in\mathcal{A}$, let $P^A_{O\mid X}$ be the conditional outcome distribution of the source population obtained by combining the populations indexed by $A$, and let $Q^{0,A}_{O,X}=P^A_{O\mid X}\times Q_X$. For a fixed target population $Q_X$ and target functional $\psi$, the selected combination is
    \begin{align*}
        A^{*}\in\underset{A\in\mathcal{A}}{\arg\min}\,
        \left|\SI\left(Q^{0,A}_{O,X},\psi\right)\right|.
    \end{align*}
    The optimization %
    will be computationally expensive for large $K$, but may be feasible for small to moderate $K$.

Second, while Section \ref{subsec:data.choice.source} considered which source population is most robust for a fixed target population, Table~\ref{tab:data.appI} can also be used to assess the robustness of generalizing from a fixed source population to different target populations. This comparison is made among entries within the same row in Table~\ref{tab:data.appI}. When India is the source, Honduras has the lowest SLOPE, followed closely by Ghana, whereas Pakistan and Peru have higher SLOPEs. For the remaining source countries, SLOPE varies little across target countries. These patterns are similar between the mean and median: for each fixed source, the two estimands yield broadly consistent rankings of the target countries and therefore lead to the same conclusions about their  sensitivity for generalization.

\subsection{Extended Remarks on the Robustness of a Low Conditional Variance}\label{supp.subsec:disc.slope.practice}

For the mean, the dependency of its SLOPE on the conditional variance $\sigma^2(X)$ advocates source populations with a lower $\sigma^2(X)$; this principle holds  generally for location parameters in that SLOPE advocates a more homogeneous design.
Since $\sigma^2(X)$ is smaller when $X$ contains less information, SLOPE seems to advocate a homogeneous $X$ (e.g., $X=x$ almost surely and therefore $\sigma^2(X)=0$), which contradicts with the existing understanding that $X$ should be sufficiently rich for the overlap condition (Assumption \ref{assump:overlap}) to hold.
To understand why such contradiction appears, we elaborate on the comments noted in  Section \ref{sec:discussion}.

First, SLOPE is proposed and applied in cases when the overlap condition holds, meaning that covariates in the source population cannot contain less information than the covariates in the target population. This aligns with most works in sensitivity analysis for the conditional exchangeability assumption in different contexts. We refer readers to \citet{huang2025overlap} for a sensitivity analysis solely for the overlap condition. Considering violations to both assumptions is a valuable future direction (see \citet{bonvini2022sensitivity} and \citet{cui2025robust}) and is beyond the scope of this work.

Second, as mentioned in the main text, we echo \citet{tipton2018review} and \citet{degtiar2023review} in the importance of a careful data collection for the conditional exchangeability to hold, or, for the violation to be as small as possible.
In this sense, it's more preferable to collect more common characteristics in source and target populations, namely a richer $X$.
Secondary to that, SLOPE is a useful tool to assess sensitivity/robustness when it is unrealistic to meet the conditional exchangeability with the observed set of $X$. This happens, for example, when (i) it is infeasible to randomize units into the source or target population, (ii) not all covariates can be measured in both populations, which is common if the investigator opts in after data collection, and/or (iii) there exist unobservable differences between populations (e.g., sites) even under a careful design \citep{allcott2015site,jin2024beyond}.

\subsection{SLOPE in Other Contexts of Conditional Exchangeability}\label{supp.sec:slope.other.contexts}

While this paper focuses on the conditional exchangeability assumption in transportability/generalizability, SLOPE can be used to study the sensitivity of other types of conditional exchangeability typed assumptions.
With different meanings of $P$ and $Q$, SLOPE can  measure the sensitivity/robustness of other types of conditional exchangeability assumptions, including the no unmeasured confounding assumption in causal inference and the missing at random assumption in missing data. We describe these settings in the following subsections.

\subsubsection{SLOPE for Unmeasured Confounding}
Let $A\in\{0,1\}$ be the treatment and $Y(A)$ be the potential outcome under treatment $A$, where $A=1$ means intervention and $A=0$ means control. We suppose SUTVA (Assumption \ref{assump:sutva}) holds; i.e., when $A=a$, the observed outcome $Y$ is $Y(a)$, for $a=0,1$. To fix ideas, suppose we are interested in the average treatment effect on the treated, a popular estimand in causal inference,
\begin{align}
    \psi^{\att} &= \E\{Y(1)-Y(0)\mid A=1\}\nonumber\\
    &= \E\{Y(1)\mid A=1\} - \E\{Y(0)\mid A=1\}\nonumber\\
     &= \E\{Y\mid A=1\} - \E\{Y(0)\mid A=1\}.\label{eq:att}
\end{align}
where the third line follows from SUTVA. From \eqref{eq:att}, the key challenge in identifying $\psi^{\att}$ lies in the challenge of identifying the second term, $\E\{Y(0)\mid A=1\}$, since it involves the potential outcome $Y(0)$. One common strategy for identifying $\psi^{\att}$ is through the conditional exchangeability assumption (Assumption \ref{assump:exchange}) and the overlap condition (Assumption \ref{assump:overlap}). To demonstrate that, we define $P$ and $Q$ as follows.

Suppose $P$ represents the population of units with $A=0$  and $Q$ represents the population of units with $A=1$. Therefore, $\E_{P}(\cdot)=\E(\cdot\mid A=0)$ and $\E_{Q}(\cdot)=\E(\cdot\mid A=1)$. Suppose $X$ contains pre-treatment covariates (i.e., measured confounders) and $O=Y(0)$ is the potential outcome under control. Then the conditional exchangeability assumption (Assumption \ref{assump:exchange}), i.e.,  $\QOmidX(\cdot\mid x) = \POmidX(\cdot\mid x)$ almost everywhere $\QX$, implies that
\begin{align*}
    \left(\E_{\QOmidX}[O\mid X] := \right)\quad  \E\{Y(0)\mid X,A=1\}  = \E\{Y(0)\mid X,A=0\}\quad \left(:=\E_{\POmidX}[O\mid X]\right).
\end{align*}
Then $\psi^{\att}$ in \eqref{eq:att} can be identified, since
\begin{align*}
   \E\{Y(0)\mid A=1\} &= \E\left[\E\{Y(0)\mid X,A=1\}\mid A=1\right]&\\
   &=  \E\left[\E\{Y(0)\mid X,A=0\}\mid A=0\right] &\text{(by conditional exchangeability)}\\
   &= \E\left[\E\{Y\mid X,A=0\}\mid A=0\right]&\text{(by SUTVA)}
\end{align*}
no longer involves potential outcomes.

In this context,  the conditional exchangeability assumption is also referred to as no unmeasured confounding. In observational studies where unmeasured confounding is highly likely, there is an extensive literature that develops sensitivity analysis methods to study the consequence of unmeasured confounding. With the above defined $P$ and $Q$, SLOPE can be naturally applied to study the sensitivity of violation to no unmeasured confounding. For example, the SLOPE for $\psi^{\att}$ as defined above is
\begin{align*}
    &\SI(\QOXzero, \psi^{\att}) \\
    =& \E_{\QX} \var_{\POmidX}(O\mid X) & \text{(by Theorem \ref{thm:slope.mean})} \\
    =& \E_{X\mid A=1}\left[\var_{Y(0)\mid X,A=0}(Y(0)\mid X,A=0) \mid A=1\right] & \text{(by definitions of }P\text{ and }Q\text{)}.
\end{align*}

\begin{remark}[SLOPE for Other Causal Estimands]
We note that $P$ and $Q$ may need to be defined differently for other causal estimands. For example, when the target estimand becomes the average treatment effect on the control, then the roles of $P$ and $Q$ need to be swapped. Nevertheless, at a high level, the definition of SLOPE remains consistent, i.e., the derivative of the target causal estimand with respect to $\gamma$ at $\gamma=0$, where $\gamma$ is such that
\begin{align}\label{eq:sensitivity.y(a)}
    \dfrac{[(Y(a)\mid X,A=1-a]}{[Y(a)\mid X,A=a]}\propto \exp\{\gamma\cdot Y(a)\}.
\end{align}
where we use $[\cdot\mid X,A]$ to represent conditional densities (provided exist with respect to some measure) on the underlying population.
For more explanations on the sensitivity model \eqref{eq:sensitivity.y(a)}, see \citet{scharfstein1999adjusting} and \citet{franks2019flexible}.
\end{remark}

Finally, we make a clarification on this setting.
The current section should not be confused with Section \ref{supp.sec:causal.functional}. In Section \ref{supp.sec:causal.functional}, we transport treatment effect from a source population to a target population where we have assumed that the treatment effect can be identified within the source; in other words, the ``causal'' assumptions already hold in the source population. However, in this section, we no longer consider any transfer learning setting but instead focus on a causal setting with potential unmeasured confounding. Here, one may still think of $P$ (i.e., population of control units) as the ``source'' population and $Q$ (i.e., population of treated units) as the ``target'' population, while keeping in mind the difference in the meaning of conditional exchangeability assumption as well as the different interpretations of the corresponding SLOPEs.

\subsubsection{SLOPE for Non-Ignorable Missingness}
Consider a missing data problem where we have observed covariates $X$ on all units, but only observe the outcome variable $\O$ on a subset of the units. Let $P$ be the population of units with complete data and $Q$ be the population of units with a missing $\O$. Then the conditional exchangeability assumption (Assumption \ref{assump:exchange}) is the missing at random assumption which is commonly adopted in the literature of missing data. In this context, the SLOPE can be defined similarly as the main text and it can be interpreted as the sensitivity/robustness of the target estimand with respect to non-ignorable missingness.

\subsection{Connections to Marginal Interventional Effect}\label{supp.subsec:MIE}
Broadly speaking, the concept and the mathematical format of the SLOPE are connected to the marginal interventional treatment effect \citep{zhou2022marginal} with the incremental propensity score intervention \citep{kennedy2019nonparametric}. In this section, we elaborate on this connection.

We begin with a brief introduction to the marginal interventional treatment effect; see \citet{zhou2022marginal} for details.
In causal inference (see notation in Section \ref{subsec:causal.notation}), an interventional effect  quantifies the change in outcome when the propensity of a proportion of units changes. This is in contrast to conventional estimands, like the average treatment effect and the quantile treatment effect, which compare the outcome when all units receive treatment with the outcome  when all receive placebo. To define an interventional effect, suppose the propensity score is $\pi(Z)=P(A=1\mid Z)$ and consider changing the propensity in a way of $\pi^{\gamma}(Z)$ as a function of $\pi(Z)$ with $\pi^0(X)=\pi(Z)$. Then under assumptions in Section 3 of \citet{zhou2022marginal}, the interventional effect (IE) is
\begin{align*}
    \text{IE}^{\gamma} &=
    \E\left[
    \dfrac{\pi^{\gamma}(Z)-\pi^0(Z)}{\E\left[\pi^{\gamma}(X)-\pi^0(Z)\right]\tau(X)}
    \right],
\end{align*}
where $\tau(Z)=\E\{Y(1)-Y(0)\mid X\}$ is the conditional average treatment effect (CATE). The marginal interventional effect (MIE), as proposed by \citet{zhou2022marginal}, describes the marginal change in the interventional effect by taking the limit of $\gamma$ of IE going to zero:
\begin{align*}
    \text{MIE} &= \E\left[\dfrac{\dot\pi^0(Z)}{\E\left\{\dot\pi^0(Z)\right\}}\tau(Z)\right],
\end{align*}
where $\dot\pi^0(X)$ is the derivative of $\pi^{\gamma}(Z)$ with respect to $\gamma$ at $\gamma=0$.
When the propensity score shift $\pi^{\gamma}(Z)$ is determined by the incremental propensity score interventions  (IPSI) proposed by \citet{kennedy2019nonparametric}, i.e.,
\begin{align}\label{eq:ipsi}
    \dfrac{\pi^{\gamma}(Z)/\{1-\pi^{\gamma}(Z)\}}{\pi^0(Z)/\{1-\pi^0(Z)\}} = \exp(\gamma),
\end{align}
the MIE is
\begin{align}\label{eq:mie.ipsi}
      \text{MIE}^{\text{IPSI}} &= \dfrac{\E\left[\pi^0(Z)\{1-\pi^0(Z)\}\tau(Z)\right]}
      {\E\left[\pi^0(Z)\{1-\pi^0(Z)\}\right]}.
\end{align}

The MIE with IPSI is connected to SLOPE for two reasons. First, the IPSI shifts the intervention $\pi^{\gamma}(Z)$ in the same exponential tilting model as in the sensitivity model \eqref{eq:sensitivity}; see below for details. Second, the concept of MIE, which defines a treatment effect in terms of the change in the outcome with an infinitesimal change in propensity, is broadly connected to the concept of SLOPE. To see the connections more explicitly, we next re-express the SLOPE under the regime of incremental propensity score intervention. We note that although the resulting quantity no longer represents the ``sensitivity to local perturbations from exchangeability'', we still refer to it as SLOPE for ease of communication, keeping in mind that we temporarily treat SLOPE as a mathematical quantity instead of a sensitivity/robustness measure.

Let $O=A$ be the binary treatment.
Suppose $P$ is the population of interest and $P_{Z,A,Y}$ is the joint distribution of pre-treatment covariate $Z$, binary treatment $A$ and outcome $Y$. We adopt the potential outcome framework (Section \ref{subsec:causal.notation}) and suppose  SUTVA (Assumption \ref{assump:sutva}) and strong ignorability (Assumption \ref{assump:strong.ign}) hold.

Let  $Q$ be a hypothetical population with the same distribution as $P$ except that the intervention mechanism has been changed by $\gamma$. In specific, let $O=A$, then the sensitivity model \eqref{eq:sensitivity} implies that the odds of receiving treatment in $Q$ is $\exp(\gamma)$ times the odds of receiving treatment in $P$:
\begin{align*}
 \exp(\gamma)= \dfrac{Q^{\gamma}(A=1\mid Z)/\{1-Q^{\gamma}(A=1\mid Z\}}
  {P(A=1\mid Z)/\{1-P(A=1\mid Z)\}}.
\end{align*}
Note that in this case, $X=(A,Z)$ includes the intervention and the pre-treatment covariates.
We let $\pi^{\gamma}(Z) =  Q^{\gamma}(A=1\mid Z)$ and therefore $\pi^0(Z)=Q^{0}(A=1\mid Z)=P(A=1\mid Z)$ and $\pi^{\gamma}(Z) = \dfrac{\exp(\gamma)\pi^0(X)}{1-\pi^0(Z) + \exp(\gamma)\pi^0(Z)}$.

To define SLOPE, let $Q^{\gamma}_{Z,A,Y}=P_{Z}\times Q^{\gamma}_{A\mid Z}\times P_{Y\mid Z,A}$ and let the target functional be the mean outcome, i.e.,
\begin{align*}
    \psi(\QOXgamma) & = \E_{Q_{Y}^{\gamma}}(Y)\\
    & =\E_{\QX}\left[\pi^{\gamma}(Z) \E_{P_{Y\mid Z,A=1}}\{Y\mid Z,A=1\} +
        \{1-\pi^{\gamma}(Z)\} \E_{P_{Y\mid Z,A=0}}\{Y\mid Z,A=0\}\right]\\
      &= \E_{\QX} \left[\pi^{\gamma}(Z)\tau(Z)  \right] + \E_{\QX}\left[\left\{\E_{P_{Y\mid Z,A=0}}(Y(0)\mid Z,A=0\right\}\right],
\end{align*}
where again $\tau(X)$ is the CATE.

Noticing $\partial\pi_{\gamma}(X)/\partial\gamma\vert_{\gamma=0}=\pi^0(Z)\{1-\pi^0(Z)\}$, we have the SLOPE as
\begin{align}\label{eq:slope.ie}
    \E_{\QX} \left[\dfrac{\partial\pi^{\gamma}(Z)}{\partial\gamma}\Bigg\vert_{\gamma=0}\tau(Z)  \right]
    &= \E_{\QX} \left[
    \pi^0(Z)\{1-\pi^0(Z)\}\tau(X)
    \right].
\end{align}
We make two remarks on \eqref{eq:slope.ie}. First, it resembles the general form of the SLOPE of an outcome mean, since
\begin{align*}
    \E_{\QX} \left[
    \pi^0(X)\{1-\pi^0(X)\}\tau(Z)
    \right] =
    \E_{\QX} \left\{
    \cov[A,Y(1)-Y(0)\mid Z]
    \right\},
\end{align*}
where the $\cov[\cdot\mid X]$ represents conditional covariance under $P$. Note that here SLOPE is the expectation of a conditional \emph{covariance} instead of a conditional \emph{variance} because the ``shift'' is on the treatment (i.e., $O=A$) rather than the outcome.
Second, \eqref{eq:slope.ie} can be viewed as the non-standardized version of the MIE in \eqref{eq:mie.ipsi}. It is also a non-standardized version of the average treatment effect for the overlap population (ATO) \citep{li2018balancing} since the two quantities (ATO and MIE) are identical  in this specific regime \citep{zhou2022marginal}.

\section{Supplementary Materials for SLOPE Based on Set-Identified Sensitivity Models}\label{supp.sec:slope.bound}
This section provides supplementary materials for Section \ref{subsubsec:slope.bound}, where SLOPE was defined under set-identified sensitivity models. Section \ref{supp.subsec:slope.msm.condition} provides deferred conditions and statements for SLOPE under the  marginal sensitivity model (Example \ref{ex:msm}).
Section \ref{supp.subsec:slope.msm.proof} contains proofs.

\subsection{Conditions and Restatement of Conclusions}\label{supp.subsec:slope.msm.condition}
We first restate the sensitivity model and the conditions needed for the
proof. Fix $\gamma\geq 0$ and recall that
$\QOXzero=\POmidX\times\QX$. The marginal sensitivity model is
\begin{align*}
    \calQ^{\text{MSM}}(\gamma)
    =
    \left\{
    \QOmidX\times\QX:
    \exp(-\gamma)
    \leq
    \dfrac{f_{\QOmidX}(\O,X)}
          {f_{\POmidX}(\O,X)}
    \leq
    \exp(\gamma)
    \text{ almost surely in }\QOXzero
    \right\}.
\end{align*}
In particular, every $\QOmidX\times\QX\in
\calQ^{\text{MSM}}(\gamma)$ has the same marginal distribution
$\QX$, and $\QOmidX$ is absolutely continuous with respect to
$\POmidX$ almost surely in $\QX$.

\begin{condition}[Conditions for the Sharp Bounds]
\label{condition:regular.msm}
Suppose the following conditions hold: (i) $\E_{\QOXzero}(|\O|)<\infty$. (ii) For $\QX$-almost every $x$, the conditional c.d.f.
$F_{\POmidX}(\cdot\mid x)$ is continuous. (iii) The conditional quantiles
\begin{align*}
F_{\POmidX}\inv\left(
\dfrac{\exp\{\gamma\}}{1+\exp\{\gamma\}}
\,\middle|\,X
\right)
\quad\text{and}\quad
F_{\POmidX}\inv\left(
\dfrac{1}{1+\exp\{\gamma\}}
\,\middle|\,X
\right)
\end{align*}
are measurable as functions of $X$.
\end{condition}

\begin{proposition}[Sharp Bounds Under the Marginal Sensitivity Model]
\label{prop:msm}
Suppose Condition \ref{condition:regular.msm} holds. Then the
sharp upper and lower bounds of the mean functional over
$\calQ^{\text{MSM}}(\gamma)$ are as stated in Example \ref{ex:msm}, i.e.,
    \begin{align}
        \psi_+^{\mean}(\calQ(\gamma)) =& \E_{\QOXzero}
	        \left[O\cdot \exp\left(\gamma\cdot \sign\left\{O-F_{\POmidX}\inv\left(\dfrac{\exp\{\gamma\}}{1+\exp\{\gamma\}}\,\mid X\right)\right\}\right)\right],
            \label{eq:msm.upper}\\
	       \psi_-^{\mean}(\calQ(\gamma)) =& \E_{\QOXzero}
	        \left[O\cdot \exp\left(\gamma\cdot \sign\left\{F_{\POmidX}\inv\left(\dfrac{1}{1+\exp\{\gamma\}}\,\middle|\,X\right)-O\right\}\right)\right].
            \label{eq:msm.lower}
    \end{align}
\end{proposition}

\begin{condition}[Conditions for the Upper and Lower SLOPEs]
\label{condition:regular.msm.slope}
Suppose Condition
\ref{condition:regular.msm}(i)--(ii) holds.
Suppose there exists $\varepsilon>0$ such that, for every
$\gamma\in[0,\varepsilon]$, the two conditional quantiles in
Condition \ref{condition:regular.msm}(iii) are measurable functions
of $X$. In addition, suppose that for
$\QX$-almost every $x$, the conditional quantile function $ p\mapsto F_{\POmidX}\inv(p\mid x)$
is continuous at $p=1/2$.
\end{condition}

\begin{proposition}[Upper and Lower SLOPEs Under the Marginal Sensitivity Model]
\label{prop:msm.upper.slope}
Suppose Condition \ref{condition:regular.msm.slope} holds.
The upper and lower SLOPEs of the mean functional are as stated in Example \ref{ex:msm} of the main text.
\end{proposition}

\subsection{Proofs}\label{supp.subsec:slope.msm.proof}
\subsubsection{Proof of Proposition \ref{prop:msm}}
\begin{proof}[Proof of Proposition \ref{prop:msm}]
   We prove the upper bound using Condition \ref{condition:regular.msm}(i)--(iii). The proof for the lower bound follows analogously by reversing which tail receives the largest allowable density ratio.

   The proof proceeds in four steps. First, we rewrite the optimization
over $\calQ^{\text{MSM}}(\gamma)$ as a conditional optimization over
density ratios. Second, we construct a feasible density ratio by
placing the largest allowable weight above a particular conditional
quantile and the smallest allowable weight below it. Third, we show
that the constructed density ratio maximizes the conditional mean and
therefore provides an upper bound after averaging over $\QX$. Finally,
we construct a conditional distribution that attains this upper bound
and express the resulting density ratio in the form stated in
\eqref{eq:msm.upper}.

\noindent\ul{Step 1: Reformulate the problem as a conditional
optimization over density ratios.}
For any $\QOmidX\times\QX\in
\calQ^{\text{MSM}}(\gamma)$, write its conditional density ratio
with respect to $\POmidX$ as
\begin{align*}
    \lambda(\O,X)
    =
    \dfrac{f_{\QOmidX}(\O,X)}
          {f_{\POmidX}(\O,X)}.
\end{align*}
Because $\QOmidX$ is a conditional probability distribution,
\begin{align}
    \E_{\POmidX}\{\lambda(\O,X)\mid X\}=1
    \quad\text{almost surely in }\QX.
    \label{eq:msm.normalization}
\end{align}
Also, the marginal sensitivity model implies
\begin{align}
    \exp(-\gamma)
    \leq \lambda(\O,X)
    \leq \exp(\gamma)
    \quad\text{almost surely in }\QOXzero.
    \label{eq:msm.ratio.bound}
\end{align}
The mean under $\QOmidX\times\QX$ can therefore be written as
\begin{align}
    \psi^{\mean}(\QOmidX\times\QX)
    =
    \E_{\QX}
    \left[
    \E_{\POmidX}
    \{\O\lambda(\O,X)\mid X\}
    \right].
    \label{eq:msm.mean.ratio}
\end{align}

\noindent\ul{Step 2: Construct a feasible density ratio using a
conditional quantile.}
In this step, we construct a candidate density ratio that assigns the largest allowable weight to large values of $\O$.
Define
\begin{align*}
    q_\gamma(X)
    =
    F_{\POmidX}\inv\left(
    \dfrac{\exp\{\gamma\}}{1+\exp\{\gamma\}}
    \,\middle|\,X
    \right).
\end{align*}
Motivated by the quantile-balancing argument of
\citet{dorn2023sharp}, consider the conditional density ratio
\begin{align}
    \lambda_+(\O,X)
    =&\
    \exp(-\gamma)\ind\{\O\leq q_\gamma(X)\}
    +
    \exp(\gamma)\ind\{\O>q_\gamma(X)\}.
    \label{eq:msm.lambda.upper}
\end{align}
By Condition \ref{condition:regular.msm}, for almost every $X$
in $\QX$,
\begin{align*}
    F_{\POmidX}\{q_\gamma(X)\mid X\}
    =\
    \dfrac{\exp\{\gamma\}}{1+\exp\{\gamma\}},
    \qquad
    1-F_{\POmidX}\{q_\gamma(X)\mid X\}
    =\
    \dfrac{1}{1+\exp\{\gamma\}}.
\end{align*}
It follows that
\begin{align*}
    \E_{\POmidX}\{\lambda_+(\O,X)\mid X\}
    =
    \exp(-\gamma)
    \dfrac{\exp\{\gamma\}}{1+\exp\{\gamma\}}
    +
    \exp(\gamma)
    \dfrac{1}{1+\exp\{\gamma\}}
    = 1.
\end{align*}
Thus, $\lambda_+$ satisfies both the normalization condition
\eqref{eq:msm.normalization} and the density-ratio restriction
\eqref{eq:msm.ratio.bound}.

\noindent\ul{Step 3: The constructed density ratio is
optimal.}
In this step, we compare the candidate density ratio from Step 2 with
an arbitrary feasible density ratio. The precise claim is that every
conditional density ratio $\lambda$ satisfying
\eqref{eq:msm.normalization} and \eqref{eq:msm.ratio.bound} obeys
\begin{align}
    \E_{\POmidX}
    \{\O\lambda(\O,X)\mid X\}
    \leq
    \E_{\POmidX}
    \{\O\lambda_+(\O,X)\mid X\}
    \quad \QX\text{-almost surely}.
    \label{eq:msm.conditional.upper}
\end{align}
Thus, for $\QX$-almost every $x$, $\lambda_+(\cdot,x)$ maximizes the
conditional mean over all conditional density ratios whose
expectation under $\POmidX(\cdot\mid x)$ is one and whose values lie in
$[\exp(-\gamma),\exp(\gamma)]$.

To prove the claim, fix any feasible $\lambda$. The key observation is
that $\lambda_+$ places the largest allowable weight above
$q_\gamma(X)$ and the smallest allowable weight below
$q_\gamma(X)$. Consequently, for $\QOXzero$-almost every $(\O,X)$,
\begin{align}
    \{\O-q_\gamma(X)\}
    \{\lambda(\O,X)-\lambda_+(\O,X)\}
    \leq 0.
    \label{eq:msm.pointwise}
\end{align}
To verify \eqref{eq:msm.pointwise}, consider three cases.
On $\{\O>q_\gamma(X)\}$, the first factor is positive and
$\lambda-\lambda_+\leq0$ because $\lambda_+=\exp(\gamma)$. On
$\{\O<q_\gamma(X)\}$, the first factor is negative and
$\lambda-\lambda_+\geq0$ because $\lambda_+=\exp(-\gamma)$. On
$\{\O=q_\gamma(X)\}$, the first factor is zero. Thus,
\eqref{eq:msm.pointwise} holds without needing to exclude the equality
event.

The quantile $q_\gamma(X)$ is finite $\QX$-almost surely because its
probability index lies strictly between zero and one. The boundedness
of $\lambda$ and $\lambda_+$, together with
$\E_{\QOXzero}(|\O|)<\infty$, therefore ensures that the conditional
expectations below are finite for $\QX$-almost every $x$. Moreover,
both density ratios satisfy the normalization condition. Hence, the
conditional difference between their induced means is
\begin{align*}
    &\E_{\POmidX}
    \left[
    \O\{\lambda(\O,X)-\lambda_+(\O,X)\}
    \mid X
    \right]\\
    =&\
    \E_{\POmidX}
    \left[
    \{\O-q_\gamma(X)\}
    \{\lambda(\O,X)-\lambda_+(\O,X)\}
    \mid X
    \right]\\
    &+
    q_\gamma(X)
    \E_{\POmidX}
    \{\lambda(\O,X)-\lambda_+(\O,X)\mid X\}\\
    =&\
    \E_{\POmidX}
    \left[
    \{\O-q_\gamma(X)\}
    \{\lambda(\O,X)-\lambda_+(\O,X)\}
    \mid X
    \right]\\
    \leq&\ 0,
\end{align*}
in which the final inequality follows from \eqref{eq:msm.pointwise}. This
proves \eqref{eq:msm.conditional.upper}.

Finally, integrate \eqref{eq:msm.conditional.upper} with respect to
$\QX$. For every
$\QOmidX\times\QX\in\calQ^{\text{MSM}}(\gamma)$, we obtain
\begin{align}
    \psi^{\mean}(\QOmidX\times\QX)
    =&\
    \E_{\QX}
    \left[
    \E_{\POmidX}
    \{\O\lambda(\O,X)\mid X\}
    \right]\notag\\
    \leq&\
    \E_{\QX}
    \left[
    \E_{\POmidX}
    \{\O\lambda_+(\O,X)\mid X\}
    \right].
    \label{eq:msm.global.upper}
\end{align}
This establishes the desired upper bound. Step 4 shows that the bound
is attained.

\noindent\ul{Step 4: Establish sharpness.}
In this step, we use the candidate density ratio to construct a conditional distribution in the sensitivity model. This construction
shows that the upper bound is attained, and conditional continuity
then gives the sign-function representation in
\eqref{eq:msm.upper}.
It remains to show sharpness. Define a conditional distribution
$Q_{O\mid X}^+$ by
\begin{align*}
    dQ_{O\mid X}^+(\O\mid X)
    =
    \lambda_+(\O,X)d\POmidX(\O\mid X).
\end{align*}
The normalization of $\lambda_+$ shows that $Q_{O\mid X}^+$ is a
conditional probability distribution, and
\eqref{eq:msm.ratio.bound} shows that
$Q_{O\mid X}^+\times\QX\in
\calQ^{\text{MSM}}(\gamma)$. Hence, the upper bound is
attained.

Finally, conditional continuity implies $\lambda_+(\O,X)
    =
    \exp\left(
    \gamma\cdot
    \sign\{\O-q_\gamma(X)\}
    \right)$
almost surely in $\QOXzero$. Consequently,
\begin{align*}
    \psi_+^{\mean}
    =&\
    \E_{\QX}
    \left[
    \E_{\POmidX}
    \{\O\lambda_+(\O,X)\mid X\}
    \right]\\
    =&\
    \E_{\QOXzero}
    \left[
    \O\cdot
    \exp\left(
    \gamma\cdot
    \sign\left\{
    \O-
    F_{\POmidX}\inv\left(
    \dfrac{\exp\{\gamma\}}{1+\exp\{\gamma\}}
    \,\middle|\,X
    \right)
    \right\}
    \right)
    \right].
\end{align*}
This proves \eqref{eq:msm.upper}.
\end{proof}

\subsubsection{Proof of Proposition \ref{prop:msm.upper.slope}}
\begin{proof}[Proof of Proposition \ref{prop:msm.upper.slope}]
We prove the formula for the upper SLOPE. The formula for the lower
SLOPE follows analogously by using the lower sharp bound in
\eqref{eq:msm.lower}.

The proof proceeds in three steps. First, we express the difference
quotient defining the upper SLOPE using the sharp upper bound from
Proposition \ref{prop:msm}. Second, we establish the almost-sure limit
of its integrand as $\gamma\downarrow0$. Finally, we apply the
dominated convergence theorem to exchange the limit and expectation.

\noindent\ul{Step 1: Rewrite the difference quotient using the sharp
upper bound.}
 For $\gamma\in[0,\varepsilon]$, define
\begin{align*}
    q_\gamma(X)
    =&
    F_{\POmidX}\inv\left(
    \dfrac{\exp\{\gamma\}}{1+\exp\{\gamma\}}
    \,\middle|\,X
    \right), \text{ and}\\
    s_\gamma(\O,X)
    =&
    \ind\{\O>q_\gamma(X)\}
    -
    \ind\{\O\leq q_\gamma(X)\}.
\end{align*}
By Proposition \ref{prop:msm},
\begin{align*}
    \psi_+^{\mean}(\gamma)
    =
    \E_{\QOXzero}
    \left[
    \O\exp\{\gamma s_\gamma(\O,X)\}
    \right].
\end{align*}
At $\gamma=0$,  $\psi_+^{\mean}(0)
    =
    \psi^{\mean}(\QOXzero)
    =
    \E_{\QOXzero}(\O)$.
Consequently, for $\gamma>0$,
\begin{align*}
    \dfrac{
    \psi_+^{\mean}(\gamma)-\psi^{\mean}(\QOXzero)
    }{\gamma}
    =
    \E_{\QOXzero}
    \left[
    \O
    \dfrac{
    \exp\{\gamma s_\gamma(\O,X)\}-1
    }{\gamma}
    \right].
\end{align*}

\noindent\ul{Step 2: Establish the almost-sure limit of the
integrand.}
Since $ \exp\{\gamma\}/[1+\exp\{\gamma\}]
    \longrightarrow 1/2$
as $\gamma\downarrow0$, the quantile-continuity requirement in
Condition \ref{condition:regular.msm.slope} implies that $ q_\gamma(X)
    \longrightarrow
    F_{\POmidX}\inv(1/2\mid X)$
$\QX$-almost surely. Moreover, Condition \ref{condition:regular.msm}(ii) implies
\begin{align*}
    \E_{\POmidX}
    \left[
    \ind\left\{
    \O=F_{\POmidX}\inv(1/2\mid X)
    \right\}
    \,\middle|\,X
    \right]
    =0
\end{align*}
$\QX$-almost surely. It follows that, for
$\QOXzero$-almost every $(\O,X)$,
\begin{align*}
    s_\gamma(\O,X)
    \longrightarrow&
    \ \ind\left\{
    \O>F_{\POmidX}\inv(1/2\mid X)
    \right\}
    -
    \ind\left\{
    \O\leq F_{\POmidX}\inv(1/2\mid X)
    \right\}.
\end{align*}
Both $s_\gamma(\O,X)$ and its limit lie in $\{-1,1\}$. Thus, at every
point where the preceding convergence holds, $s_\gamma(\O,X)$
eventually equals its limit. Combining this fact with
$\{\exp(\gamma s)-1\}/\gamma\to s$ for each $s\in\{-1,1\}$ yields
\begin{align*}
    \dfrac{
    \exp\{\gamma s_\gamma(\O,X)\}-1
    }{\gamma}
    \longrightarrow&
    \ \ind\left\{
    \O>F_{\POmidX}\inv(1/2\mid X)
    \right\}
    -
    \ind\left\{
    \O\leq F_{\POmidX}\inv(1/2\mid X)
    \right\}
\end{align*}
$\QOXzero$-almost surely.

\noindent\ul{Step 3: Exchange the limit and expectation.}
In this step, we verify an integrable dominating function and apply
the dominated convergence theorem. For
$\gamma\in(0,\varepsilon]$,
\begin{align*}
    \left|
    \dfrac{
    \exp\{\gamma s_\gamma(\O,X)\}-1
    }{\gamma}
    \right|
    \leq \exp(\varepsilon).
\end{align*}
Thus, the absolute value of the integrand in the difference quotient
from Step 1 is bounded by $\exp(\varepsilon)|\O|$, which is integrable
under $\QOXzero$ by Condition \ref{condition:regular.msm}(i).
Applying the dominated convergence theorem gives
\begin{align*}
    \SI_+^{\msm}(\QOXzero,\psi^{\mean})
    =&\
    \lim_{\gamma\downarrow0}
    \dfrac{
    \psi_+^{\mean}(\gamma)-\psi^{\mean}(\QOXzero)
    }{\gamma}\\
    =&\
    \E_{\POmidX\times\QX}
    \left(
    \left[
    \begin{aligned}
    &\ind\left\{
    \O>F_{\POmidX}\inv(1/2\mid X)
    \right\}\\
    &\quad-
    \ind\left\{
    \O\leq F_{\POmidX}\inv(1/2\mid X)
    \right\}
    \end{aligned}
    \right]\O
    \right),
\end{align*}
which is the upper-SLOPE formula in Example \ref{ex:msm}.

For the lower SLOPE, one instead uses the cutoff
$F_{\POmidX}\inv\{1/[1+\exp\{\gamma\}]\mid X\}$ from
\eqref{eq:msm.lower}. This cutoff also converges to
$F_{\POmidX}\inv(1/2\mid X)$, while the limiting sign is the negative
of the one above. Repeating the three steps therefore gives
$\SI_-^{\msm}(\QOXzero,\psi^{\mean})
=-\SI_+^{\msm}(\QOXzero,\psi^{\mean})$, as claimed.
\end{proof}

\end{document}

%% file: preamble.tex
\usepackage{amssymb,amsmath,amsfonts,amsthm}
\usepackage{authblk}% author affiliations

\usepackage{algorithm}
\usepackage{algpseudocode}

\usepackage{nicematrix} % table cells to block some cells a
\usepackage{comment}
\usepackage{url}
\usepackage[dvipsnames]{xcolor}
\usepackage[colorlinks=true]{hyperref}
\usepackage{dsfont}% indicator function using \mathds{1}
\hypersetup{
  colorlinks=true,
  citecolor=BrickRed,
  linkcolor=BrickRed,
  urlcolor=BrickRed}
\usepackage{physics}
\usepackage{soul}

\usepackage{caption}
\usepackage{subcaption}
\usepackage{multirow}

\def\B{\mathcal{B}}
\def\calA{\mathcal{A}}
\def\calH{\mathcal{H}}

\def\calT{\mathcal{T}}

\def\cor{\textnormal{Corr}} % correlation
\def\cov{\textnormal{Cov}} % covariance

\def\E{\textnormal{E}}
\def\EIF{\textnormal{EIF}}

\def\inv{^{-1}}
\def\ind{\mathds{1}}% indicator function
\def\IF{\textnormal{IF}} % influence function

\def\P{\mathbb{P}}
\def\pr{\textnormal{pr}}

\def\R{\mathbb{R}}
\def\sign{\textnormal{sign}}
\def\S{\mathcal{S}}
\def\SC{\textnormal{SC}}
\def\SX{\mathcal{S}_X}
\def\SO{\mathcal{S}_O}
\def\SOX{\mathcal{S}_{O,X}}
\def\star{^{*}}

\def\trans{^{\intercal}}
\def\var{\textnormal{Var}} % variance

\def\wh{\widehat}
\def\wt{\widetilde}
\def\X{X}
\def\Xsub{X_{\textnormal{Sub}}}% subset of design matrix

\def\0{\boldsymbol{0}}

\def\att{\textnormal{ATT}}
\def\mad{{\textnormal{MAD}}}
\def\mean{{\textnormal{mean}}}
\def\med{{\textnormal{med}}}
\def\trim{{\alpha\textnormal{-trim}}}
\def\ols{{\textnormal{OLS}}}
\def\olssub{\textnormal{OLS,Sub}}

\def\adj{_\textnormal{a}}
\def\unadj{_\textnormal{u}}

\newcommand{\pf}[1]{\noindent \textit{Proof of #1}. }

\newtheorem{theorem}{Theorem}
\newtheorem{definition}{Definition}
\newtheorem{corollary}{Corollary}
\newtheorem{lemma}{Lemma}
\newtheorem{proposition}{Proposition}
\newtheorem{remark}{Remark}

\newtheorem{condition}{Condition}
\newtheorem{assumption}{Assumption}

\newtheorem{example}{Example}
{\end{example}}

\def\np{n_p}
\def\nq{n_q}
\def\meannp{\dfrac{1}{n_p}\sum_{i=\nq+1}^{n}}
\def\meannq{\dfrac{1}{n_q}\sum_{i=1}^{\nq}}
\def\meann{\dfrac{1}{n}\sum_{i=1}^n}
\def\sumnp{\sum_{i=\nq+1}^{n}}
\def\sumnq{\sum_{i=1}^{\nq}}

\def\O{O}
\def\o{o}
\def\POmidX{P_{O\mid X}}% P_{D|X}
\def\PYamidX{P_{Y(a)\mid X}}% P_{D|X}
\def\POX{P_{O,X}}% P_{D,X}
\def\PX{P_{X}}% P_{D,X}

\def\QO{Q_{O}} % Q_{D}
\def\QOgamma{Q_{O}^{\gamma}} % Q_{D}
\def\QOzero{Q_O^0}
\def\QOmidX{Q_{O\mid X}} % Q_{D|X}

\def\QOmidXgamma{Q_{O\mid X}^{\gamma}} % Q_{D|X}^{\gamma}
\def\QOmidXzero{Q_{O\mid X}^0} % Q_{D|X}^0
\def\QOX{Q_{O,X}}% Q_{D,X}
\def\QYaX{Q_{Y(a),X}}% Q_{Y(a),X}
\def\QYamidXgamma{Q_{Y(a)\mid X}^\gamma}% Q_{Y(a)|X}
\def\QYaXzero{Q_{Y(a),X}^0}
\def\QYazero{Q_{Y(a)}^0}
\def\QOXzero{Q_{O,X}^0}
\def\QOXgamma{Q_{O,X}^{\gamma}}

\def\QX{Q_X}% Q_X

\def\wt{\textnormal{W}}
\def\reg{\textnormal{R}}

\def\SI{\textnormal{SLOPE}}
\def\NA{\textnormal{NA}}

\def\LR{\textnormal{LR}}
\def\EB{\textnormal{EB}}

\def\calQ{\mathcal{Q}}

\def\msm{\textnormal{MSM}}
\def\cbm{\textnormal{CBM}}